\documentclass[
 amsmath,amssymb,
 aps,
]{revtex4-2}

\usepackage{graphicx}
\usepackage{dcolumn}
\usepackage{bm}
\usepackage{hyperref}
\usepackage{caption}
\usepackage{subcaption}
\usepackage{adjustbox}
\usepackage{ulem}
\usepackage{amsmath}
\usepackage{multirow}

\begin{document}


\title{Point-particle drag, lift, and torque closure models using machine learning:\\ hierarchical approach and interpretability}

\author{B. Siddani}
\email{siddanib@ufl.edu}
\author{S. Balachandar}%
\email{bala1s@ufl.edu}
\affiliation{Department of Mechanical \& Aerospace Engineering, University of Florida, Gainesville, FL 32611, USA}%


\begin{abstract}
Developing deterministic neighborhood-informed point-particle closure models using machine learning has garnered interest in recent times from dispersed multiphase flow community. The robustness of neural models for this complex multi-body problem is hindered by the availability of particle-resolved data. The present work addresses this unavoidable limitation of data paucity by implementing two strategies: (i) by using a rotation and reflection equivariant neural network and (ii) by pursuing a physics-based hierarchical machine learning approach. The resulting machine learned models are observed to achieve a maximum accuracy of 85\% and 96\% in the prediction of neighbor-induced force and torque fluctuations, respectively, for a wide range of Reynolds number and volume fraction conditions considered. Furthermore, we pursue force and torque network architectures that provide universal prediction spanning a wide range of Reynolds number ($0.25 \leq Re \leq 250$) and particle volume fraction ($0 \leq \phi \leq 0.4$). The hierarchical nature of the approach enables improved prediction of quantities such as streamwise torque, by going beyond binary interactions to include trinary interactions.
\end{abstract}

\maketitle


\section{Introduction} \label{intro}
Dispersed multiphase flows can be described as flow systems that contain discrete elements scattered within a continuous phase. In particle-laden flows, which is one such multiphase flow system, the particles are dispersed in a carrier fluid \citep{balachandar_eaton, doi:10.1146/annurev-fluid-030121-021103}. 
Particle-laden flows of practical interest often involve billions of particles and more. Particle-resolved (PR) simulations of such large particle-laden systems is out of current computational reach. However, only by resolving the interface between the particles and the surrounding fluid, we can obtain the force and torque that couple the two phases. 
The two widely used alternate approaches are the Euler-Lagrange (EL) and Euler-Euler (EE) methodologies. In both these approaches, the microscale details of the flow on the scale of the particle diameter and inter-particle separation distance are not resolved. The advantage is that by sacrificing the information at the small scales, computational attention could be focused on resolving the larger scales and their dynamics. However, the challenge is that first-principle-based access to hydrodynamic force and torque on the particles is lost. EL and EE approaches must therefore rely upon empirical force and torque closure relations. 

In the present work, we will mainly focus attention on the EL approach where all the particles within the system are tracked and therefore the averaging is only over the small scales of the fluid phase. How well the dynamics of the flow and the particles in an EL simulation duplicate the true dynamics of a PR simulation depends only on the accuracy of the force and torque closure models employed in the EL simulation. Furthermore, we will restrict attention to a steady static system, where all the particles are held stationary within the system and the Reynolds number is sufficiently small (smaller than a few hundred) that the flow remains time independent. While there are some direct applications of this restricted configuration in porous media flows and slow-moving inertia-dominated particle-laden flows, our intent is to use the static system as a starting point. Successful demonstration of a machine learning-based framework for the accurate prediction of force and torque in the static system can be extended to more relevant non-static conditions.

Consider a steady static system consisting of $N+1$ particles within a triply periodic domain with a one-dimensional mean flow of Reynolds number $Re$ (based on particle diameter and mean fluid velocity within domain) driven by a steady uniform streamwise pressure gradient. Particle-resolved simulations of this static configuration with $N \sim O(10^2)$ to $O(10^3)$ have been performed by a number of research groups \citep{Beetstra2007, tenneti_drag, bogner2015drag, tang_drag,  akiki_ibm, tafti_ibm, wachs_prs} with the primary purpose of understanding the complex three dimensional flow at the microscale and the modeling of force and torque on the particles.

The modeling of force (and similarly torque) on a particle, which will henceforth be called as the {\it {reference particle}}, has been approached at two different levels. In the traditional {\it {mesoscale approach}}, starting from the pioneering work of \citet{richardson_zaki}, the non-dimensional force is modeled as a function of the mesoscale Reynolds number $Re$ and particle volume fraction $\phi$. Here $Re$ serves to account for the intensity of the flow and $\phi$ accounts for the number density of neighbors as seen by the reference particle. Both $Re$ and $\phi$ are averaged over a large enough volume that particle-scale variations are erased, while retaining mesoscale variation in flow velocity and particle volume fraction. The advantage of the mesoscale approach to modeling is its simplicity - force and torque depend only on the two parameters $Re$ and $\phi$ \citep{Beetstra2007,tenneti_drag,bogner2015drag,tang_drag}. But the simplicity comes at a price. An ensemble of reference particles, all having the same $Re$ and $\phi$, will experience a distribution of forces/torques, and the mesoscale model can only predict the average. Particle-to-particle variation of force can be stochastically accounted for in the mesoscale approach with additional knowledge of the force variance as a function of $Re$ and $\phi$ \citep{akiki_force_variation,wachs_prs,lattanzi_force_variation_static,lattanzi_force_variation_moving}.

The alternate is the {\it{microscale approach}}, where force and torque are modeled in terms of not only the mesoscale variables $Re$ and $\phi$, but also taking into account where the neighbors are located. The advantage of the microscale approach is that we can deterministically account for particle-to-particle variation in force and torque, without just relying on statistical variation. This is accomplished with the additional burden of making the model dependent on the relative location of the neighbors. The number of neighbors $M$ being considered in the microscale model is an important parameter. The complexity of the model as well as its predictive capability will increase with the number of neighbors being considered. 

The microscale modeling approach must properly recognize the multi-body nature of the problem. Even in the Stokes regime, the perturbing influence of any one neighbor on the reference particle is affected by the presence of all other neighbors. At finite $Re$, the interaction between the particles is further complicated by nonlinear effects. Thus, in the microscale approach, the perturbing influence of all the $M$ neighbors must ideally be considered together as an $(M+1)$-body problem. Accounting for such complete multi-body interaction is often hard to accomplish in a model. Following the hierarchical approach pursued in classical physics and molecular dynamics \citep{n_body_water,many_body_potential}, here we hierarchically account for the effect of the $M$ neighbors by (i) considering the perturbing effect of each neighbor individually, (ii) considering the perturbing effect of each pair of neighbors (there are ${}_M C_2$ pairs of neighbors), (iii) considering the perturbing effect of three neighbors at a time (there are ${}_M C_3$ combinations of neighbors), and so on. Since at each level we are concerned with the interaction of one, two, three, or more neighbors with the reference particle, these levels will be termed ``binary'', ``trinary'', ``quaternary'', and so on.

First and foremost, in microscale modeling, force or torque are expressed as functions of $Re$, $\phi$, and $M$ neighbor locations, i.e., depends on $3M+2$ variables. Simple curve fits of force as a function of $Re$ and $\phi$ were sufficient in the case of mesoscale models. The number of independent variables rapidly increases with $M$ and here we resort to modern machine learning techniques to extract the proper statistical relation between force/torque and the independent variables. There has been considerable interest in the deterministic prediction of force and torque using data-driven and deep learning methods.

\citet{akiki_jfm} presented pairwise interaction extended point-particle (PIEP) model that approximates undisturbed flow of a reference particle using superposition of perturbation flow fields induced by the $M$ neighbors taken one at a time. As the name suggests, the PIEP model accounts for only the binary interactions, because it considers each neighbor individually. The undisturbed fields obtained with the PIEP model are then utilized to evaluate particle force using the Fax\'{e}n form of the force relation. The force and torque models under this framework are presented in \citet{akiki_jcp}. The performance of this model decreases considerably at high particle volume fractions. This shortcoming was addressed in \citet{moore_jcp} through a hybrid model that combined physical PIEP method with a data-driven approach, which relies on PR simulations and nonlinear regression. \citet{wachs_prs} also presented a binary interaction approach called microstructure-informed probability-driven point-particle (MPP) model that leverages neighbors' conditional probability density function (PDF) to approximate force and torque fluctuations.

\citet{tafti_ann} used a multilayer perceptron (MLP) that takes $Re$, $\phi$, and relative position of 15 nearest neighbors together as inputs ((15+1)-body model) to predict the drag force acting on a reference particle. \citet{phynet_tafti} showed that inclusion of physical intermediates such as pressure, velocity, shear and pressure forces, and physics-guided loss terms significantly improves network generalizability for this (15+1)-body problem. More recently, \citet{wachs_piep} presented a neural network model that is similar in approach to PIEP model \citep{akiki_jfm}. The neighbors are considered individually instead of the multi-body approach. However, they developed their model to have neighbor-order dependency. The method ensured generalizability in capturing force and torque fluctuations for data paucity applications unlike the multi-body model. \citet{tafti_non_spherical} showed that deep learning can also be successfully applied to predict drag forces on nonspherical particles. 

\citet{moore_superposable} defined the perturbation flow induced by particles at a given $Re$ and $\phi$ to be the \textit{superposable wake}. The flow field around any reference particle can be recreated by adding its own superposable wake and those of each neighbor. \citet{siddani_gan} pursued this flow prediction task using conditional generative adversarial network (GAN) \citep{gan,cgan}. The volumetric representation of the $M$ neighbors ($(M+1)$-body input) and the convolutional neural network (CNN) architecture enabled the approach in reconstructing the complex flow through the random bed of particles far more accurately than the superposable wake approach. The higher accuracy can be attributed to the multi-body nature of GAN prediction over the binary approximation of the PIEP model. This suggests the potential advantage of going beyond binary interactions and including trinary and higher-order interactions among the particles. Similarly, \citet{phyflownet_tafti} presented a physics-guided architecture to predict the flow fields. The task of achieving particle force/torque upon the availability of surface resolved flow field is a linear process. Therefore, an additional neural network (NN) was used to evaluate drag force using these flow fields. The results indicate an improvement over previous methods \citep{tafti_ann,phynet_tafti}.   

The following two additional factors pertaining to machine learning must be considered in the microscale modeling of force and torque. First, the NN's architechture is determined by many attributes, including the number of hidden layers (i.e., the depth), the width of the hidden layers, the form of nonlinearity enforced, and other hyperparameters used in the training of the NN. Collectively we refer to all these details as the {\it{neural network details}}. Second, for any chosen architecture, the neural network must be trained appropriately with ground truth. In the present case of steady stationary particle-laden flow, the ground truth for each reference particle consists of its $Re$, $\phi$, and local information of its $M$ nearest neighbors as input and the hydrodynamic force and torque as the output. The number of reference particles $P$ used in the training process becomes an important parameter. Only with a very large training dataset the network can be properly trained without overfitting. However, since the training data comes from either PR simulations or high-quality experimental measurements, $P$ is typically limited. 

In summary, the performance of a deterministic NN-based microscale model, in terms of predicting the force and torque on individual particles, can be expected to be substantially better than the mesoscale models. Nevertheless, based on above discussion, we can identify four reasons why the predicted force or torque by a microscale model may not perfectly yield the corresponding exact values as obtained in a PR simulation. (i) {\underline{Neighbor-truncation error:}} In the present geometry of a triply periodic box, each reference particle's neighborhood at the microscale is fully determined by the location of its $N$ neighbors. If the NN model is developed based on only the $M$ nearest neighbors, there will be error associated with the truncation $M < N$. (ii) {\underline{Interaction-truncation error:}} The NN model based on $M$ nearest neighbors may not attempt to solve the $(M+1)$-body problem. The model truncation at binary, trinary, or quatenary level will contribute to additional error. (iii) {\underline{Network error:}} The third source of error comes from any limitations of the network architecture and optimization.
We admit that the network architecture and the chosen hyperparameters (including the optimization algorithm) that are being used for training may not be perfect, and the difference contributes to network error. (iv) {\underline{Generalization error:}} Even if all the other errors are eliminated by choosing $M=N$, an $(N+1)$-body interaction, and a perfect NN, the resulting trained model will approach perfection only if there is enormous amount of training data. With a finite training set of $P$ realizations, there will be generalization error.

The above errors are interconnected. For example, as the complexity of the model increases, either with increasing number of neighbors $M$ or with increasing level of multi-body interaction, one needs a more complex network architecture in order to be able to appropriately map the force and torque on the expanding set of input parameters. As the depth, width, and the complexity of the NN increases, the amount of training data $P$ required to train the expanding number trainable-weights increases as well. Thus, with the value of $P$ dictated by the amount of available ground truth, one may have to limit $M$, restrict attention to only binary and trinary interactions, and consider simpler NN architectures. As a result, the predicted model will not be perfect and all four errors will contribute to the overall error.

The present work aims to add to the existing body of knowledge on the development of robust force and torque coupling models appropriate for particle-laden flows using machine learning with the following contributions:

\begin{itemize}
\item The deterministic microscale model, when interpreted as a mapping between the input features consisting of $Re$, $\phi$, and the relative position ${\bf{r}}_j$ of the $j = 1, 2, \cdots, M$ neighbors, must obey the following fundamental symmetries: Galelian and translation invariance, rotation and reflection equivariance. By formulating the problem in the frame attached to the reference particle, Galilean and translation invariance are automatically satisfied. Rotation and reflection equivariance imply that as the configuration of the $M$ neighbors rotate or reflect about the center of the reference particle, the corresponding predicted force and torque must equivalently rotate or reflect as well. Past machine learning models of force and torque have not enforced rotation and reflection equivariance. In the present work, with the use of an equivariant NN framework \citep{e3nn_lib}, we will strictly enforce these symmetries. As demonstrated in the context of flow prediction around a random distribution of particles \citep{siddani_rotation}, imposition of rotation and reflection equivariance not only enforces these fundamental symmetries, but has the great advantage of amplifying the effectiveness of available training data. Thus, enforcing rotation and reflection symmetries in the present problem is of fundamental importance due to the limited number of training samples for the complex modeling task at hand.
\item Past binary interaction modeling efforts have been at fixed values of $Re$ and $\phi$ \citep{moore_jcp,wachs_piep}. Models developed at varying discrete combinations of $Re$ and $\phi$ are then interpolated to make predictions at any intermediate values of $Re$ and $\phi$. The advantage of the machine learning approach is that $Re$ and $\phi$ can be used as parameters in order to develop a universal model that spans a wide range of Reynolds number and volume fraction. The only requirement is that the training data spans the required range of $Re$ and $\phi$.
\item Following the theoretical PIEP modeling effort of \citet{akiki_jfm}, the machine learning modeling effort of \citet{wachs_piep} has clearly demonstrated that force modeling as a complete $M$-body problem is not possible due to limited availability of training data from PR simulations. They further showed the viability of force modeling at the level of binary interaction with available data. Here we will develop a rigorous hierarchical multi-body force/torque modeling framework, following a similar approach pursued in molecular dynamics \citep{n_body_water,many_body_potential}, and use the framework to go beyond binary interaction to consider trinary interactions. We quantitatively illustrate the importance of trinary interactions, especially through the prediction of streamwise torque component, which cannot be captured using a binary model.  
\item The present work will also analyze the limitations of the deterministic microscale modeling arising from limited training data availability. In particular, we will expound on the four different sources of error. Attention will also be paid to establishing the adequacy of network architecture.
\end{itemize}

The rest of the manuscript is arranged as follows: Section \ref{enn} explains mathematical details regarding the strict imposition of rotational and reflectional symmetries in the utilized neural networks. Section \ref{pr_data} entails discussion regarding the PR simulation-data and its preparation for model development. The proposed hierarchical machine learning framework is thoroughly detailed in Section \ref{ml_framework}, and the associated neural network details are mentioned in Section \ref{nn_details}. The performance and analysis of developed models is carried out in Section \ref{results_discussion}. Finally, the interpretability attained through the pursued systematic method is demonstrated in Section \ref{interpret}.

\section{Equivariant neural networks: Symmetry-preserving models} \label{enn}
It is well understood that physical systems, and governing equations describing them satisfy certain symmetries. A closure model that is used to approximate any physical phenomenon is also expected to satisfy these underlying symmetries. Equivariant neural networks are a particular type of NN that by construction preserve certain transformations, for example, rotational and/or reflectional symmetries \citep{tensor_field_networks,3d_steerable_cnn,cg_net_kondor}. As mentioned earlier, the imposition of these symmetries into neural closure models significantly improves the generalizability through innate data-amplification, especially for three-dimensional (3D) spaces. Many such implementations across different applications achieved superior performance \citep{rose_yu_2dsymmetry,siddani_rotation,molecular_wavefunctions,nequip,remus_gnn_flow}. 

The current work leverages basic MLP architecture in developing the particle force and torque models. Furthermore, the input information for the problem at hand is only scalars ($Re$ and $\phi$) and 3D vectors (relative position of neighbors and streamwise direction). Nonlinearity can be applied without any restrictions to scalars as they are rotation and reflection invariant. On the other hand, rotational and reflectional equivariance for vectors (or pseudo-vectors) reduces to ensuring that nonlinearity is applied only to their magnitude and not applied separately to each component. The following constitutive block, shown in \ref{eq:norm-act}, details how $m$ input vectors denoted as $\bm{a}$ are nonlinearly combined into $n$ output vectors ($\bm{b}$) using the so-called norm-based activation \citep{e3nn_lib}
\begin{equation}\label{eq:norm-act}
    \bm{b}_{n} = \frac{\mathcal{F}(|w_{nm}\bm{a}_m|+ d_{n})}{|w_{nm}\bm{a}_m|} \,\, w_{nm}\bm{a}_m \, ,
\end{equation}
where $\bm{w}$, $\bm{d}$, and $\mathcal{F}$ are weights, bias, and activation function, respectively. The NN architectures utilized in this work are based on the above idea and hence preserve rotational and reflectional symmetries.

\section{Particle-resolved simulation-data and curation}\label{pr_data}
Particle-resolved simulation-data used in the present work is gathered from two different sources that generate the random particle distributions and simulate the flow using different computational approaches. The first 6 datasets presented in Table \ref{tab:datasets} are produced using NekIBM solver \citep{nekibm, wang_nekibm}. The remaining 10 datasets are obtained from \citet{wachs_piep}. The incompressible Navier-Stokes equations with no-slip and no-penetration boundary conditions on particle surface are solved to steady state in a triply-periodic box. The reader is directed to \citet{nekibm,wachs_prs} and references therein for a thorough explanation of the considered numerical simulations and setups.   

These PR simulations are performed in a large computational domain of size $O(10d_{p})$, where $d_p$ is particle diameter, in order to consider $O(10^3)$ particles within the domain. 
Due to the triply periodic boundary condition, all the $N$ particles within the computational domain are statistically equivalent. Thus, the first step of the curation process is to make each of the $N$ particle within the domain to be the reference particle and consider a {\it{local volume}} around it for the purposes of microscale modelling. 
The local volume is typically much smaller than the entire computational box and determines the number of $M$ closest neighbors that are within which will be considered in the microscale modeling \citep{wachs_prs, siddani_rotation, siddani_gan}. The location of the $M$ neighbors within the local volume define the microscope environment of the reference particle, while averages within the local volume are used to define the mesoscale $Re$ and $\phi$. Thus, from each steady state solution of the PR simulation, $N$ curated samples centered around each of the particle is extracted.

For the PR simulations of \citet{wachs_piep} a larger sphere of diameter $4d_p$ for $\phi=0.1,0.2$ and $3d_p$ for $\phi=0.4$ was chosen as the local volume to evaluate volume-averaged fluid velocity around each reference particle. For the NekIBM simulations, the local volume is a cube with size $L_{A}$, where $L_{A} = \left[ \frac{(M+1)\pi}{6\phi} \right]^{1/3} d_p$. This local cube size $L_A$ was evaluated using $M=26$ for NekIBM simulated datasets presented in Table \ref{tab:datasets}. The chosen value of $M=26$ is within the range of influential neighbors that make a significant contribution for the considered datasets as mentioned in previous studies \citep{moore_jcp,wachs_prs,wachs_piep}.
In a random distribution of particles, the number of nearest neighbors $M$ can be chosen without restriction. Whereas in simple cubic (SC) and body-centered cubic (BCC) arrangements, to preserve the spatial symmetries, only special values of $M$ are allowed. $M=26$ satisfies this requirement for first, second, and third nearest neighbors in SC and BCC arrangements. The streamwise direction in all these PR simulations is along $+x$ axis. Therefore, the local volume-averaged fluid velocity in every dataset is also approximately along $+x$ with minor deviations in direction and magnitude due to the unique particulate microstructure occurring in each individual local volume.

The gathered data is further curated in the following manner before it is used in the data-driven approach. Particle diameter is chosen as the length scale. The local volume-averaged fluid velocity, $\overline{\bm{u}}_f$, is non-dimensionalized using $\mu/(\rho \: d_{p})$, where $\mu$ is dynamic viscosity, and $\rho$ is fluid density, and therefore, the non-dimensional velocity is the Reynolds number, $Re$. Particle forces are scaled by the Stokes drag, $3\pi \mu \: d_p |\overline{\bm{u}}_f|$ and  particle torques are non-dimensionalized using $\mu\: d_{p}^{2} \: |\overline{\bm{u}}_f|$.

Generally, the quantities of interest (QoI), such as $Re$, $\phi$, and non-dimensional forces and torques, are reported by averaging over the entire computational domain. In contrast, the current work evaluates these QoI based on local volume averages for microscale modeling. Hence, the statistics presented in this work are based on local volume averages. Thus, every reference particle in each PR simulation has its own unique set of QoI: $Re_i$, $\phi_{i}$, $\bm{F}_i$, and $\bm{T}_i$ are respectively the Reynolds number, particle volume fraction, non-dimensional force, and non-dimensional torque of the $i^{th}$ reference particle evaluated based on its local volume-averaged fluid velocity.
For a given dataset, an ensemble average, $\langle \; \rangle$, is defined as an average over all the $N$ particles as follows
\begin{eqnarray}
    & \langle Re \rangle  = \frac{1}{N} \sum\limits_{i}^{N} Re_{i} \, , \quad 
    \langle \phi \rangle  = \frac{1}{N} \sum\limits_{i}^{N} \phi_{i} \, , \nonumber \\
    & \langle \bm{F} \rangle  = \frac{1}{N} \sum\limits_{i}^{N} \bm{F}_{i} \, , \quad 
    \langle \bm{T} \rangle = \frac{1}{N} \sum\limits_{i}^{N} \bm{T}_{i} \, .
\end{eqnarray}
Note that $\langle Re \rangle$ and $\langle \phi \rangle$ will be slightly different from the Reynolds number and volume fraction defined based on the entire computational volume. This is due to the nonuniform sampling of the entire domain using local volumes around the randomly distributed particles.

Particle ensemble averaged quantities for every dataset are presented in Table \ref{tab:datasets}. The statistics along the mean flow direction is different from that along the transverse directions.
The standard deviation of non-dimensional force along the streamwise direction, denoted by $\sigma_{\mathrm{Drag}}$, measures the deviation of the individual particle's non-dimensional drag force from the ensemble average. Also shown in the table are the corresponding standard deviation of force along the transverse direction, represented by $\sigma_{\mathrm{Lift}}$, and the mean transverse force is zero. Similarly, standard deviation of non-dimensional torque along the streamwise and transverse directions are notated as $\sigma_{\mathrm{Torque},\parallel}$ \& $\sigma_{\mathrm{Torque},\perp}$ respectively, with their corresponding mean values approaching zero.
\begin{table}
    \caption{Assimilated datasets used in the current study to develop robust universal point-particle closure models.}
    \label{tab:datasets}
   \begin{center}
   \begin{ruledtabular}
    \begin{tabular}{cccccccc}
        \rule{0pt}{4ex}$\langle Re \rangle$ & $\langle \phi \rangle$ & $| \langle \bm{F} \rangle |$ & $\sigma_{\mathrm{Drag}}$ & $\sigma_{\mathrm{Lift}}$ & $\sigma_{\mathrm{Torque},\parallel}$ & $\sigma_{\mathrm{Torque},\perp}$ & $N$ \\[2ex]
        \hline
        \rule{0pt}{4ex}9.86 & 0.10 & 2.76 & 0.7823 & 0.4661 & 0.2407 & 0.9253 & 1000 \\
        121.36 & 0.10 & 9.55 & 3.1110 & 1.4986 & 0.5043 & 1.3689 & 1000 \\
        6.95 & 0.21 & 4.03 & 0.9581 & 0.7454 & 0.4195 & 1.2987 & 1000\\
        73.40 & 0.21 & 9.51 & 2.7987 & 1.6178 & 0.6441 & 1.7103 & 1000\\
        27.81 & 0.40 & 12.68 & 2.0613 & 1.5028 & 0.6734 & 2.0589 & 1000\\
        73.42 & 0.40 & 19.14 & 3.6087 & 2.4135 & 1.0285 & 2.5955 & 1000\\
        2.20 & 0.10 & 2.63 & 0.5425 & 0.3973 & 0.2375 & 0.9241 & 2984 \\
        10.92 & 0.10 & 3.28 & 0.7351 & 0.4839 & 0.2405 & 0.9838 & 3000\\
        165.96 & 0.10 & 9.48 & 2.4191 & 1.5662 & 0.4339 & 1.1972 & 2984\\
        0.25 & 0.20 & 4.42 & 0.7971 & 0.6737 & 0.4295 & 1.4509 & 3055\\
        2.48 & 0.20 & 4.47 & 0.8116 & 0.6707 & 0.4217 & 1.4465 & 3000\\
        49.78 & 0.20 & 7.72 & 1.8194 & 1.2608 & 0.5078 & 1.6549 & 3055\\
        187.25 & 0.20 & 14.24 & 3.6989 & 2.6873 & 0.8446 & 2.0219 & 3055\\
        3.25 & 0.40 & 11.75 & 2.7535 & 2.1329 & 1.0256 & 2.4789 & 2578\\
        64.35 & 0.40 & 19.67 & 5.0963 & 3.6885 & 1.3063 & 2.9899 & 2578\\
        245.65 & 0.40 & 36.62 & 9.5692 & 7.1215 & 2.0385 & 3.9776 & 2578\\
    \end{tabular}
    \end{ruledtabular}
   \end{center}
\end{table}

\section{Hierarchical Modeling Framework}\label{ml_framework}
The important takeaway from previous deep learning studies of particulate force modeling is that enormous amount of training data from PR simulations is required to achieve a generalizable multi-body model. Thus, a  physics-guided hybrid machine learning approach has the best chance of success, since we are faced with the unavoidable situation of modeling with limited amount of training data. \citet{moore_jcp,wachs_piep} presented machine learning models that consider only binary interactions among the particles and thereby incorporated some simplifying multiphase flow physics in order to cope with the limited data availability.
The current effort builds upon these physics-guided approaches and presents a hierarchical methodology that systematically considers progressively higher-order interactions between the particles in a multi-body system.
Furthermore, the present work ensures that the neural network models to be developed strictly preserve Galilean, scale, and translational invariance, and rotational and reflectional equivariance. By doing so the present models make the optimal use of the limited available data.

\subsection{Hierarchical approach}
We construct the hierarchical approach for a large homogeneous distribution of particles of uniform macroscale volume fraction $\langle \phi \rangle$ subjected to a steady uniform ambient flow of Reynolds number $\langle Re \rangle$. 
The non-dimensional hydrodynamic force acting on the $i^{th}$ reference particle considering all the $N$ neighbors\footnote{In a triply periodic domain being considered in the present work, $N+1$ represents the total number of particles (including the reference particle) within the domain and it is finite. In an unbounded homogeneous system $N \rightarrow \infty$.} It can be represented by the following series expansion:
\begin{eqnarray}\label{eq:force_overall}
    \bm{F}_{i} (\langle Re \rangle,\langle \phi \rangle,\{\bm{r}_{1},\bm{r}_{2},...,\bm{r}_{N}\}) = &  \bm{F}_{\mathrm{1}i}(\langle Re \rangle,\langle \phi \rangle) +
    \sum\limits_{j=1}^{N} \bm{F}_{\mathrm{2}i}(\langle Re \rangle,\langle \phi \rangle,\bm{r}_{j}) + \nonumber \\
   &  \sum\limits_{j=1}^{N-1} \sum\limits_{k=j+1}^{N} \bm{F}_{\mathrm{3}i}(\langle Re \rangle,\langle \phi \rangle,\bm{r}_j,\bm{r}_k ) +
    ... \, .
\end{eqnarray}
The first term on the right hand side, $\bm{F}_{1i}(\langle Re \rangle,\langle \phi \rangle)$, will be referred to as the {\it{unary term}}. It accounts for the effect of the neighboring particles only in a collective sense through the volume fraction $\langle \phi \rangle$. Furthermore, the predicted force of the unary model is identically the same for all particles in a homogeneous system, and hence is the same as the best possible mesoscale model. 

The second term, $\sum_{j=1}^{N} \bm{F}_{2i}(\langle Re \rangle,\langle \phi \rangle,\bm{r}_{j})$, is the {\it{binary interaction term}} as it considers the perturbing influence of one neighbor at a time. In the static monodispersed configuration of present interest, the $j^{th}$ neighbor is taken into account with the distance $\bm{r}_j$ from the reference particle. In a more general context, the $j^{th}$ neighbor's information will also include its size, velocity, and acceleration. The {\it{trinary interaction term}}, $\sum_{j=1}^{N-1} \sum_{k=j+1}^{N} \bm{F}_{3i}(\langle Re \rangle,\langle \phi \rangle,\bm{r}_j,\bm{r}_k)$, accounts for the influence of two neighbor pairs at a time. In the binary and trinary terms, the sums account for all two and three particle combinations that include the reference particle. Likewise, other higher-order terms until the $(N+1)$-body interaction term, $\bm{F}_{(N+1)i}(\langle Re \rangle,\langle \phi \rangle,\{\bm{r}_{1},\bm{r}_{2},...,\bm{r}_{N}\})$ which considers all the $N$ neighbors simultaneously, are considered as part of the hierarchical approach.   

The above representation is complete in the sense that it considers all the $N$ neighbors and accounts for the complete $N+1$ body interaction among the particles. If we obtain the functional dependence of all the unary, binary, trinary, and higher-order terms perfectly, then the resulting model will make perfect prediction and the deterministic estimation $\bm{F}_{i} (\langle Re \rangle,\langle \phi \rangle,\{\bm{r}_{1}, \bm{r}_{2}, \cdots,\bm{r}_{N}\})$ given in the left hand side will indeed be the force on the $i^{th}$ particle. We call this perfect model as the {\it{ideal model}}, since perfection can only be achieved in the idealized sense. In practise, the deterministic prediction will be imperfect due to the four sources of errors discussed in the introduction. Although we provide no formal proof for the existence of the ideal model based on machine learning, we point out that the PR simulations that use conventional CFD can be thought of as the ideal model, since it is capable of extracting the forces accurately given $\langle Re \rangle$, $\langle \phi \rangle$, $\{\bm{r}_{1}, \bm{r}_{2}, \cdots,\bm{r}_{N}\}$.

\subsection{Probabilistic interpretation}
We obtain further insight into the above hierarchical model with the following consideration. Let the distribution of neighbors around the reference particle be characterised by the probability $P_N(\bm{r}_{1},\bm{r}_{2},...,\bm{r}_{N})$, where $P_N(\bm{r}_{1},\bm{r}_{2},...,\bm{r}_{N}) d\bm{r}_1  d\bm{r}_{2},...,d\bm{r}_{N}$ represents the probability of finding the $N$ neighbors in a volume $d\bm{r}_1$ around $\bm{r}_1$, $d\bm{r}_2$ around $\bm{r}_2$, and so on. The unary term can then be expressed as
\begin{equation} \label{eq:4.2}
    \bm{F}_{\mathrm{1}i}(\langle Re \rangle,\langle \phi \rangle) = \int \cdots \int \bm{F}_{i} (\langle Re \rangle,\langle \phi \rangle,\{\bm{r}_{1}, \cdots ,\bm{r}_{N}\}) \, P_N(\bm{r}_{1},\cdots,\bm{r}_{N}) d\bm{r}_1  \cdots d\bm{r}_{N} \, ,
\end{equation}
where there are $N$ integrals and each extend over the entire domain to cover all possible locations of the neighboring particles. The above equation defines the unary term to be the ensemble-averaged force when particles are uniformly distributed according to the stated probability. The above expression was derived by multiplying \eqref{eq:force_overall} with $P_{N}$ and integrating over the position of all the particles. It also requires the following conditions be satisfied
\begin{equation} \label{eq:4.3}
\arraycolsep=1.4pt\def\arraystretch{2.}
\begin{array}{ll}
    \int \bm{F}_{\mathrm{2}i}(\langle Re \rangle,\langle \phi \rangle,\bm{r}_{j}) \, P_1(\bm{r}_j) \, d\bm{r}_j &= 0 \,, \\
    \int \int \bm{F}_{\mathrm{3}i}(\langle Re \rangle,\langle \phi \rangle,\bm{r}_{j}, \bm{r}_k) \, P_2(\bm{r}_j,\bm{r}_k) \, d\bm{r}_j \, d\bm{r}_k & = 0 \, , \\ &\cdots \, ,
\end{array}
\end{equation}
where $P_1(\bm{r}_j)$ is the marginal probability of one particle being at $\bm{r}_j$, while others can be anywhere, and similarly $P_2(\bm{r}_j, \bm{r}_k)$ is the marginal probability of two particles being at $\bm{r}_j$ and $\bm{r}_k$, while others can be anywhere. These two marginal probabilities are expressed as
\begin{equation}
\arraycolsep=1.4pt\def\arraystretch{2.}
\begin{array}{ll}
    P_1(\bm{r}_j) &= \int \cdots \int P(\bm{r}_1 = \bm{r}_j, \bm{r}_2, \cdots, \bm{r}_N) d\bm{r}_2 \cdots d\bm{r}_N \\
    P_2(\bm{r}_j,\bm{r}_k) &= \int \cdots \int P(\bm{r}_1 = \bm{r}_j, \bm{r}_2 = \bm{r}_k, \bm{r}_3, \cdots, \bm{r}_N) d\bm{r}_3 \cdots d\bm{r}_N \, .
\end{array}
\end{equation}

In order to obtain a similar probabilistic expression for the binary term, we first define the following probability
\begin{equation}
    P_{N-1}(\bm{r}_2, \cdots, \bm{r}_N | \bm{r}_1 = \bm{r}_j) = \frac{P(\bm{r}_1 = \bm{r}_j, \bm{r}_2, \cdots, \bm{r}_N)} {P_1(\bm{r}_j)} \, ,
\end{equation}
where the subscript ``$N-1$'' in $P_{N-1}$ indicates that the position of one of the neighbors of the reference particle has been fixed at $\bm{r}_j$ and the probability is over the position of the other $N-1$ neighbors. In the above, the denominator on the right hand side serves to normalize so that integral of $P_{N-1}$ over the domain of all its arguments becomes unity.
The binary term can then be expressed as 
\begin{equation}
\arraycolsep=1.4pt\def\arraystretch{2.2}
\begin{array}{ll}
    \bm{F}_{\mathrm{2}i}(\langle Re \rangle,\langle \phi \rangle, \bm{r}_j) &= \int \cdots \int \bm{F}_{i} (\langle Re \rangle,\langle \phi \rangle,\{\bm{r}_1 = \bm{r}_{j}, \cdots ,\bm{r}_{N}\}) \\ & P_{N-1}(\bm{r}_2, \cdots, \bm{r}_N | \bm{r}_1 = \bm{r}_j) d\bm{r}_2  \cdots d\bm{r}_{N}
    - \bm{F}_{\mathrm{1}i}(\langle Re \rangle,\langle \phi \rangle) \, .
\end{array}
\end{equation}
This offers the interpretation that ${F}_{\mathrm{2}i}(\langle Re \rangle,\langle \phi \rangle, \bm{r}_j)$ is the ensemble-averaged force on the reference particle with one of the neighbor being at $\bm{r}_j$ and the other particles being anywhere else within the domain according to stated probability minus the unary term obtained by averaging over all possible positions of all the particles. Thus, ${F}_{\mathrm{2}i}(\langle Re \rangle,\langle \phi \rangle, \bm{r}_j)$ corresponds to the added effect of fixing one of the neighbor's position to be at the fixed location of $\bm{r}_j$. The summation in the second term of equation \ref{eq:force_overall} thus accounts for the added perturbation effect of each neighbor evaluated from its known position, but taken one at a time.

For the trinary term, we first fix the position of two neighbors at $\bm{r}_j$ and $\bm{r}_k$ and define the probability $P_{N-2}$ of finding all the other neighbors at small volumes around $\bm{r}_3$, $\bm{r}_4$, and so on:
\begin{equation}
    P_{N-2}(\bm{r}_3, \cdots, \bm{r}_N | \bm{r}_1 = \bm{r}_j, \bm{r}_2 = \bm{r}_k) = \frac{P(\bm{r}_1 = \bm{r}_j, \bm{r}_2 = \bm{r}_k, \bm{r}_3, \cdots, \bm{r}_N)} {P_2(\bm{r}_j,\bm{r}_k)} \, ,
\end{equation}
We then multiply equation \ref{eq:force_overall} with the two fixed-neighbor probability $P_{N-2}$ and integrate over the position of all the other $N-2$ neighbors, which yields the relation
\begin{equation}
\arraycolsep=1.4pt\def\arraystretch{2.2}
\begin{array}{ll}
    \bm{F}_{\mathrm{3}i}(\langle Re \rangle,\langle \phi \rangle, \bm{r}_j, \bm{r}_k) &= \int \cdots \int \bm{F}_{i} (\langle Re \rangle,\langle \phi \rangle,\{\bm{r}_1 = \bm{r}_{j}, \bm{r}_2 = \bm{r}_{k},\cdots ,\bm{r}_{N}\}) \\  & P_{N-2}(\bm{r}_3, \cdots, \bm{r}_N | \bm{r}_1 = \bm{r}_j, \bm{r}_2 = \bm{r}_k) d\bm{r}_3  \cdots d\bm{r}_{N}
    - \bm{F}_{\mathrm{1}i}(\langle Re \rangle,\langle \phi \rangle) \\ 
    &- \bm{F}_{\mathrm{2}i}(\langle Re \rangle,\langle \phi \rangle, \bm{r}_j) 
    - \bm{F}_{\mathrm{2}i}(\langle Re \rangle,\langle \phi \rangle, \bm{r}_k)\, .
\end{array}
\end{equation}
The trinary term can be interpreted as the ensemble-averaged drag on the reference particle when two neighbors are fixed at $\bm{r}_j$ and $\bm{r}_k$, while the position of all others are varied according to stated probability minus the unary and binary contributions. In the subtraction, the unary contribution is independent of the known neighbor positions, while the binary contribution are based on the two known neighbor positions.

From the above discussion it is clear that each term of the hierarchical approach systematically adds a layer of additional information that was not included in the previous terms. Thus, the binary term adds the effect of binary interaction between the reference and a neighbor, taken one neighbor at a time. Similarly, the trinary term adds the effect of trinary interaction between the reference and two neighbor, taken one neighbor pair at a time, and so on. It is interesting to note that if the expansion given in equation \ref{eq:force_overall} were to be continued, the last term of the expansion will be $\bm{F}_{(N+1)i}(\langle Re \rangle,\langle \phi \rangle, \bm{r}_j, \bm{r}_k, \cdots)$, without any summation. Functionally it is exactly as complicated as the force $\bm{F}_i$ on the left hand side that we are trying to expand on the right hand side. However, the informational content of this last term of the expansion will likely be negligible, since all the prior terms have accounted for much of the physics. Thus, the real advantage of the hierarchical expansion is that the complexity of multi-particle interaction is progressively increased, so that depending on the situation it can be truncated at any level.

\subsection{Truncation and other approximations}
The ideal model presented in equation \ref{eq:force_overall} is neither necessary nor practical. First and foremost, as we will see below in the results sections, the amount of PR simulation data available for model training is sufficient for the binary term and  inadequate even for the trinary term. Therefore, additional terms in the expansion can be included in the force model only upon obtaining vast amount of additional information. The results to be discussed below show diminishing returns with the inclusion of each additional term and thus for any desired level of accuracy going beyond the first few terms may not be necessary.

Based on these considerations, we introduce the following truncated hierarchical expansion, which will be used for machine learning model
\begin{eqnarray}\label{eq:force_overall-truc}
    \bm{F}_{i}[M_2,M_3,3] = &  \underbrace{\bm{F}_{\mathrm{1}i}(Re_i,\phi_i) +
    \sum\limits_{j=1}^{M_2} \bm{F}_{\mathrm{2}i}(Re_i,\phi_i,\bm{r}_{j})}_{\bm{F}_{i}[M_2,2]} + \nonumber \\
   &  \sum\limits_{j=1}^{M_3-1} \sum\limits_{k=j+1}^{M_3} \bm{F}_{\mathrm{3}i}(Re_i,\phi_i,\bm{r}_j,\bm{r}_k ) \, ,
\end{eqnarray}
where on the left hand size we have dropped presenting the dependence on $(\langle Re \rangle,\langle \phi \rangle,\{\bm{r}_{1}, \bm{r}_{2}, \cdots ,\bm{r}_{N}\})$ explicitly. The notation $\bm{F}_{i}[M_2,M_3,3]$ indicates that it is a trinary model where the binary term is based on $M_2$ neighbors and the trinary term is based on $M_3$ neighbors. Here we recognize the fact that each term of the hierarchical expansion can be based on different set of nearest neighbors. It can be noted that if the trinary term is neglected then the binary model can be denoted as $\bm{F}_{i}[M_2,2]$. 

Clearly, $\bm{F}_{i}[M_2,2]$ and $\bm{F}_{i}[M_2,M_3,3]$ are increasing level of approximations to the ideal model $\bm{F}_{i}$. When compared to the ideal model, the above summation involves two truncations: (i) instead of all the $N$ neighbors, the binary and trinary terms are based on only subsets of $M_2$ and $M_3$ neighbors, and (ii) the expansion itself is truncated, since the quaternary and higher-order terms are neglected. These two truncations introduce error and they can be separated as shown below for the trinary model
\begin{equation}
    \underbrace{\bm{F}_{i}[M_2,M_3,3] - \bm{F}_{i}}_{\mbox{Truncation error}} = \underbrace{\bm{F}_{i}[N,N,3] - \bm{F}_{i}}_{\mbox{Interaction-truncation}} + 
    \underbrace{\bm{F}_{i}[M_2,M_3,3] - \bm{F}[N,N,3]_{i}}_{\mbox{Neighbor-truncation}} \, .
\end{equation}
It can be seen that the interaction-truncation error is due to limiting the expansion to only the first three terms and the neighbor-truncation error is due to limiting the number of neighbors in the binary and trinary terms. As discussed in the introduction, when developing machine learning models of the binary and trinary terms two additional errors, namely network and generalization errors, arise, which we shall consider in the following sections. 

One additional important modification has been made in equation \ref{eq:force_overall-truc} compared to the ideal model given in \ref{eq:force_overall}. On the right hand side, $\langle Re \rangle$ has been replaced by $Re_i$ and $\langle \phi \rangle$ has been replaced by $\phi_i$. The rationale for this change is as follows. The Reynolds number $Re_i$ and volume fraction $\phi_i$ evaluated within the averaging volume around the reference particle will differ from the ensemble-averaged values. The difference depends on the size of the averaging volume chosen around the reference particle. With increasing averaging volume, the difference decreases. This dependence of $Re_i$ and $\phi_i$ on averaging volume has been discussed by \citet{siddani_gan}. The resulting particle-to-particle variation in $Re_i$ and $\phi_i$ in some sense accounts for the interaction of the reference particle with its neighbors in a back handed way. As a result, even the unary mode $\bm{F}_{1i}$ with its dependence on $Re_i$ and $\phi_i$ has in it some aspects of the effect of the specific neighbor interactions with the reference particle. This replacement is clearly ad hoc and has been made mainly as a result of extensive experimentation which consistently showed slight improvement in predictive capability. It should also be pointed out that the dependence of the binary term can be expanded to include $Re_j$ and $\phi_j$ and the dependence of the trinary term can be expanded to include $Re_j$, $\phi_j$, $Re_k$, and $\phi_k$. We have tried this modeling option and with the available data it has not been possible to extract generalizable models and therefore we restrict attention to the model given in equation \ref{eq:force_overall-truc}.

The above force model is duplicated to obtain the corresponding torque model, which is given as
\begin{eqnarray}\label{eq:torque_overall-truc}
    \bm{T}_{i}[M_2,M_3,3] = &  \underbrace{\bm{T}_{\mathrm{1}i}(Re_i,\phi_i) +
    \sum\limits_{j=1}^{M_2} \bm{T}_{\mathrm{2}i}(Re_i,\phi_i,\bm{r}_{j})}_{\bm{T}_{i}[M_2,2]} + \nonumber \\
   &  \sum\limits_{j=1}^{M_3-1} \sum\limits_{k=j+1}^{M_3} \bm{T}_{\mathrm{3}i}(Re_i,\phi_i,\bm{r}_j,\bm{r}_k ) \, .
\end{eqnarray}

\subsection{Unary model}
The current work utilizes the mesoscale force model developed by \citet{tenneti_drag} as the unary force model, which can be expressed as
\begin{eqnarray}\label{eq:force_unary}
    \bm{F}_{\mathrm{1}i}(Re_i,\phi_i) = &  \left[\frac{1+0.15Re_{m,i}^{0.687}}{(1-\phi_i)^2} + \right. \frac{5.81\phi_{i}}{(1-\phi_{i})^2}+\frac{0.48\phi_{i}^{1/3}}{(1-\phi_{i})^3} \nonumber \\
    & \left. +\phi_{i}^{3}Re_{m,i} \left(0.95(1-\phi_i)+\frac{0.61\phi_{i}^{3}}{(1-\phi)}\right)\right] \widehat{\overline{\bm{e}}}_i
\end{eqnarray}
where $Re_{m,i} = Re_{i}(1-\phi_i)$, and $\widehat{\overline{\bm{e}}}_i$ denotes unit vector along the mean flow direction for the $i^{th}$ reference particle. Unary torque model is trivial because the PR simulations performed in a triply-periodic domain is not subjected to any external moments. As a result the mean zero torque has not been presented in Table \ref{tab:datasets}. The unary force model used in this work is thus an empirical fit and does not involve any machine learning. 

An important point about the unary model must be discussed. The above unary model by \citet{tenneti_drag} is based on a certain $N$-particle uniform probability distribution $P_N$. Although it is not explicit known, we assume that there exists an $N$-particle distribution that corresponds to the manner in which the particles were distributed in their PR simulations. If we call this $N$-particle distribution to be $P_{N,Tn}$, then only for this distribution and its marginal one- and two-particle distributions the relations \eqref{eq:4.2} and \eqref{eq:4.3} apply. I.e., only for this particular uniform distribution of particles that obey the probability $P_{N,Tn}$ the ensemble averaged-drag will be the same as the unary model given above. For any other distribution of particles, uniform or otherwise, whose $P_N \ne P_{N,Tn}$, the relation \eqref{eq:4.3} apply and the ensemble-averaged drag will receive contributions from binary, trinary, and higher-order interactions, and thus will differ from the above unary model. 

\section{Neural networks}\label{nn_details}
A \textit{sequential learning process} is adopted to obtain the binary and trinary terms. This means we first develop the best possible binary term before moving onto similar strategy in search of the best trinary term. This process along with other network related aspects used to produce the described results are detailed in this section.

It was stated earlier that a single global microscale model that can accurately predict forces/torques in the presence of substantial mescoscale variation will be pursued. The Reynolds number of the training dataset varies between [0.25, 245.65] and the volume fraction extends over the range [0, 0.4]. The mean drag as represented by the unary model increases with increasing $Re$ and $\phi$. Particle-to-particle force variation, measured as rms drag, has been observed to scale as the mean force \citet{akiki_force_variation, siddani_gan, wachs_prs}. Furthermore, the number of training samples available for each combination of Reynolds number and volume fraction varies. Therefore, we choose the following force loss function ($\mathcal{L}_{\mathrm{Force}}$) for training the global force models.
\begin{equation} \label{eq:force_loss}
    \mathcal{L}_{\mathrm{Force}} = \sum_{\mathrm{datasets}} \frac{\sum\limits_{\mathrm{samples}}\frac{|\bm{F}_{i,\mathrm{PR}}-\bm{F}_{i,\mathrm{NN}}|^2}{|\langle \bm{F} \rangle|^2}} {N}
\end{equation}
where $\bm{F}_{i,\mathrm{PR}}$ denotes the curated ground truth of the $i^{th}$ reference particle obtained from PR simulations and $\bm{F}_{i,\mathrm{NN}}$ is a corresponding neural network prediction, using the unary, binary, and trinary truncation. 

As part of sequential learning process we first set out to achieve the best binary interaction force term, $\bm{F}_{2i}$. Hence, we substitute $\bm{F}_{i}[M_2,2]$ for $\bm{F}_{i,\mathrm{NN}}$ in \ref{eq:force_loss} during the training stage of the binary model.
Once a satisfactory binary term is produced, we set out to achieve the best trinary interaction force term, $\bm{F}_{3i}$ by substituting $\bm{F}_{i}[M_2,M_3,3]$ for $\bm{F}_{i,\mathrm{NN}}$ in \ref{eq:force_loss}. In the modeling of the trinary term, only the machine learning model for $\bm{F}_{3i}$ is optimized, while the already optimized binary model of the previous step is utilized for the binary term.

A similar strategy is employed for the torque models as well. However, the torque loss function ($\mathcal{L}_{\mathrm{Torque}}$) is slightly different. Torque standard deviation along transverse directions, $\sigma_{\mathrm{Torque},\perp}$, which is the larger component is used for normalization.
\begin{equation} \label{eq:torque_loss}
    \mathcal{L}_{\mathrm{Torque}} = \sum_{\mathrm{datasets}} \frac{\sum\limits_{\mathrm{samples}}\frac{|\bm{T}_{i,\mathrm{PR}}-\bm{T}_{i,\mathrm{NN}}|^2}{\sigma_{\mathrm{Torque},\perp}^{2}}} {N}
\end{equation}

\subsection{Binary force network details}
\begin{figure}
    \centering
    \includegraphics[width=0.7\textwidth,keepaspectratio=true]{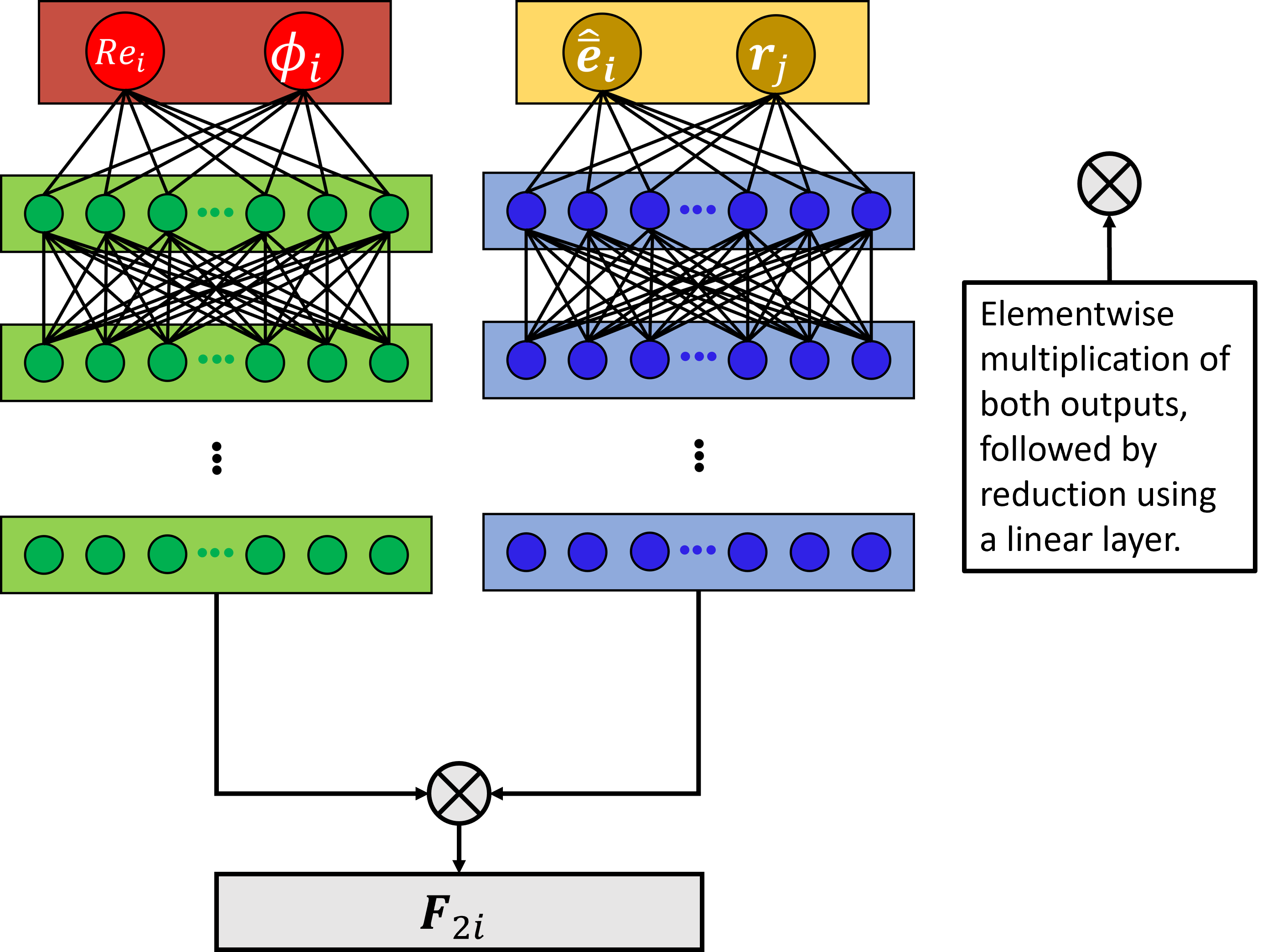}
    \caption{Binary-interaction force model network architecture.}
    \label{fig:binary_force}
\end{figure}

The architectures used in the present work are inspired by \citet{deeponet} for their ability to accurately predict multiscale bubble growth dynamics \citep{deeponet_bubble_1,deeponet_bubble_2}. The architecture of binary force model is shown in Figure \ref{fig:binary_force}. The mesoscale information ($Re_i$, $\phi_i$) is passed to left sub-network. The unit vector along the mean flow direction ($\widehat{\overline{\bm{e}}}_i$) and the neighbor's relative location ($\bm{r}_j$) are passed to the right sub-network. The appropriate mixing of these sub-networks' outputs occurs only at the final stage. As the left sub-network only deals with scalars no special effort is required to preserve rotational and reflectional symmetries. An unbounded activation function, rectified linear unit (ReLU), is used in the left sub-network based on the reasoning that the mean and standard deviations of force are directly proportional to Reynolds number and particle volume fraction. On the contrary, the right sub-network takes in the two vectors ($\widehat{\overline{\bm{e}}}_i, \bm{r}_j$) as input. Thus, the constitutive block given by \ref{eq:norm-act} is implemented to preserve equivariance. Hyperbolic tangent is used as the activation function because of its proven success in achieving accurate results at fixed mesoscale variables \citep{wachs_piep}. The final outputs of left and right sub-networks are scalars and vectors respectively. However, we only need a single force contribution vector, $\bm{F}_{2i}$, as the final binary force model output. Hence, we create equal number of output scalars and vectors from both sub-networks to perform elementwise multiplication \citep{deeponet}. This leaves us with only vectors that are finally combined into a single vector using a linear (vector summation) layer, which is also optimizable.

The following training details are same for every model considered in this work. All models are implemented using PyTorch \citep{pytorch} and \href{https://github.com/e3nn/e3nn}{e3nn} libraries \citep{e3nn_lib} for rotational and reflectional equivariance. A mini-batch size of 100 and ADAM optimizer \citep{adam} with an initial learning rate of 0.001 are used. Learning rate scheduler of type \href{https://pytorch.org/docs/stable/generated/torch.optim.lr_scheduler.ReduceLROnPlateau.html}{ReduceLROnPlateau}, which reduces the learning rate by a factor of 0.1 if the total training loss does not reduce in 10 consecutive epochs, is also availed. Default weight initialization of PyTorch and e3nn is used for the network layers, and these computations were carried out on a single NVIDIA A100 GPU.
The process of $k$-fold cross-validation \citep{k_fold}, with $k=5$, was adopted to produce the model test performance. In a single round of the $k$-fold cross-validation procedure, 20\% of every dataset will used in creating the total \textit{test samples}. Out of the remaining 80\% in each dataset 64\% is used in training a network and the remaining 16\% is used as part of \textit{validation samples}. Thus, for each $\langle Re \rangle$ and $\langle \phi \rangle$ combination given in Table \ref{tab:datasets}, the total number of data samples $N$ was divided into training, validation, and testing data samples (i.e, into $N_{trn}$, $N_{val}$, and $N_{tst}$). In the 5-fold cross-validation, by using different 20\% as test sample in each fold, the entire dataset is included in the testing process.  Every network is trained for a minimum of 100 epochs and the maximum number of epochs is limited to 1000. The training process is stopped (early stopping) if the total validation loss does not reduce further in 50 consecutive epochs.              

\begin{table}
    \caption{Grid search hyperparameter tuning for $\bm{F}_{2i}$ with $M_2 = 26$: Force loss ($\mathcal{L}_{\mathrm{Force, test}}$) evaluated for the test data using different network architectures.}
    \label{tab:binary_force_hypertune}
    \begin{center}
    \begin{ruledtabular}
    \begin{tabular}{c|cc}
       \rule{0pt}{4ex}Nonlinear layers ($L_2$)& \multicolumn{2}{c}{Width of each layer ($W_2$)}\\
          & 50 & 100 \\[2ex] \hline
        \rule{0pt}{4ex}2 & 0.6447 & 0.5688 \\
        3 & 0.5614 & 0.5285  \\
        4 & 0.5330 & 0.5134 \\
        5 & 0.5165 & 0.5066 \\
        6 & 0.5164 & 0.5104 \\
    \end{tabular}
    \end{ruledtabular}
    \end{center}
\end{table}

Table \ref{tab:binary_force_hypertune} shows the performance of different binary force networks on the test samples measured in terms of the force loss. The reported values are a collective representation of five independent versions of the network, which is a consequence of the $(k=5)$-fold cross-validation procedure. The rows indicate the number of nonlinear layers ($L_2$) used in each sub-network and the columns correspond to the width ($W_2$) of the nonlinear layer. For a layer of $W_2 = 50$ the left sub-network (Figure \ref{fig:binary_force}) has 50 scalars and the right sub-network has 50 vectors. From the performed grid search hyperparameter tuning, $(L_2,W_2) = (6,50)$ was chosen for the binary force model as its performance is comparable with that of (5,100) and (6, 100). The difference in $\mathcal{L}_{\mathrm{Force, test}}$ is less than 0.01. Furthermore, the selected model has significantly fewer parameters. For instance, the (6,50) network has 25850 trainable parameters whereas (5,100) network has 81500 trainable parameters. These tests were performed by including 26 neighbors in the binary force model (i.e., $M_2 = 26$).

\subsection{Trinary force network details} 
\begin{figure}
    \centering
    \includegraphics[width=0.7\textwidth,keepaspectratio=true]{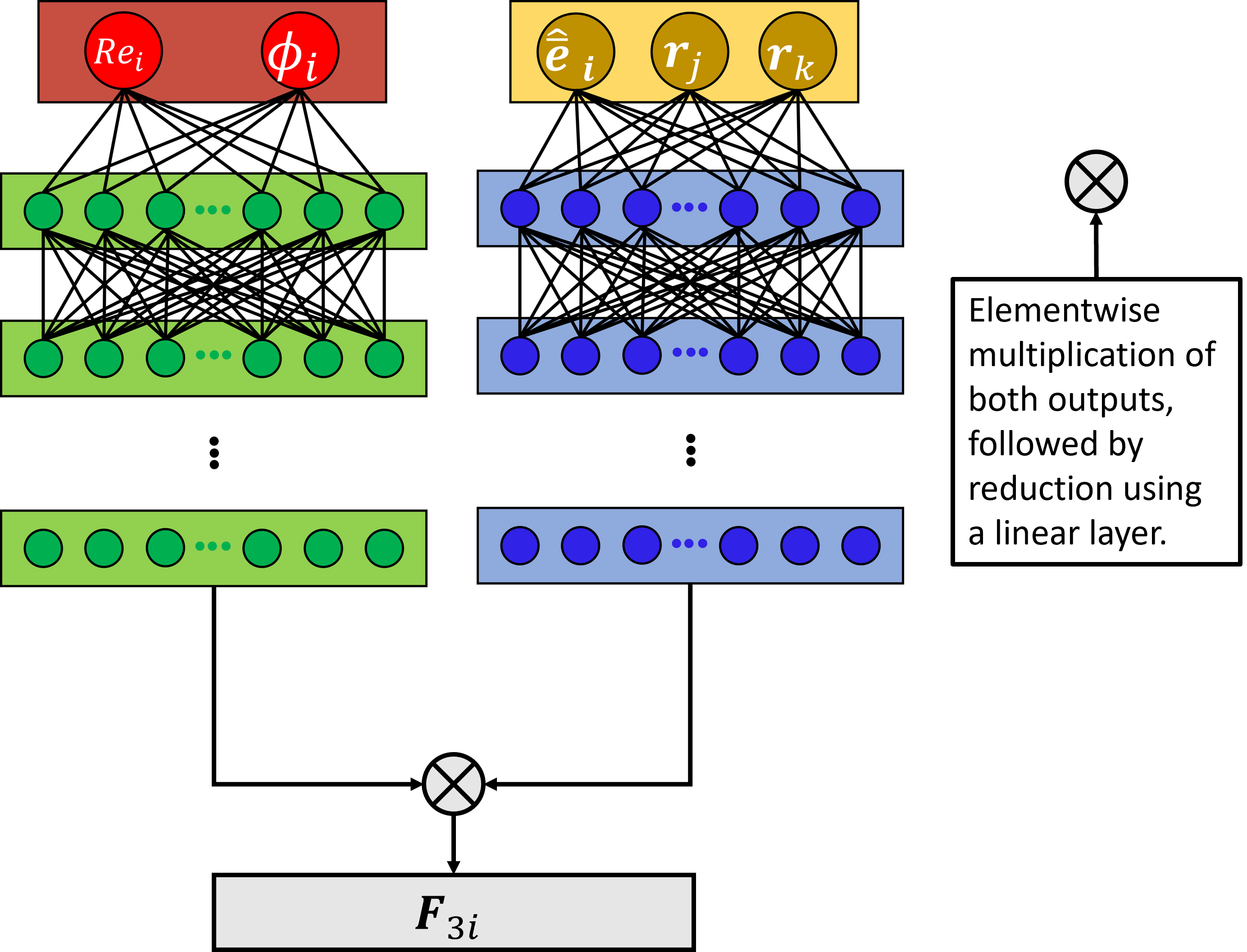}
    \caption{Trinary-interaction force model network architecture.}
    \label{fig:trinary_force}
\end{figure}

The training, validation, and testing partitioning of the dataset for the trinary model is the same as that used in the binary force model. The architecture of the trinary force model (Figure \ref{fig:trinary_force}) is similar to its binary counterpart. With an intention to have a trinary network that has equal or higher expressiveness, the hyperparameter tuning, shown in Table \ref{tab:trinary_force_hypertune}, is started with the number of layers and their width to be $(L_3,W_3) = (6,50)$. Moreover, the hyperparameter tuning is performed using only 10 nearest neighbors (i,e, $M_3 = 10$), which yields ${}_{10} C_2 = 45$ pairs of neighbor combinations. This decision has been taken to substantially reduce the computational cost, since including 26 neighbors as in the binary model would result in ${}_{26} C_2 = 325$ neighbor pair combinations. Finally, network with $(L_3,W_3) = (10,50)$ is chosen as the optimal trinary force model for the given training data.

\begin{table}
    \caption{Grid search hyperparameter tuning for $\bm{F}_{3i}$ with $M_3 = 10$: Test samples force loss ($\mathcal{L}_{\mathrm{Force, test}}$) values for different architectures.}
    \label{tab:trinary_force_hypertune}
    \begin{center}
    \begin{ruledtabular}
    \begin{tabular}{c|cc}
       \rule{0pt}{4ex}Nonlinear layers ($L_3$) & \multicolumn{2}{c}{Width of each layer ($W_3$)}\\
        & 50 & 100 \\[2ex] \hline
       \rule{0pt}{4ex}6 & 0.4891 & 0.4868 \\
       10 & 0.4767 & 0.4965 \\
       15 & 0.4858 & 0.4961 \\
    \end{tabular}
    \end{ruledtabular}
    \end{center}
\end{table}

\subsection{Torque network details}
\begin{figure}
    \centering
    \includegraphics[width=0.9\textwidth,keepaspectratio=true]{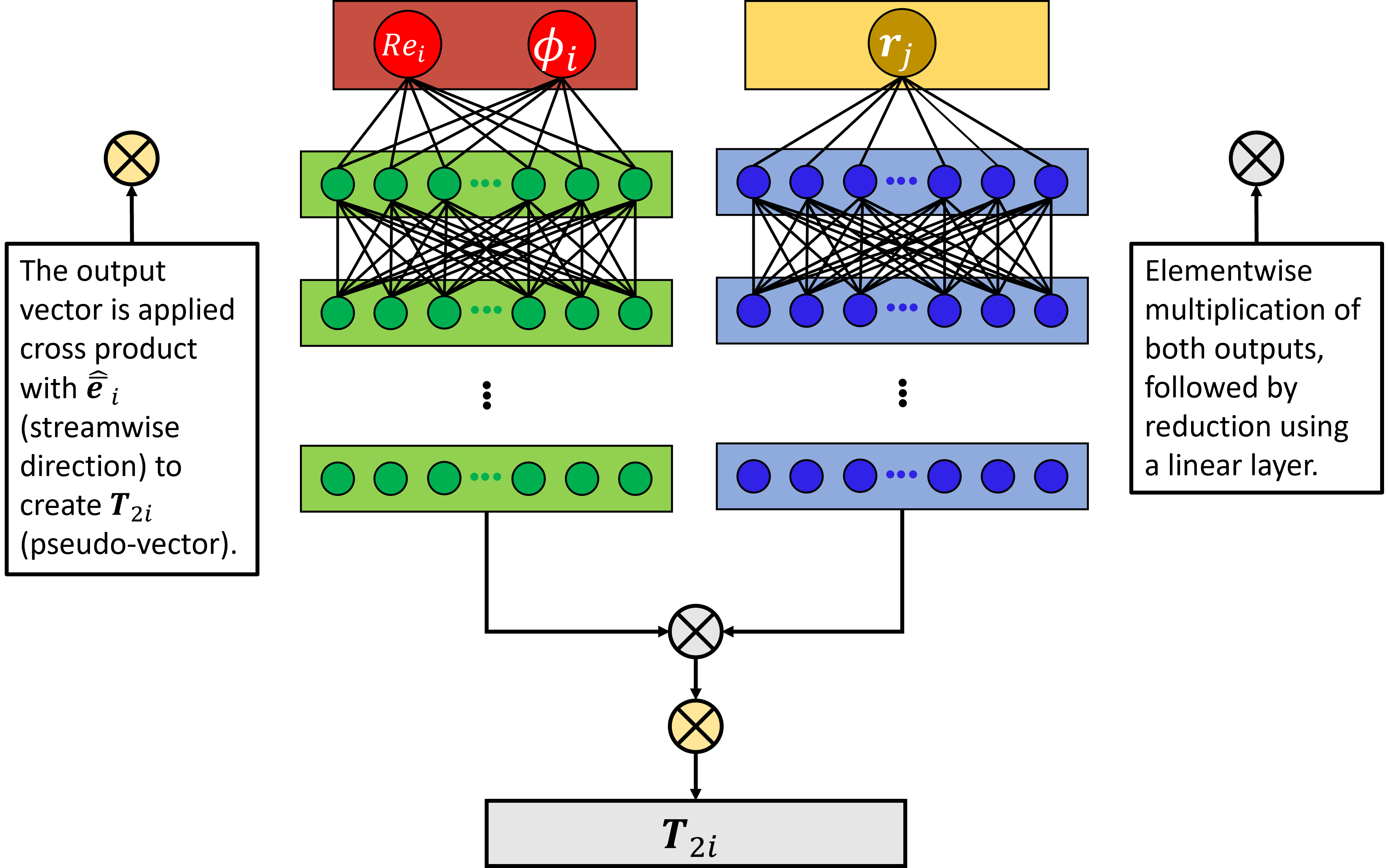}
    \caption{Binary-interaction torque model network architecture.}
    \label{fig:binary_torque}
\end{figure}

While the input Reynolds number, volume fraction, and relative location of neighbor information comprise only of scalars and vectors, the torque on the reference particle as one of the desired outputs is a pseudo-vector. Therefore, special manipulation is necessary to preserve rotational and reflectional equivariance. Binary torque architecture presented in Figure \ref{fig:binary_torque} illustrates the methodology used in the creation of a rotation and reflection equivariant pseudo-vector output. The left sub-network takes $Re_i$ \& $\phi_i$ as inputs and it uses ReLU as the activation function. The right sub-network takes the position of a neighbor as input and uses hyperbolic tangent for nonlinear manipulation of intermediate vector magnitudes. Similar to the force models, the sub-networks' outputs are combined at the final stage to create a single vector. Cross product between two vectors results in a pseudo-vector. This operation also preserves rotational and reflectional symmetries. Therefore, the mean flow direction unit vector, $\widehat{\overline{\bm{e}}}_i$, is involved in cross product with the output vector to create binary torque contribution, $\bm{T}_{2i}$. Based on hyperparameter tuning tabulated in Table \ref{tab:binary_torque_hypertune}, a network with $(L_2,W_2) = (5,100)$ is chosen as the final binary torque model. 

\begin{table}
    \caption{Grid search hyperparameter tuning for $\bm{T}_{2i}$ with $M_2=26$: Test samples torque loss ($\mathcal{L}_{\mathrm{Torque,test}}$) values for different networks.}
    \label{tab:binary_torque_hypertune}
    \begin{center}
    \begin{ruledtabular}
    \begin{tabular}{c|cccc}
       \rule{0pt}{4ex}Nonlinear Layers ($L_2$) & \multicolumn{4}{c}{Width of each layer ($W_2$)}\\
        & 50 & 100 & 150 & 200 \\[2ex] \hline
       \rule{0pt}{4ex}2 & 13.4651 & 12.9478 & 12.9741 & 13.1131 \\
       3 & 12.9238 & 12.3638 & 12.4767 & 12.4124 \\
       4 & 13.5391 & 12.3857 & 12.4419 & 12.4785\\
       5 & 14.7822 & 12.3424 & 12.3664 & 12.4092\\
       6 & 15.3737 & 12.4346 & 12.6441 & 12.6697\\
    \end{tabular}
    \end{ruledtabular}
    \end{center}
\end{table}

\begin{figure}
    \centering
    \includegraphics[width=\textwidth,keepaspectratio=true]{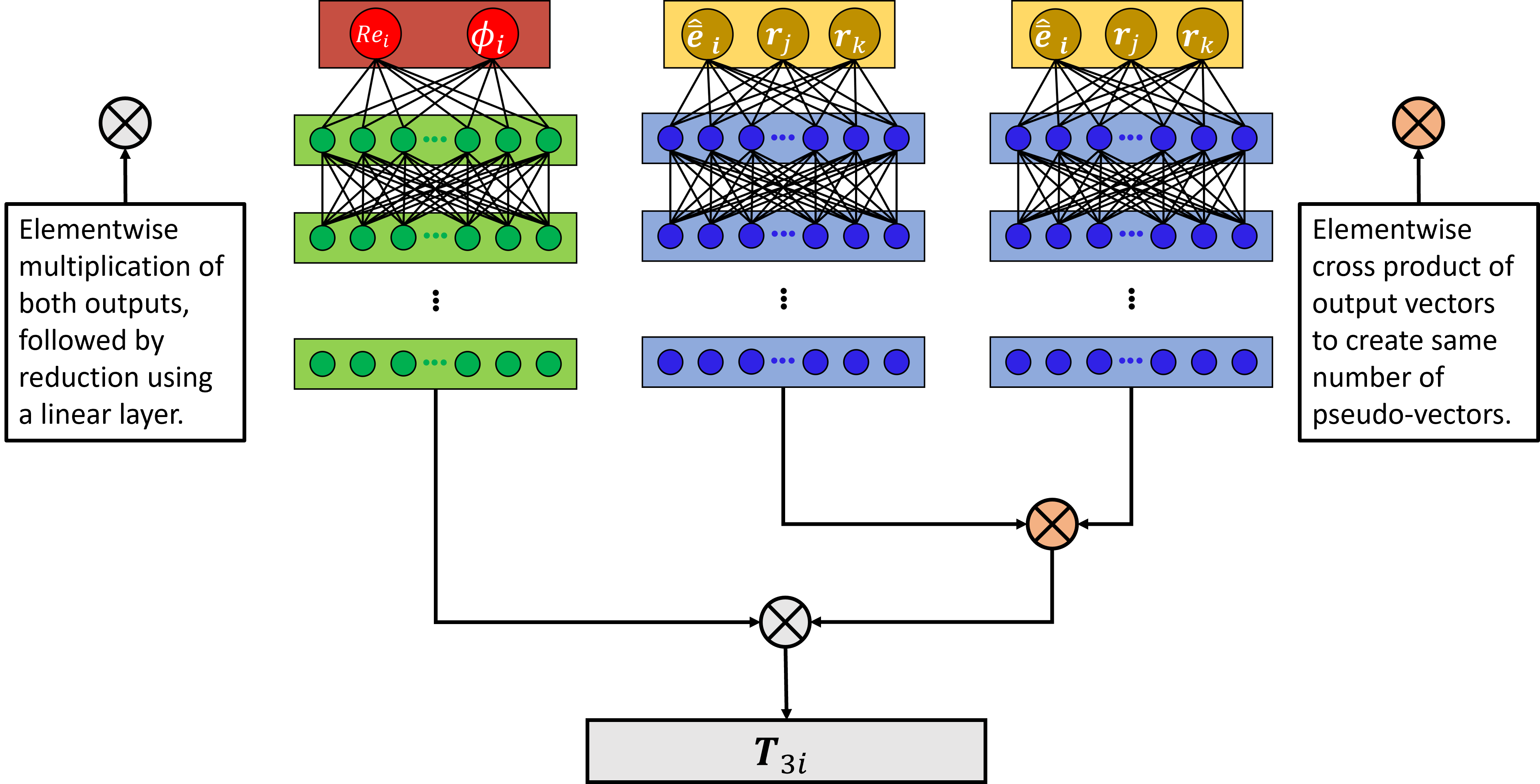}
    \caption{Trinary-interaction torque model network architecture.}
    \label{fig:trinary_torque}
\end{figure}

A binary torque model will not accurately predict the streamwise component of torque \citep{akiki_jcp,wachs_piep} due to the two-body flow symmetry. This signifies the importance of including trinary interactions in the torque model. The architecture of the trinary torque model considered in this work is illustrated in Figure \ref{fig:trinary_torque}. The network contains three different sub-networks. The left sub-network takes mesoscale information (scalars) as inputs and produces scalars as outputs. The activation function used to apply nonlinearity is ReLU. The middle and right sub-networks are identical in terms of their architecture. The inputs are the vectors $\widehat{\overline{\bm{e}}}_i, \bm{r}_j,$ and $\bm{r}_k$ and hyperbolic tangent is used as the activation function. Elementwise cross product is applied among the final output vectors of middle and right sub-networks, thereby creating pseudo-vectors. The resulting pseudo-vectors are involved in an elementwise multiplication with the scalar outputs of the left sub-network in the succeeding step. These final pseudo-vectors are then combined together with the aid of a linear (pseudo-vector summation) layer to produce the trinary torque contribution, $\bm{T}_{3i}$. Similar to trinary force model, the hyperparameter tuning was only performed using $M_3=10$ neighbors. Based on results presented in Table \ref{tab:trinary_torque_hypertune}, a network with $(L_3, W_3) = (5,150)$ was selected as the trinary torque model.

\begin{table}
    \caption{Grid search hyperparameter tuning for $\bm{T}_{3i}$ with $M_3=10$: Test samples torque loss ($\mathcal{L}_{\mathrm{Torque,test}}$) values for different network architectures.}
    \label{tab:trinary_torque_hypertune}
    \begin{center}
    \begin{ruledtabular}
    \begin{tabular}{c|cccc}
       \rule{0pt}{4ex}Nonlinear layers ($L_3$) & \multicolumn{4}{c}{Width of each layer ($W_3$)}\\
       & 50 & 100 & 150 & 200\\[2ex] \hline
       \rule{0pt}{4ex}5 & 6.4323 & 6.4714 & 6.4048 & 6.6055\\
       10 & 6.5684 & 6.6677 & 6.6741 & 6.7255\\
       15 & 6.5978 & 6.8225 & 6.7167 & 6.8780\\
    \end{tabular}
    \end{ruledtabular}
    \end{center}
\end{table}

\subsection{Errors in machine learning process} 
In order to quantify the network and generalization errors incurred during the training and deployment of the binary and trinary machine learning models, we consider $\bm{F}[M_2,2]$ and $\bm{F}[M_2,M_3,3]$ to be the best-trained binary and trinary terms that result in zero network and generalization errors. In other words, $\bm{F}[M_2,2]$ and $\bm{F}[M_2,M_3,3]$ are the best-trained models that result from perfect neural networks that were trained with unlimited training data. However, as discussed in the above subsections, actual implementations are limited both in terms of the network architecture and the amount of training data. The number of layers and their width (i.e., $L$ and $W$) are the network architecture limitations and the number of data samples $N$ is the training data limitation. As a result of these limitations, the resulting binary and trinary models will involve additional errors and will be distinguished from the best-trained models. We now introduce the notation $\tilde{\bm{F}}[M_2,2;L_2,W_2;N]$ and $\tilde{\bm{F}}[M_2,M_3,3;L_2,W_2,L_3,W_3;N]$, where the tilde denotes the fact that the force prediction is based on a finite training dataset and network architecture and thus involves additional errors. Furthermore, the details of the network and the number of data samples used in training/validation/testing is also included. 

With this notation, the network and generalization errors of the trinary model can be expressed as
\begin{equation}
\arraycolsep=1.4pt\def\arraystretch{2.2}
\begin{array} {ll}
    &\underbrace{ \bm{F}[M_2,M_3,3] - \tilde{\bm{F}}[M_2,M_3,3;L_2,W_2,L_3,W_3,\infty]}_{\mbox{Network error}} \\
    &\underbrace{ \tilde{\bm{F}}[M_2,M_3,3;L_2,W_2,L_3,W_3,\infty] - \tilde{\bm{F}}[M_2,M_3,3;L_2,W_2,L_3,W_3,N]}_{\mbox{Generalization error}}
    \end{array}
\end{equation}
According to this notation, $\bm{F}[M_2,M_3,3]$ can be interpreted as $\tilde{\bm{F}}$ in the limit $N \rightarrow \infty$, and with infinitely deep and wide (expressive) network layer(s). We now conclude this section with the following remarks. The above notations and errors also apply to the binary and trinary torque models.
It should also be noted that the network and the generalization errors are connected. For example, with limited availability of training data it does not make sense to employ a very complicated network architecture.
Henceforth in our discussion of the results, we will not use the complete notation presented above. We will simply denote the force and torque resulting from the binary model to be $\tilde{\bm{F}}[2]$ and $\tilde{\bm{T}}[2]$ and the force and torque resulting from the trinary model to be $\tilde{\bm{F}}[3]$ and $\tilde{\bm{T}}[3]$.

\section{Results and Discussion}\label{results_discussion}
The performance of neural network architectures described in the previous section will be examined below. Several decisions in terms of number of neighbors to be included in the binary and trinary models, the depth and width of the network layers must be optimally made. These decisions will be made based on a compromise between computational difficulty and error minimization. Particular attention will be focused on the four different sources of errors. Neighbor-truncation error will determine the values of $M_2$ and $M_3$. The choice of $M_2=26$ and $M_3=10$ will be discussed in the appendix. The interaction-truncation error will be evaluated by comparing the performance of unary, binary, and trinary models of force and torque, which will be presented in the following sections. The network error has already been considered in the previous section where we have determined the optimal values of the layer depth and width for the different neural networks.

\subsection{Accurate universal predictions}
In a typical EL simulation, even under nominally homogeneous conditions, the mesoscale quantities $Re_i$ and $\phi_i$ evaluated at the particles will show substantial variation across the computational domain. This mesoscale variation is in addition to variations seen in the random arrangement of neighbors at the microscale. Therefore, here we have developed machine learning-based deterministic models that can effectively handle such micro and mesoscale variations. When deep learning closure models are developed only at fixed ($Re$, $\phi$) there exists an additional task of switching between different networks and interpolating between the different models. A robust universal model that spans a wide range of $Re$ and $\phi$ will mitigate this issue.

\begin{figure}
    \centering
    \begin{subfigure}[b]{\textwidth}
        \centering
        \includegraphics[width=0.32\textwidth,keepaspectratio=true]{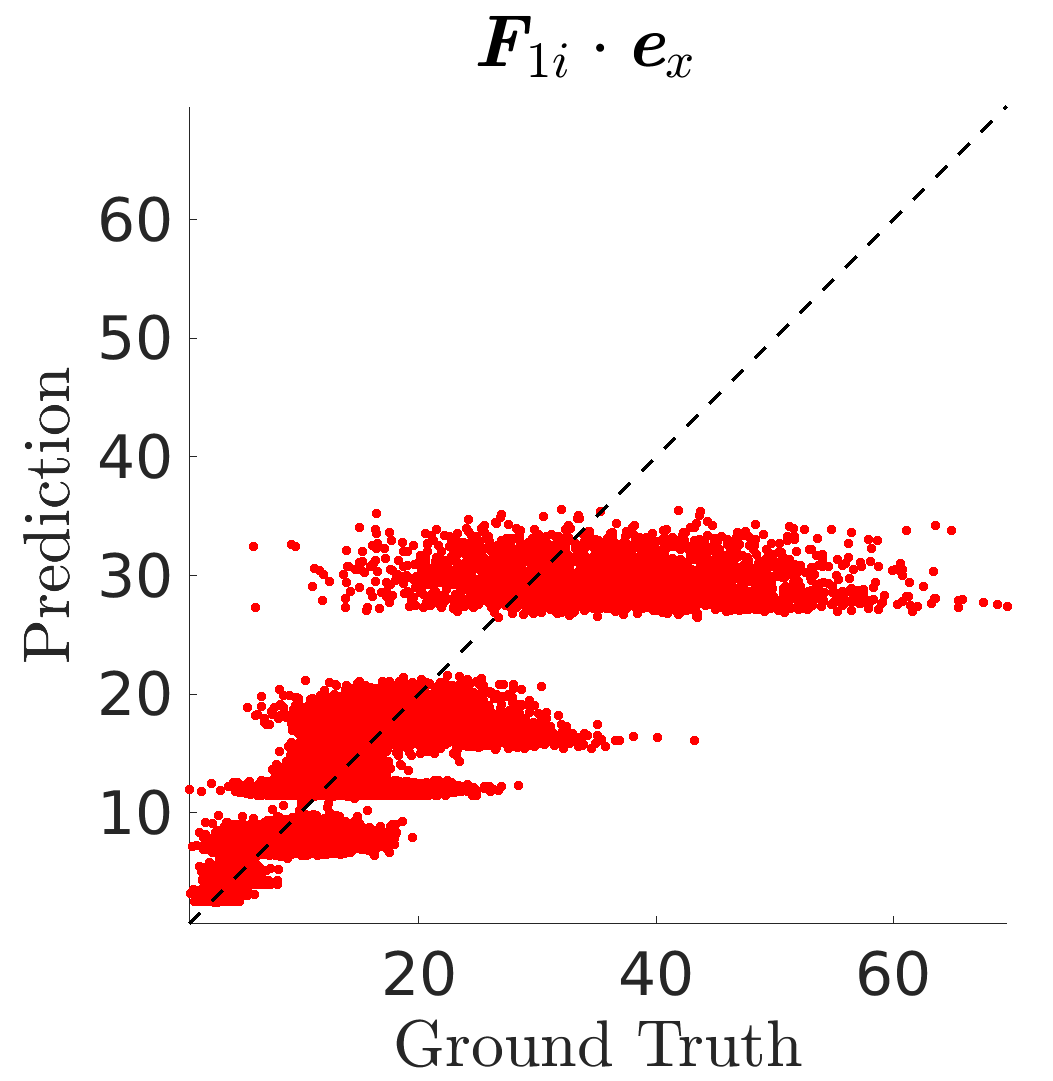}
        \includegraphics[width=0.32\textwidth,keepaspectratio=true]{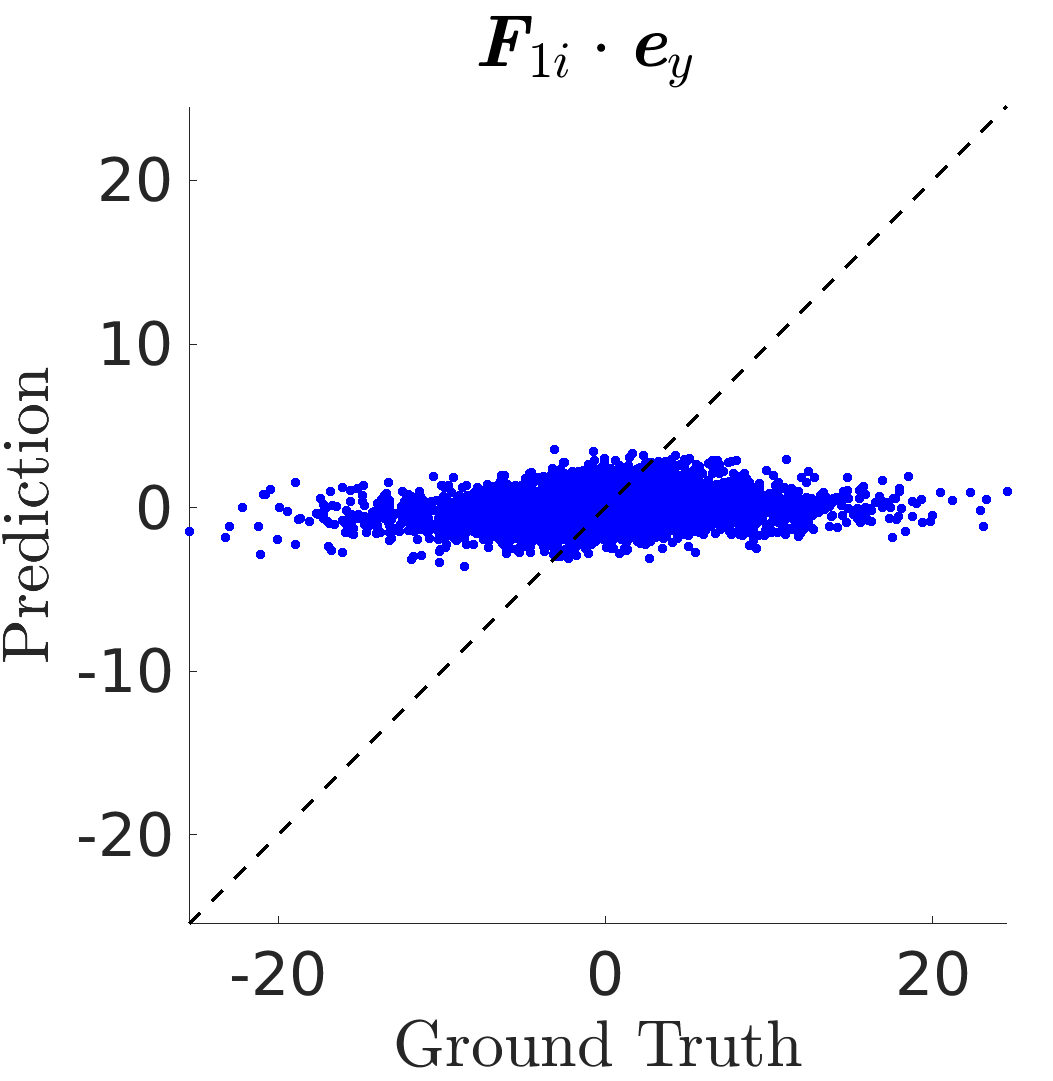}
        \includegraphics[width=0.32\textwidth,keepaspectratio=true]{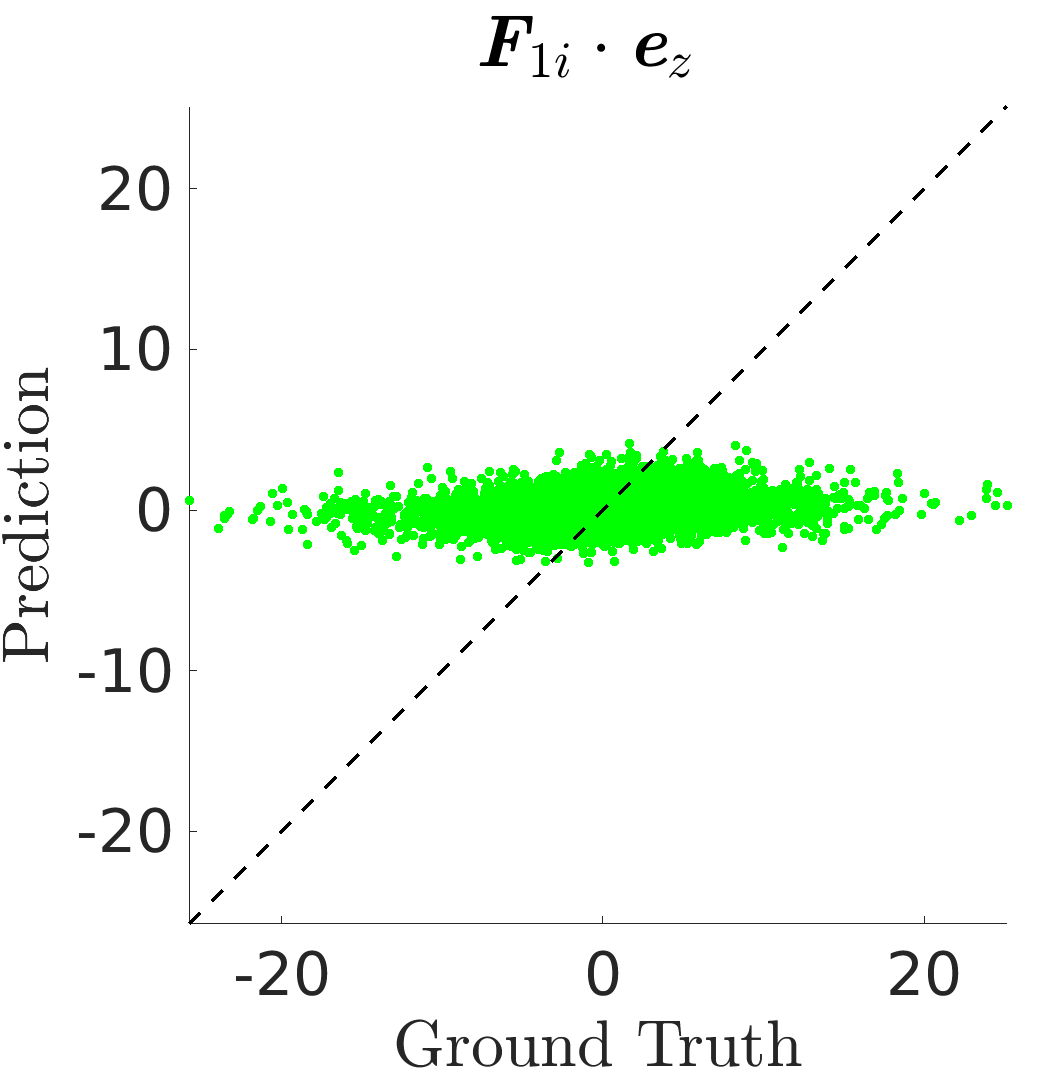}
    \end{subfigure}
    \begin{subfigure}[b]{\textwidth}
        \centering
        \includegraphics[width=0.32\textwidth,keepaspectratio=true]{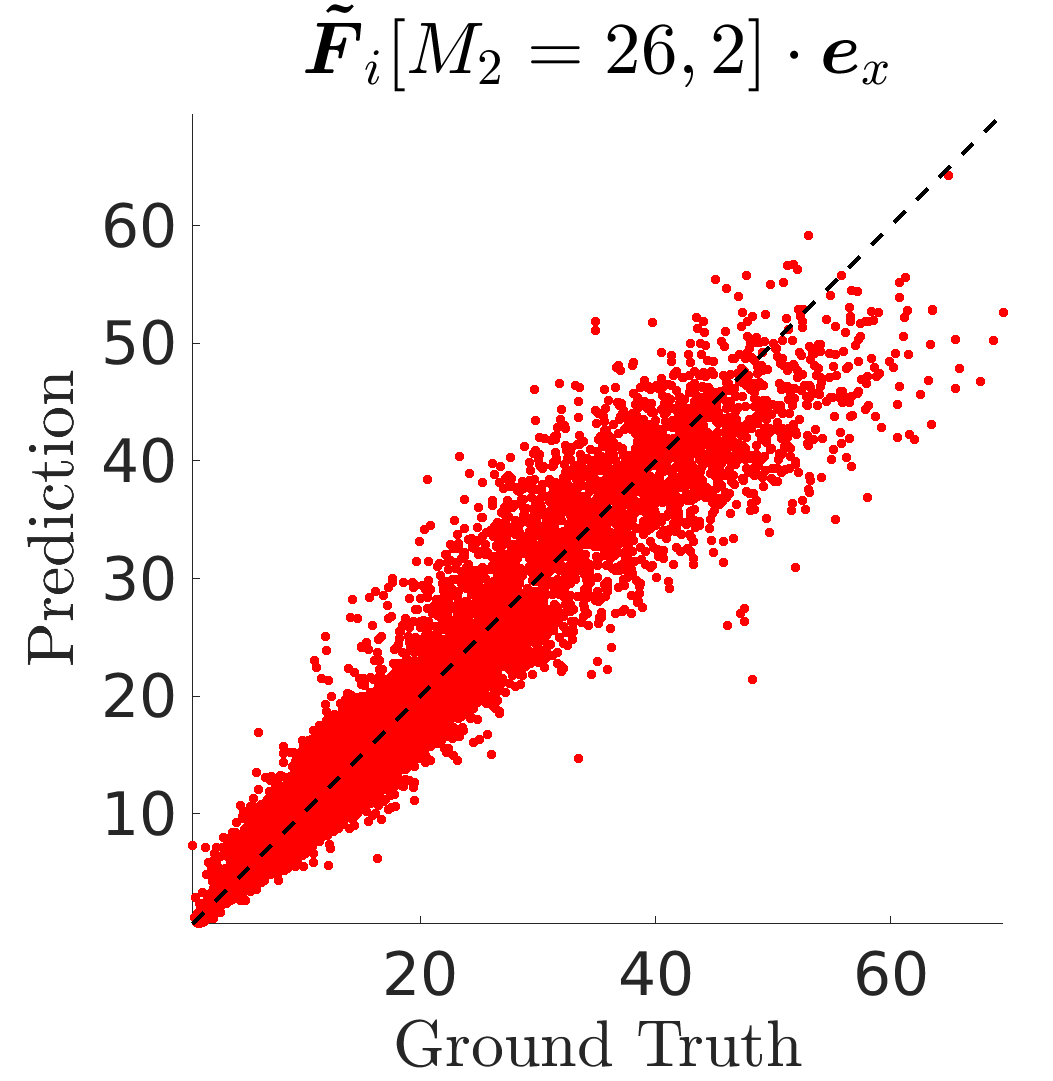}
        \includegraphics[width=0.32\textwidth,keepaspectratio=true]{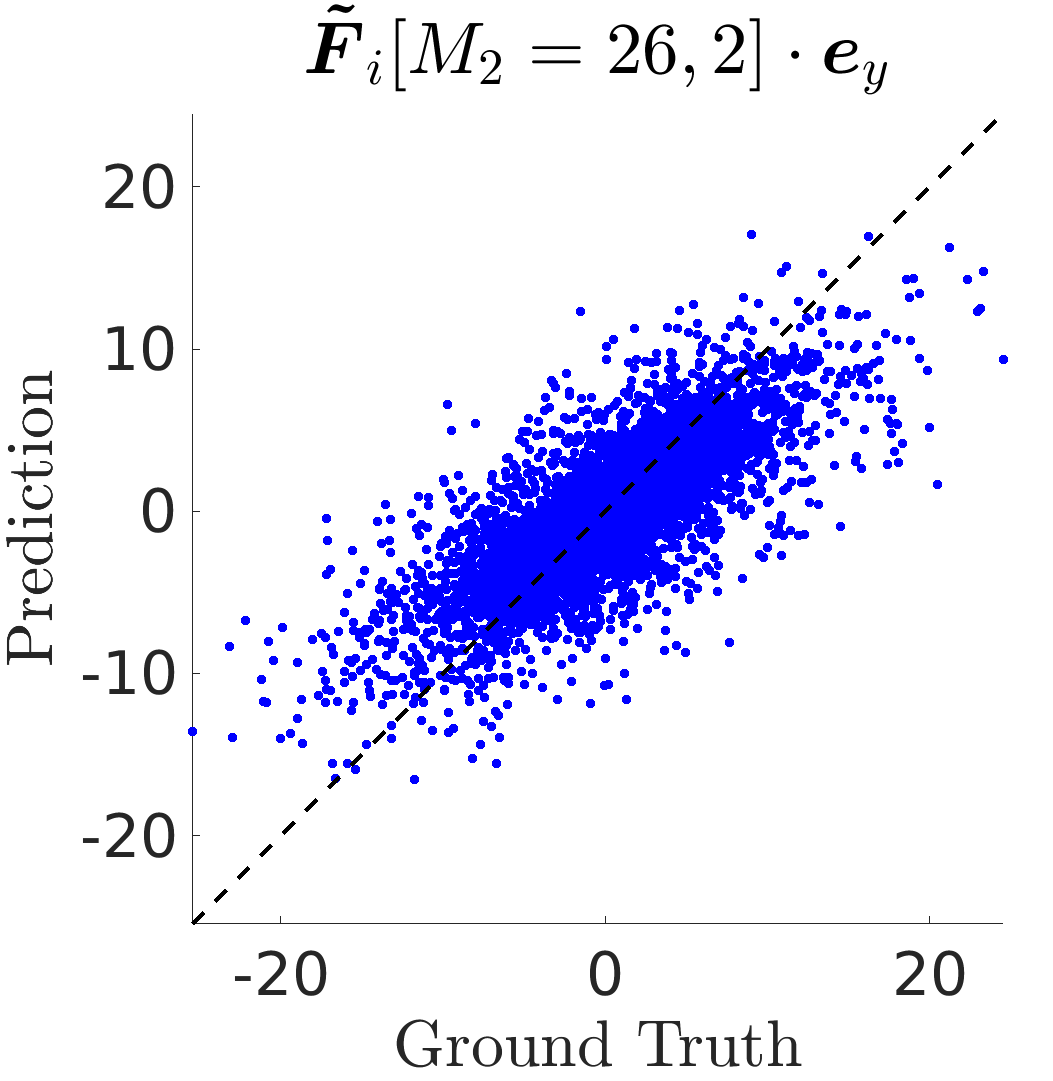}
        \includegraphics[width=0.32\textwidth,keepaspectratio=true]{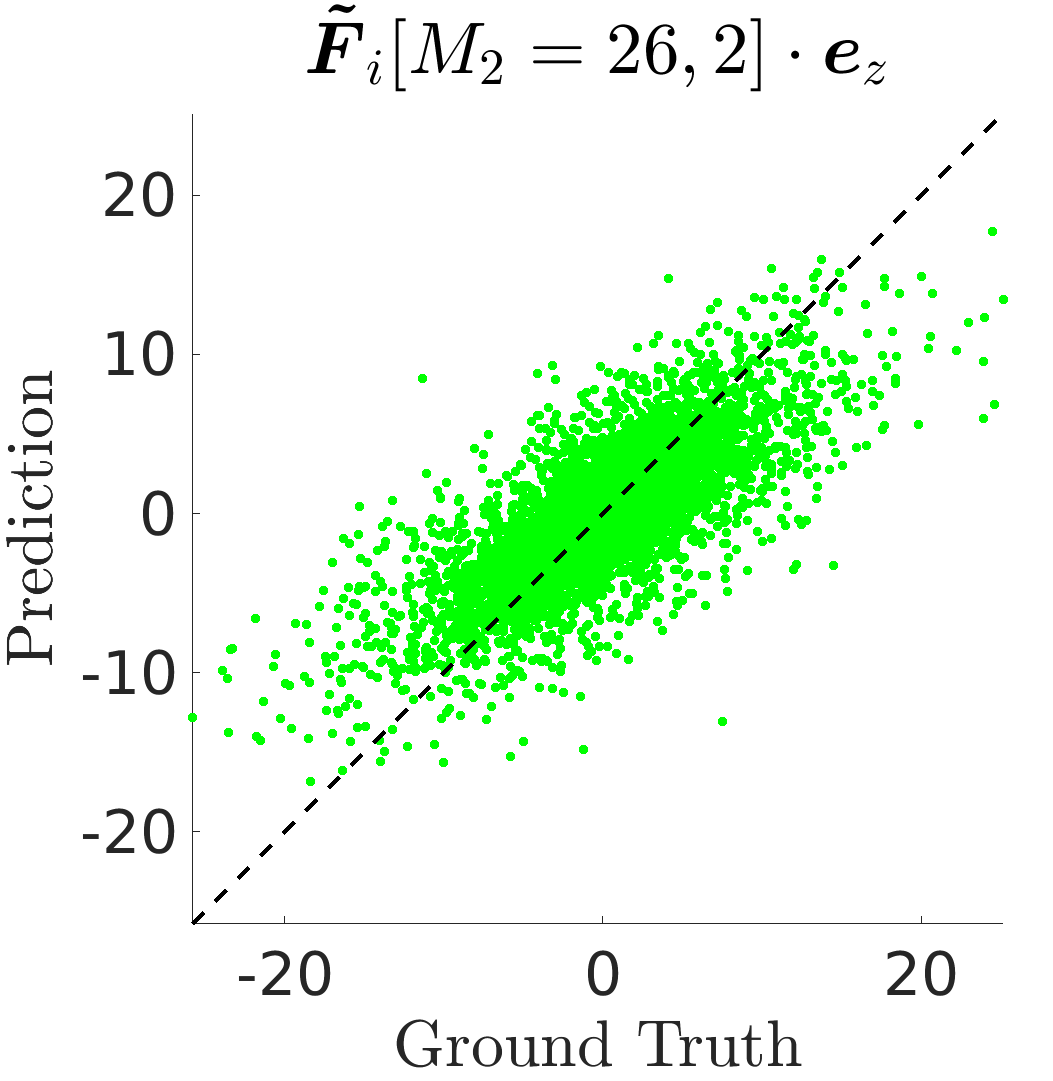}
    \end{subfigure}
    \begin{subfigure}[b]{\textwidth}
        \centering
        \includegraphics[width=0.32\textwidth,keepaspectratio=true]{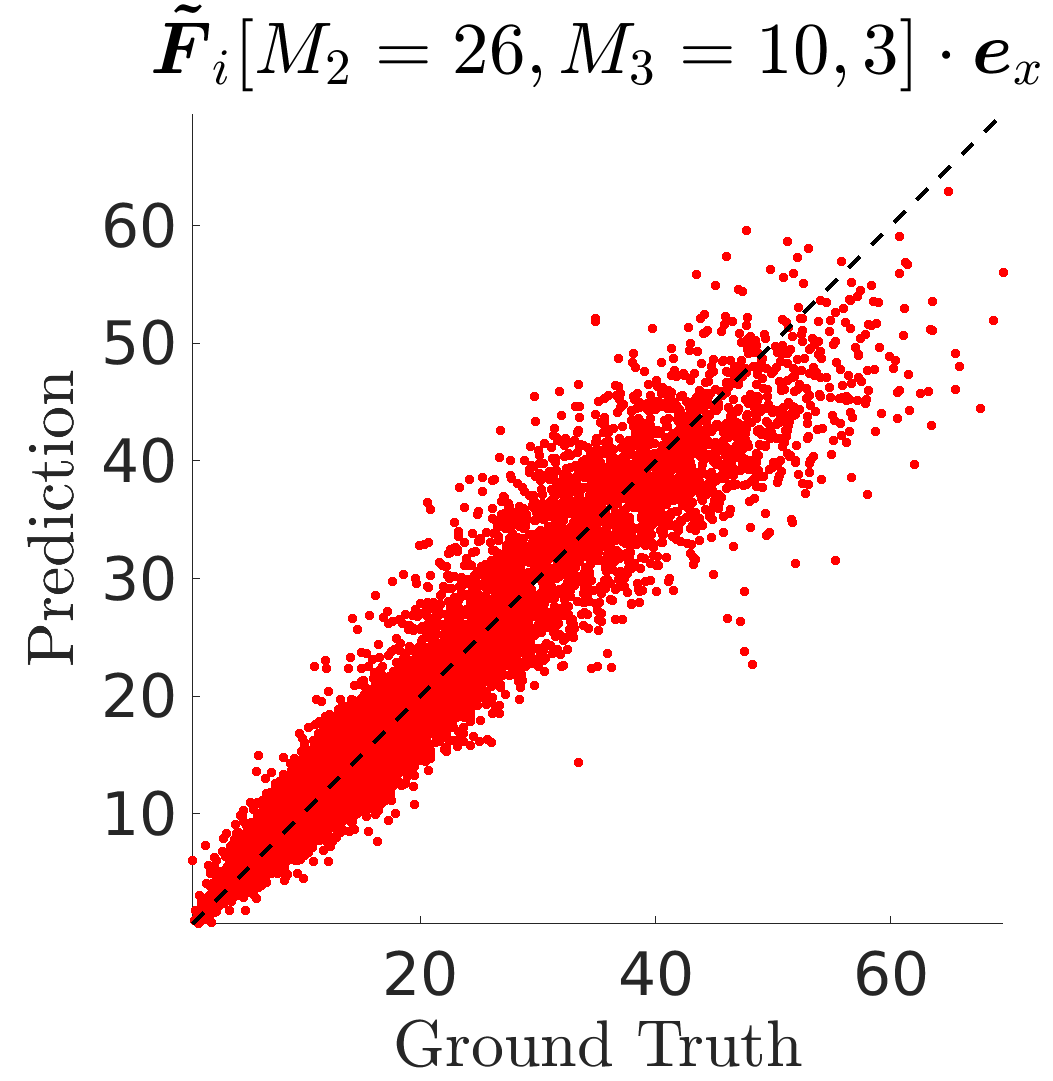}
        \includegraphics[width=0.32\textwidth,keepaspectratio=true]{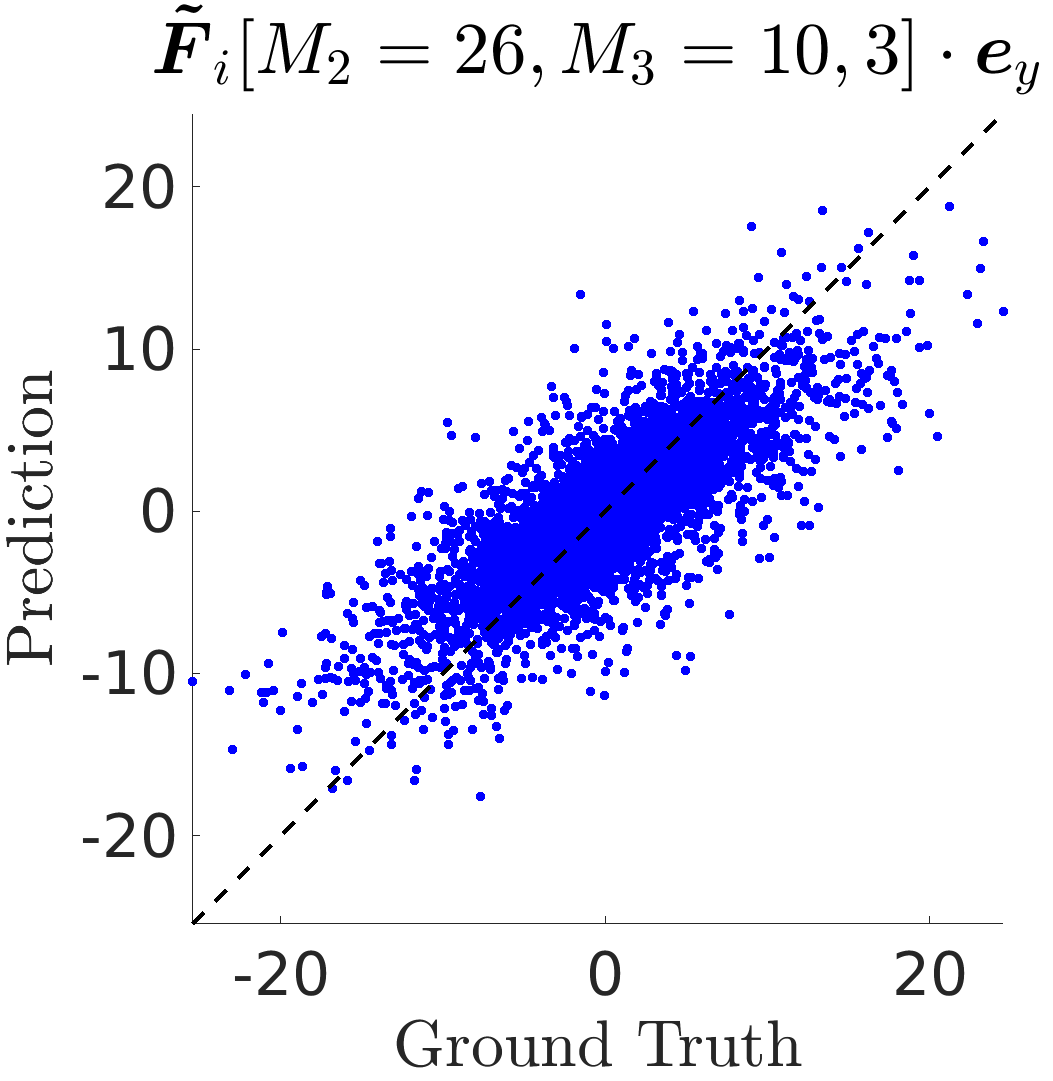}
        \includegraphics[width=0.32\textwidth,keepaspectratio=true]{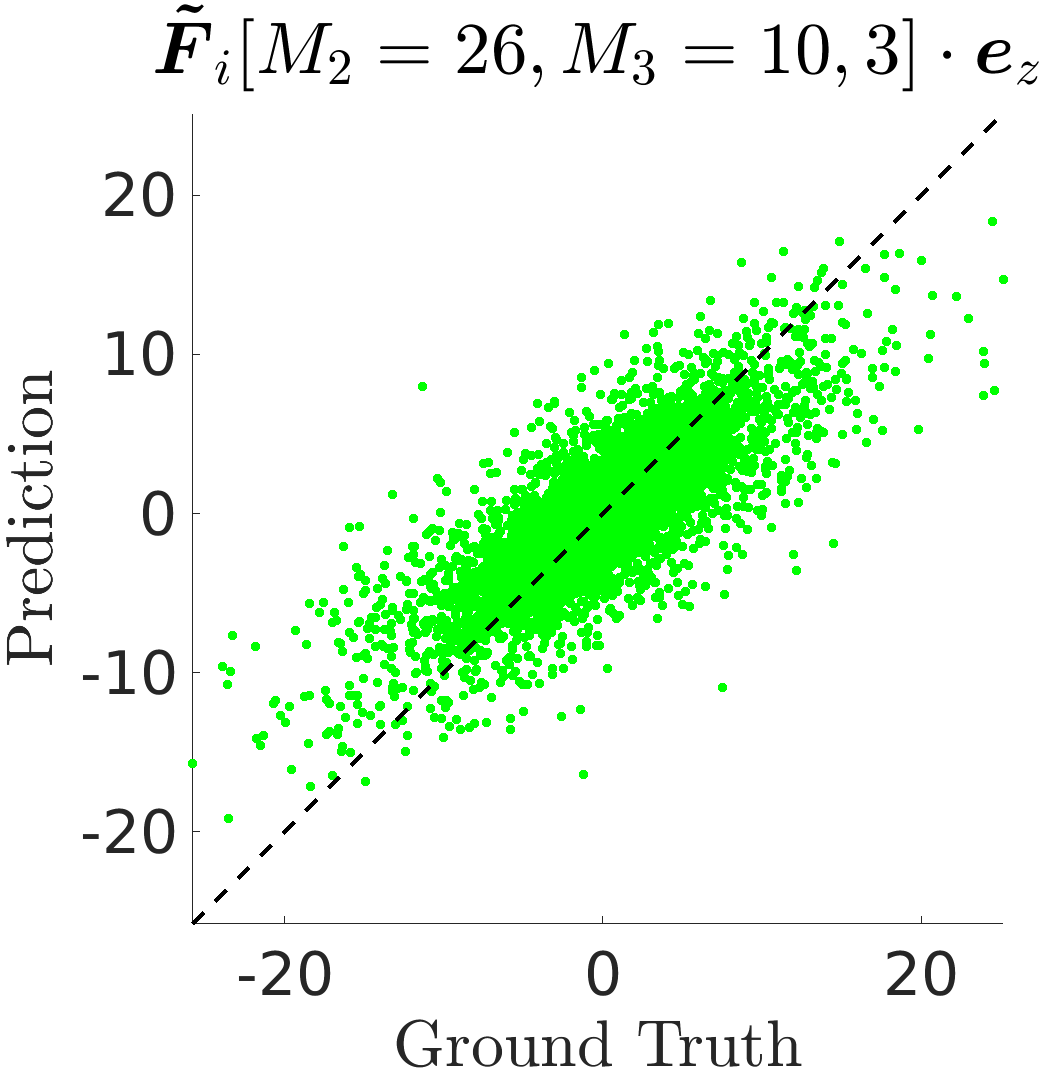}
    \end{subfigure}
    \caption{Universal force predictions using the presented hierarchical approach. The first, second and third rows corresponds to predictions obtained using $\bm{F}_{1i}$, $\tilde{\bm{F}}_i [M_2=26,2]$ and $\tilde{\bm{F}}_i[M_2=26,M_3=10, 3]$ for test samples from all datasets.}
    \label{fig:universal_forces}
\end{figure}

\begin{figure}
    \centering
    \begin{subfigure}[b]{\textwidth}
        \centering
        \includegraphics[width=0.32\textwidth,keepaspectratio=true]{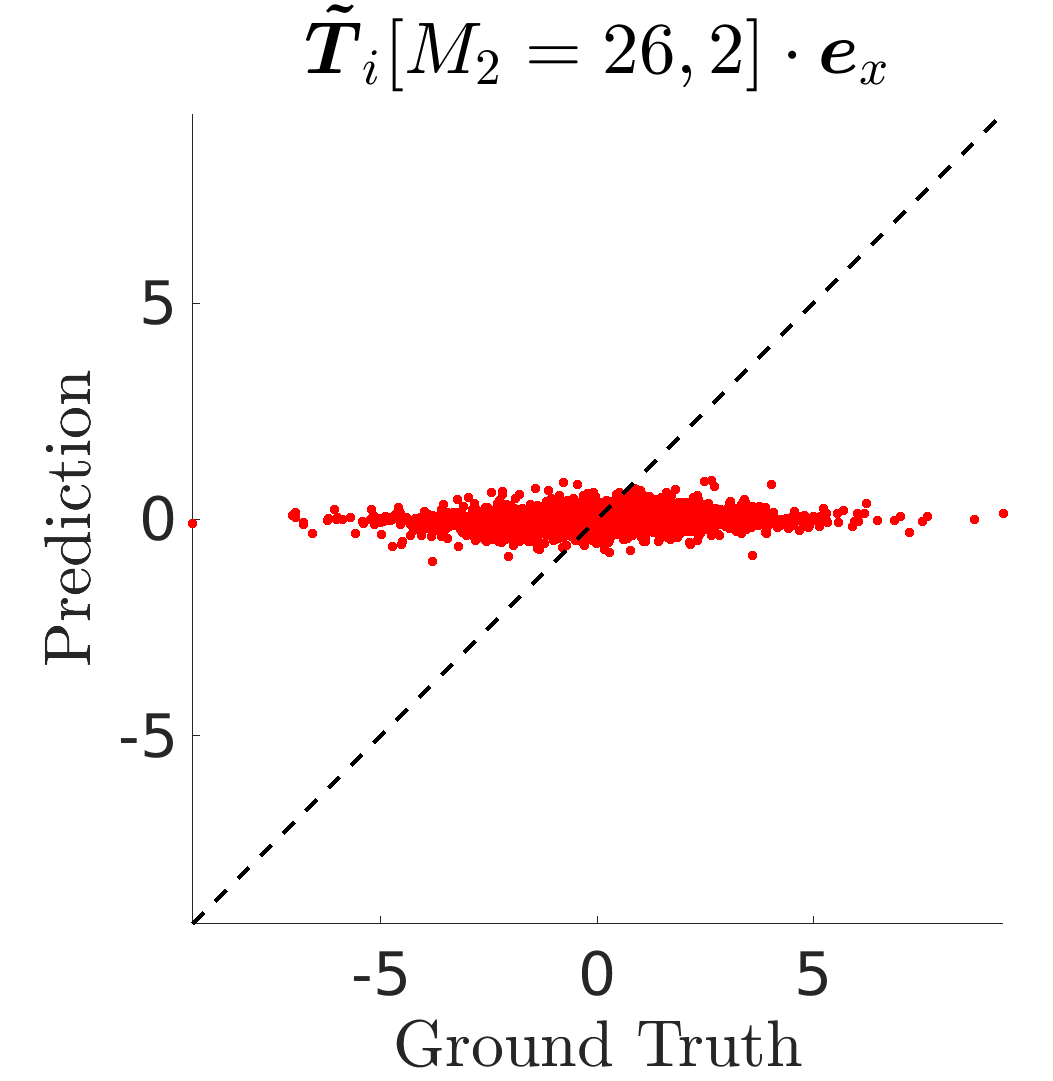}
        \includegraphics[width=0.32\textwidth,keepaspectratio=true]{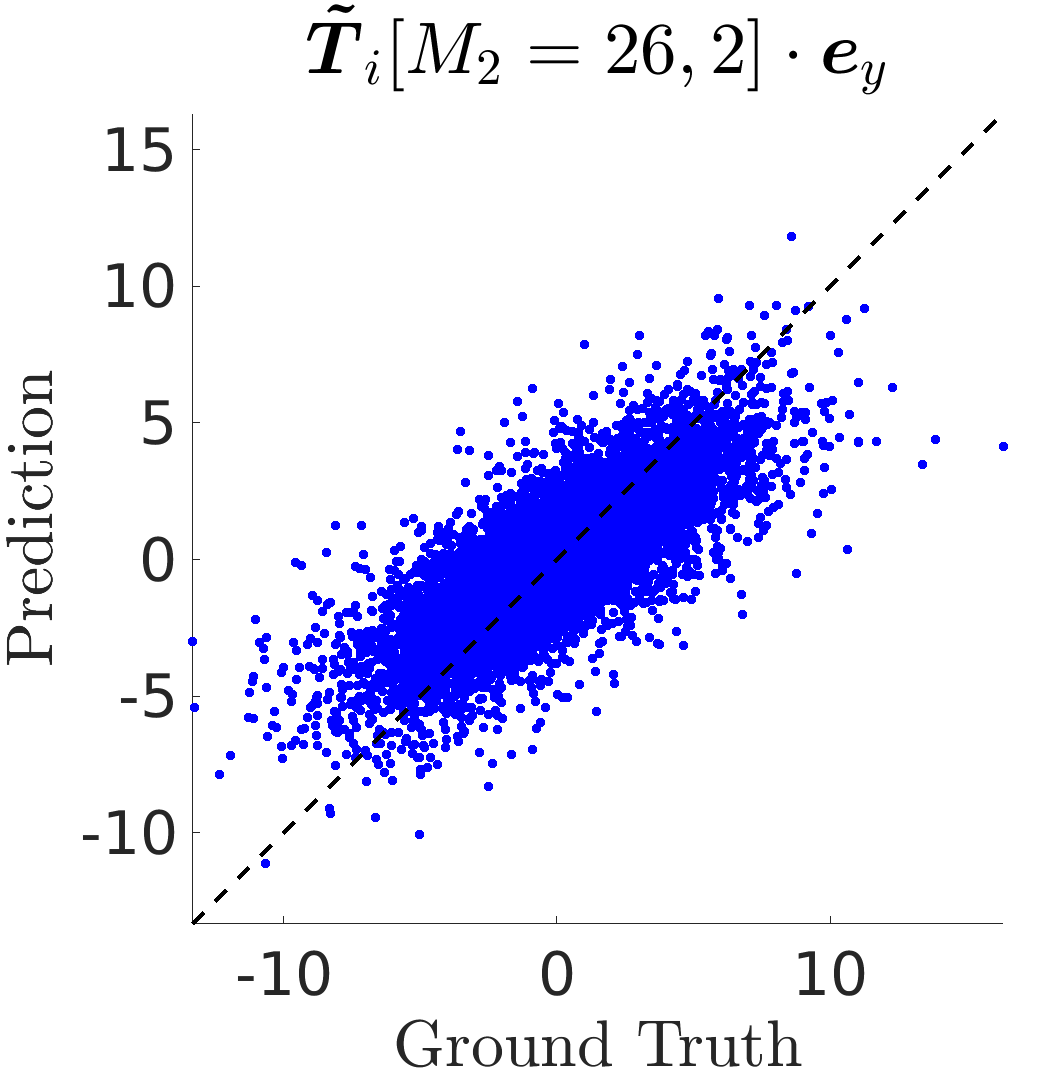}
        \includegraphics[width=0.32\textwidth,keepaspectratio=true]{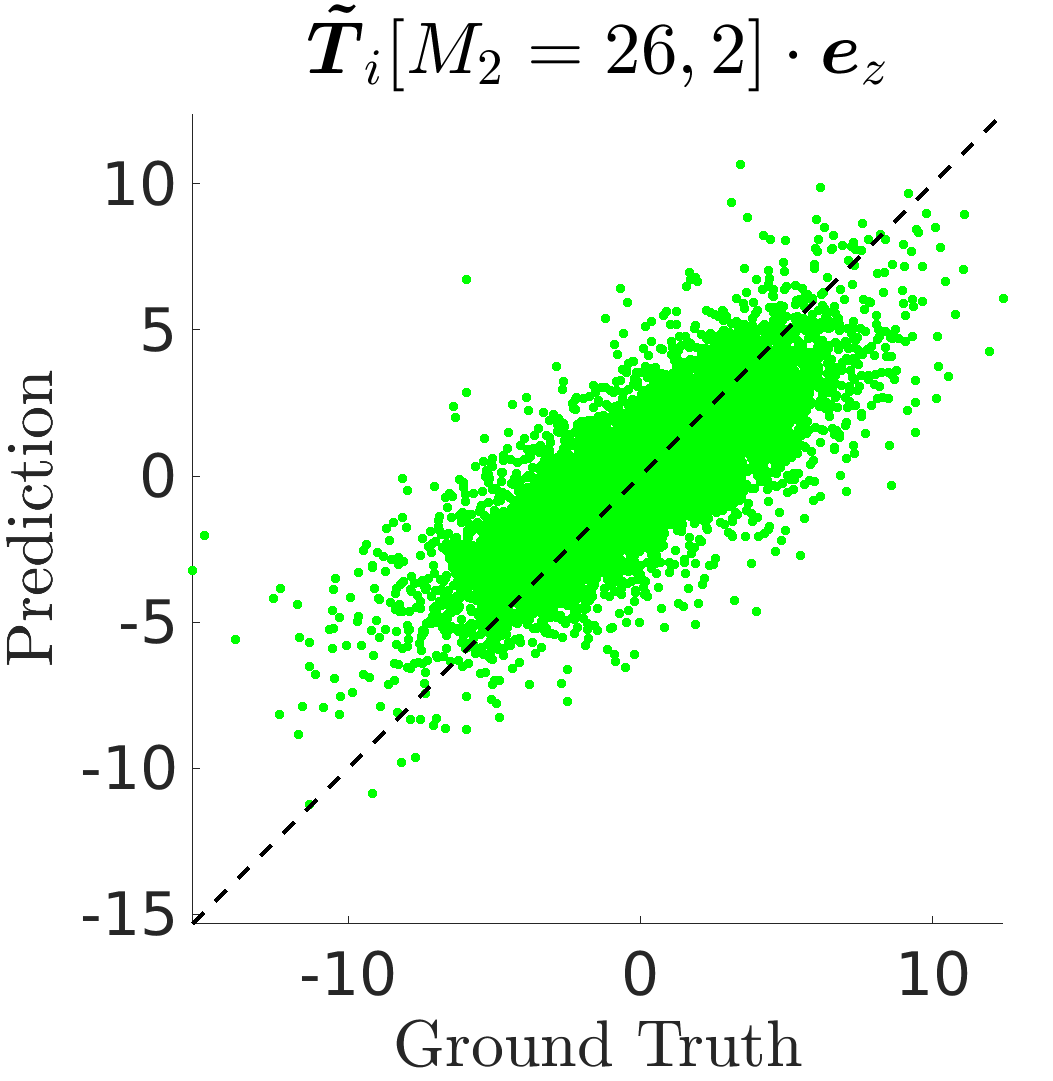}
    \end{subfigure}
    \begin{subfigure}[b]{\textwidth}
        \centering
        \includegraphics[width=0.32\textwidth,keepaspectratio=true]{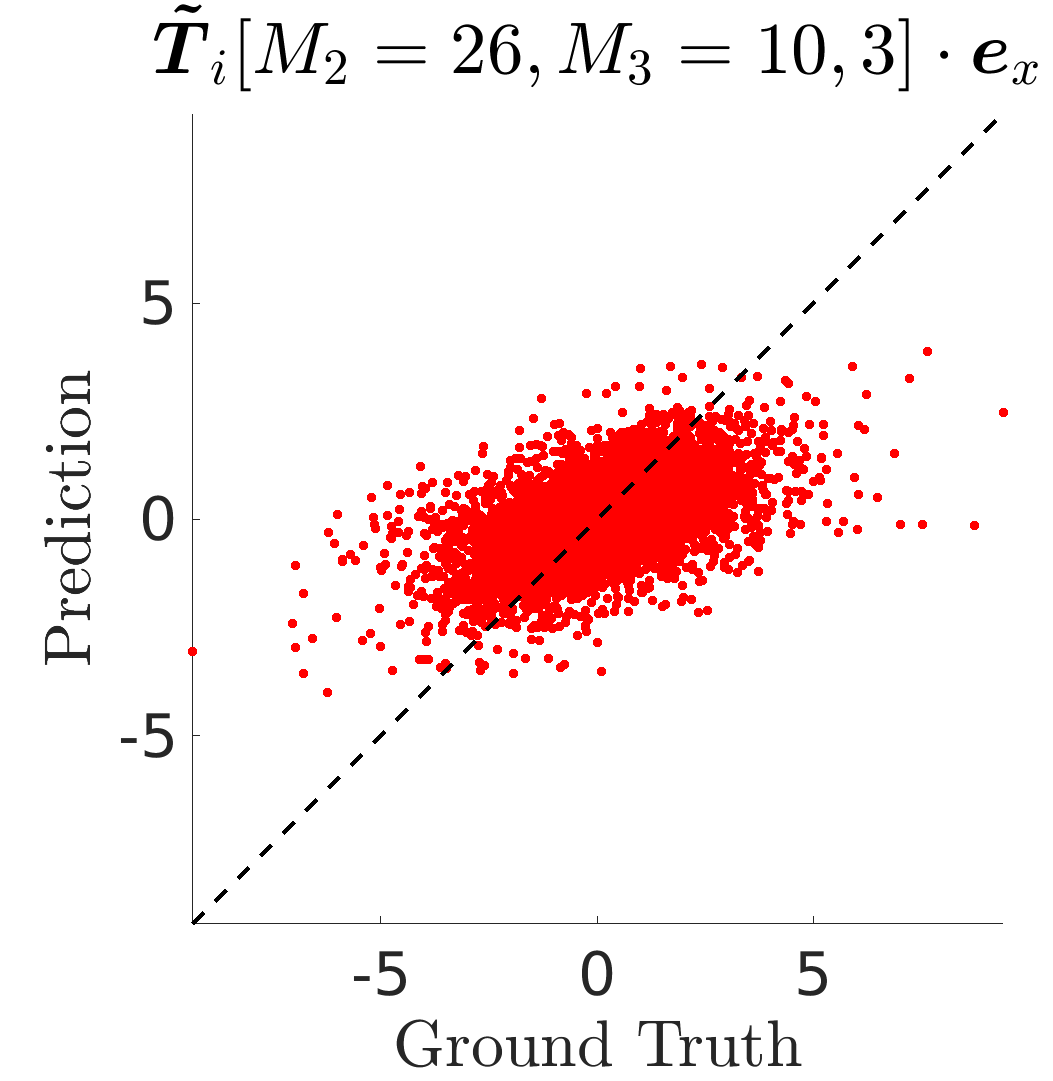}
        \includegraphics[width=0.32\textwidth,keepaspectratio=true]{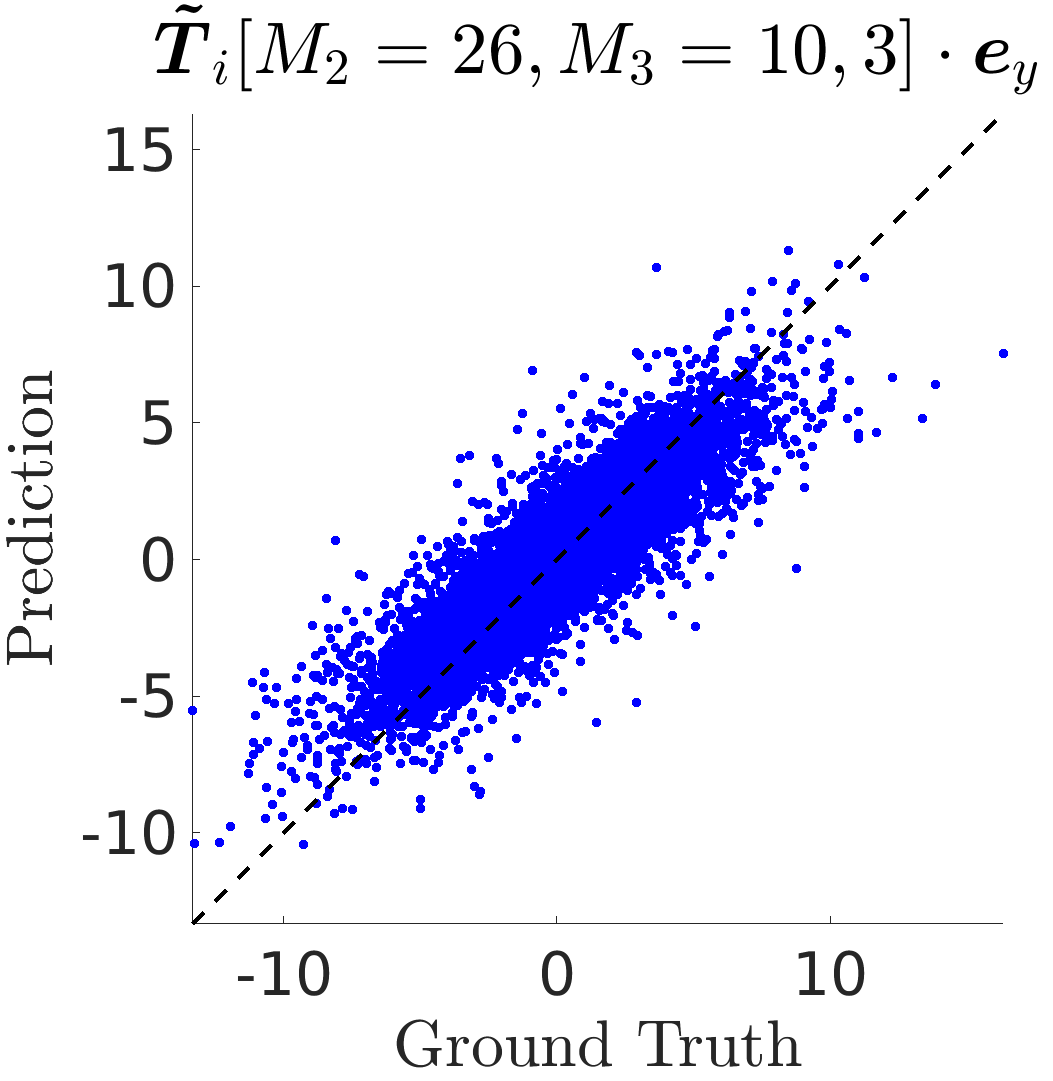}
        \includegraphics[width=0.32\textwidth,keepaspectratio=true]{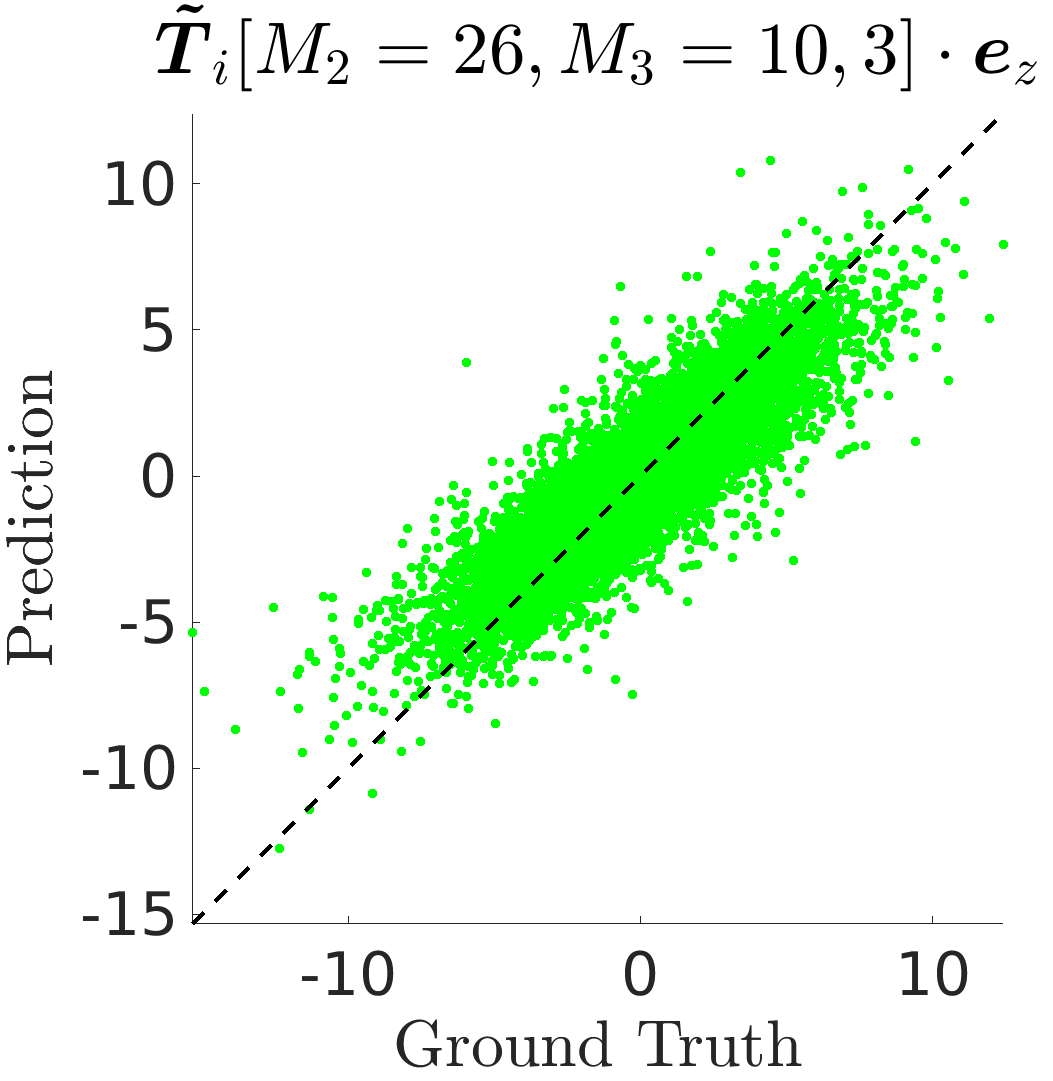}
    \end{subfigure}
    \caption{Universal torque predictions for test samples from all considered datasets. The first row depicts predictions obtained using $\tilde{\bm{T}}_i[M_2=26, 2]$, while the second row corresponds to predictions using $\tilde{\bm{T}}_i[M_2=26,M_3=10,3]$.}
    \label{fig:universal_torques}
\end{figure}

A summary of force and torque predictions for the test samples for all combinations of $\langle Re \rangle$ and $\langle \phi \rangle$ for the unary, binary, and trinary truncations are shown in Figures \ref{fig:universal_forces} \& \ref{fig:universal_torques} respectively. The three columns of these figures correspond to $x$, $y$ and $z$ components respectively. It can be observed that the characteristics of the $y$ \& $z$ components in these figures are similar but they are distinct from those of the $x$ component, which is the direction of mean flow.
The different rows of Figure \ref{fig:universal_forces} represent different hierarchical force models with increasing complexity from top to bottom. In each frame, the ground truth obtained from the PR simulations is plotted along the horizontal coordinate and the corresponding model prediction is plotted along the vertical coordinate. A perfect model will lead to a scatter plot that aligns with the diagonal dash line.

The first row illustrates the predictions of the unary force model. As expected, a unary model that only depends on mesoscale information is able to {capture} mean drag of the different datasets, but it is not able to capture the particle-to-particle variation within the different datasets. This explains the horizontal slab-like predictions for all three components. A closer look at the slabs indicates that they are positioned around the mean forces, $|\langle \bm{F} \rangle|$, presented in Table \ref{tab:datasets}. In interpreting the $y$ \& $z$ components of forces it should be noted that the the local volume-averaged fluid velocity of the particles do not perfectly align along the $x$-axis and therefore $y$ and $z$ forces cannot be equated to the lift force of the particle, as they may contain a small component of the drag force (i.e, force along the direction of local average flow). The second row shows predictions obtained with a binary force model with $M_2=26$. As the deterministic information of closest neighbors is being passed as part of the input feature, the neighbor influence in inducing higher and lower than average forces are predicted quite well. It can be seen from the last row of plots that the predictions are closer to the ground truth with the inclusion of trinary interactions, however, the improvement is not as large as that obtained from the inclusion of the binary interactions. As will be discussed below, this may be due to limited availability of data, restricting $M_3 = 10$ and the complexity of the trinary network architecture.

Similarly, the first row of Figure \ref{fig:universal_torques} shows the performance of the binary torque model with $M_2=26$ for the test samples of all the datasets. The second row corresponds to predictions achieved with the inclusion of trinary interactions with $M_3 = 10$. In the case of torque, significantly more accurate predictions can be witnessed with the trinary interactions for all three components. Especially, the streamwise component of torque on the particles can only be captured with the trinary model, although the predictive capability of the trinary model is only modest. The quantitative analysis of these predictions and the importance of binary and trinary models will be discussed in the succeeding subsection.             

\subsection{Neighbor-induced perturbation capture \& interaction-truncation error}
The ability of current models in capturing the neighbor-induced fluctuations of each individual dataset is quantified using the $R^2$ metric defined below. Although the flow averaged over the local volumes are not perfectly aligned along the $+x$ axis, the alignment is quite close. Hence, streamwise force predictions are quantified as $R^{2}_{\mathrm{Drag}}$ and the transverse force predictions are measured as $R^{2}_{\mathrm{Lift}}$. Similarly, $R^{2}_{\mathrm{Torque},\parallel}$ and $R^{2}_{\mathrm{Torque},\perp}$ measure streamwise and transverse torque predictions respectively. A value of unity for $R^2$ denotes a perfect model that is able to deterministically capture particle-to-particle variations. A value of zero indicates that the model is not capturing any variation. The value can also be negative if a model's predictions are opposite in sign to the ground truth.           

\begin{equation}
    R^{2}_{\mathrm{Drag}} =  1 - \frac{\sum_{i=1}^{N} (F_{i,(x,\mathrm{PR})}-F_{i,(x, \mathrm{NN}})^2}{\sum_{i}^{N} (F_{i,(x,\mathrm{PR})}-\langle F_{i,(x,\mathrm{PR})} \rangle)^2}
\end{equation}    
\begin{equation}
    R^{2}_{\mathrm{Lift}} =  1 - \frac{\sum_{i=1}^{N} (F_{i,(y,\mathrm{PR})}-F_{i,(y, \mathrm{NN})})^2 + (F_{i,(z,\mathrm{PR})}-F_{i,(z, \mathrm{NN})})^2}{\sum_{i}^{N} (F_{i,(y,\mathrm{PR})}-\langle F_{i,(y,\mathrm{PR})} \rangle)^2 + (F_{i,(z,\mathrm{PR})}-\langle F_{i,(z,\mathrm{PR})} \rangle)^2}
\end{equation}

\begin{equation}
    R^{2}_{\mathrm{Torque},\parallel} =  1 - \frac{\sum_{i=1}^{N} (T_{i,(x,\mathrm{PR})}-T_{i,(x, \mathrm{NN}})^2}{\sum_{i}^{N} (T_{i,(x,\mathrm{PR})}-\langle T_{i,(x,\mathrm{PR})} \rangle)^2}
\end{equation}    
\begin{equation}
    R^{2}_{\mathrm{Torque},\perp} =  1 - \frac{\sum_{i=1}^{N} (T_{i,(y,\mathrm{PR})}-T_{i,(y, \mathrm{NN})})^2 + (T_{i,(z,\mathrm{PR})}-T_{i,(z, \mathrm{NN})})^2}{\sum_{i}^{N} (T_{i,(y,\mathrm{PR})}-\langle T_{i,(y,\mathrm{PR})} \rangle)^2 + (T_{i,(z,\mathrm{PR})}-\langle T_{i,(z,\mathrm{PR})} \rangle)^2}
\end{equation}

Binary force predictions with $M_2=26$ is reported in Table \ref{tab:force_r2}. It can be seen that the learned binary force model is highly generalizable based on the small difference in training and testing performance of all datasets. The binary model indicates negligible underfitting (variation in third decimal) of transverse fluctuations for the datasets $(\langle Re \rangle, \langle \phi \rangle) = $ (165.96, 0.1), (49.78, 0.20), and (187.25, 0.20). This is probably due to us seeking a global minimum that results in the optimization process underfitting a few datasets at the expense of overfitting the other datasets. For the datasets obtained from \citet{wachs_piep}, the performance of the present model is either similar or better than those reported in their work. A consistent improvement of 6 to 11\% and 3 to 6\% for streamwise and transverse components respectively, is observed at higher particle volume fraction of  $\langle \phi \rangle = 0.4$. This additional improvement is attributed to the strict enforcement of rotational and reflectional symmetries in the present binary model.    

\begin{table}
    \caption{Individual dataset $R^2$ performance using $\tilde{\bm{F}}_i[M_2=26,2]$ \& $\tilde{\bm{F}}_i[M_2=26,M_3=10,3]$ force models for training and testing samples.}
    \label{tab:force_r2}
    \begin{center}
    \begin{ruledtabular}
    \begin{tabular}{cccccccccc}
    \rule{0pt}{4ex}$\langle Re \rangle$ & $\langle \phi \rangle$ & \multicolumn{4}{c}{$\tilde{\bm{F}}_i[M_2=26,2]$} &  \multicolumn{4}{c}{$\tilde{\bm{F}}_i[M_2=26,M_3=10,3]$} \\
     & & \multicolumn{2}{c}{Drag} & \multicolumn{2}{c}{Lift} & \multicolumn{2}{c}{Drag} & \multicolumn{2}{c}{Lift} \\
     & & Train & Test & Train & Test & Train & Test & Train & Test  \\[2ex] \hline
     \rule{0pt}{4ex}9.86 & 0.10 & 0.775 & 0.740 & 0.797 & 0.776 & 0.842 & 0.762 & 0.862 & 0.803 \\
     121.36 & 0.10 & 0.685 & 0.645 & 0.707 & 0.696 & 0.787 & 0.650 & 0.805 & 0.701\\
     6.95 & 0.21 & 0.816 & 0.809 & 0.812 & 0.800 & 0.853 & 0.817 & 0.862 & 0.827\\
     73.40 & 0.21 & 0.757 & 0.728 & 0.746 & 0.739 & 0.818 & 0.745 & 0.833 &  0.772 \\
     27.81 & 0.40 & 0.807 & 0.788 & 0.795 & 0.784 & 0.837 & 0.809 & 0.841 & 0.823 \\
     73.42 & 0.40 & 0.759 & 0.738 & 0.754 & 0.738 & 0.807 & 0.769 & 0.817 & 0.786 \\
     2.20 & 0.10 & 0.837 & 0.833 & 0.803 & 0.800 & 0.854 & 0.838 & 0.842 & 0.826 \\
     10.92 & 0.10 & 0.810 & 0.807 & 0.812 & 0.808 & 0.826 & 0.815 & 0.861 & 0.851 \\
     165.96 & 0.10 & 0.624 & 0.611 & 0.715 & 0.718 & 0.679 & 0.615 & 0.759 & 0.732 \\
     0.25 & 0.20 & 0.756 & 0.747 & 0.757 &  0.753 & 0.769 & 0.758 & 0.773 & 0.764 \\
     2.48 & 0.20 & 0.780 & 0.773 & 0.764 & 0.758 & 0.791 & 0.782 & 0.791 & 0.784 \\
     49.78 & 0.20 & 0.747 & 0.743 & 0.716 & 0.719 & 0.772 & 0.754 & 0.795 & 0.787 \\
     187.25 & 0.20 & 0.662 & 0.656 & 0.668 & 0.671 & 0.698 & 0.661 & 0.731 & 0.711 \\
     3.25 & 0.40 & 0.698 & 0.672 & 0.592 & 0.567 & 0.722 & 0.679 & 0.625 & 0.584 \\
     64.35 & 0.40 & 0.744 & 0.727 & 0.655 & 0.643 & 0.764 & 0.739 & 0.717 & 0.696 \\
     245.65 & 0.40 & 0.661 & 0.632 & 0.574 & 0.558 & 0.698 & 0.649 & 0.654 & 0.611 \\
    \end{tabular}
    \end{ruledtabular}
    \end{center}
\end{table}

The improvement achieved by the inclusion of the trinary term with $M_3=10$ can also be inferred from Table \ref{tab:force_r2}. It can be observed that the trinary force model consistently improves upon the corresponding binary prediction for all combinations of $\langle Re \rangle$ and $\langle \phi \rangle$. Furthermore, the difference between training and testing performance for both streamwise and transverse components is larger than the corresponding difference when only a binary model is used. This is in agreement with our understanding that far more training data is required to create a generalizable trinary interaction model than a binary interaction model. This also suggests that the performance of the trinary model can be further improved with the availability of more training data and thus the present models are by no means final.

The performance of the binary torque model with $M_2=26$ is presented in Table \ref{tab:torque_r2}. Only the results for the transverse torque component are presented, due to the fundamental limitation of binary interactions in capturing the streamwise torque component. The learned binary torque model is sufficiently generalizable as indicated by the consistent minor difference between training and testing performance. Surprisingly, the datasets that exhibit slight underfitting in binary force model also show such nature for the binary torque model. The model's training performance indicates that it has difficulty in learning the particular dataset $(\langle Re \rangle, \langle \phi \rangle) = (121.36, 0.10)$. 
For the datasets obtained from \citet{wachs_piep} the performance of the current binary torque model is comparable to that reported by them. {The current universal model displays affinity to low Reynolds number cases with a maximum improvement of 15\% for $(\langle Re \rangle, \langle \phi \rangle) = (0.25, 0.20)$. However, this comes at the expense of performance loss at high $\langle Re \rangle$ with the largest reduction in performance of 9\% for the dataset (165.96,0.10).}
It should be noted that the performance reported in \citet{wachs_piep} is based on training and testing carried out on each individual dataset, i.e. fixed ($\langle Re \rangle$, $\langle \phi \rangle$). Whereas the present model is trained  universally to cover a range of  $\langle Re \rangle$ and $\langle \phi \rangle$.

\begin{table}
    \caption{Individual dataset $R^2$ performance using $\tilde{\bm{T}}_i[M_2=26,2]$ \& $\tilde{\bm{T}}_i[M_2=26,M_3=10,3]$ torque models for training and testing samples.}
    \label{tab:torque_r2}
    \begin{center}
    \begin{ruledtabular}
    \begin{tabular}{cccccccc}
     \rule{0pt}{4ex}$\langle Re \rangle$ & $\langle \phi \rangle$ & \multicolumn{2}{c}{$\tilde{\bm{T}}_i[M_2=26,2]$} &  \multicolumn{4}{c}{$\tilde{\bm{T}}_i[M_2=26,M_3=10,3]$}\\
     & & \multicolumn{2}{c}{Torque,$\perp$} & \multicolumn{2}{c}{Torque,$\parallel$} & \multicolumn{2}{c}{Torque,$\perp$}\\
     & & Train & Test & Train & Test & Train & Test \\[2ex] \hline
     \rule{0pt}{4ex}9.86 & 0.10 & 0.752 & 0.751 & 0.602 & 0.469 & 0.933 & 0.899 \\
     121.36 & 0.10 & 0.308 & 0.285 & 0.467 & 0.011 & 0.835 & 0.618 \\
     6.95 & 0.21 & 0.827 & 0.825 & 0.699 & 0.610 & 0.951 & 0.929 \\
     73.40 & 0.21 & 0.496 & 0.486 & 0.458 & 0.234 & 0.868 & 0.774 \\
     27.81 & 0.40 & 0.657 & 0.647 & 0.658 & 0.544 & 0.937 & 0.897 \\
     73.42 & 0.40 & 0.435 & 0.434 & 0.571 & 0.437 & 0.864 & 0.804 \\
     2.20 & 0.10 & 0.899 & 0.898 & 0.678 & 0.630 & 0.969 & 0.962 \\
     10.92 & 0.10 & 0.798 & 0.794 & 0.583 & 0.549 & 0.928 & 0.918 \\
     165.96 & 0.10 & 0.533 & 0.541 & 0.181 & 0.047 & 0.793 & 0.741 \\
     0.25 & 0.20 & 0.896 & 0.895 & 0.673 & 0.647 & 0.954 & 0.949 \\
     2.48 & 0.20 & 0.898 & 0.896 & 0.665 & 0.653 & 0.958 & 0.954 \\
     49.78 & 0.20 & 0.713 & 0.717 & 0.415 & 0.363 & 0.876 & 0.864 \\
     187.25 & 0.20 & 0.595 & 0.595 & 0.192 & 0.120 & 0.793 & 0.758\\
     3.25 & 0.40 & 0.820 & 0.813 & 0.661 & 0.598 & 0.916 & 0.886\\
     64.35 & 0.40 & 0.694 & 0.685 & 0.451 & 0.408 & 0.806 & 0.777 \\
     245.65 & 0.40 & 0.565 & 0.564 & 0.305 &  0.208 & 0.768 & 0.715\\
    \end{tabular}
    \end{ruledtabular}
    \end{center}
\end{table}

The added benefit of including the trinary term in the torque model is also quantified in Table \ref{tab:torque_r2}. The trinary model offers a significant improvement over the binary model for the transverse torque component. It can also be observed that the learning deficiency of the binary model for certain datasets, particularly $(\langle Re \rangle, \langle \phi \rangle) = (121.36, 0.10)$, is addressed by the trinary model. As mentioned earlier the trinary torque model shows promise in capturing the streamwise torque component. This component is learned and predicted better at low $\langle Re \rangle$ and it perhaps requires more training data to achieve robust predictions at higher $\langle Re \rangle$. Furthermore, even higher order interactions may be needed for the prediction of streamwise torque.

\subsection{Statistical analysis of predictions}
The ability of presented force and torque models in capturing important statistical quantities (shown in Table \ref{tab:datasets}) is evaluated here. The force statistics presented in Table \ref{tab:force_stats} illustrate the disparity between mean drag ($|\langle \bm{F} \rangle|$) obtained in the PR simulation and that predicted by the unary force model. As explained earlier, this difference can be partly attributed to the fact that the uniform distribution of $N$-particles in the present datasets were not strictly the same as those employed in the generation of the unary model. I.e., the uniform probability distribution $P_N$ of the current datasets differs from $P_{N,Tn}$. The difference is also partly due to limited samples used for calculating the force average both in the present work as well in the construction of the unary model by \citet{tenneti_drag}. 

The binary-interaction model ($\tilde{\bm{F}}_{2i}$) ensures that the mean drag is brought substantially close to the actual value for every considered dataset. It is important to realize that the binary model achieves this trait for random particle distributions generated in different ways. As the mean drag values are already well captured by the binary force model, the added benefit of the trinary-interaction model ($\tilde{\bm{F}}_{3i}$) is generally quite small. 

By definition, the unary model predicts the same force for all the particles, and therefore its standard deviation is identically zero. It can be observed that the standard deviation of both the drag and lift forces predicted by the binary model are noticeably smaller than the PR simulation results. As a result, including trinary interactions does lead to improvement in the prediction of standard deviations of streamwise and transverse force components. This property of standard deviation improvement using trinary-interaction torque model is also observed for torque statistics evaluated in Table \ref{tab:torque_stats}. Despite the observed improvement, there still exists a substantial deficit in the standard deviation when compared to the ground truth. We associate this deficit to several factors including not sufficient training of the trinary interactions and the neglect of even higher-order interactions. Furthermore, the intertwined errors elaborated in the introduction of the manuscript and also the lower emphasis placed by the loss function on extreme/rarely occurring configuration samples in the datasets may contribute to the difference.

The force perturbation induced by the binary-interaction model for a sample is given by $\tilde{\bm{F}}_{i}[M_2,2]-\bm{F}_{1i}$, and similarly the added perturbation induced by the trinary-interaction model can be expressed as $\tilde{\bm{F}}_{i}[M_2,M_3,3]-\tilde{\bm{F}}_{i}[M_2,2]$. 
A statistical measure of these force perturbations learned through $\tilde{\bm{F}}_{2i}$ and $\tilde{\bm{F}}_{3i}$ of each dataset is obtained as their rms value. These rms values in turn denote how much the binary and trinary terms contribute to the overall force and torque values. Thus, the relative importance of the trinary term compared to the binary term can be measured from the ratio of rms of trinary-interaction to rms of binary-interaction. This ratio for the streamwise force of the different cases are in the range [0.13, 0.26], and the corresponding range for the transverse force is [0.16, 0.38]. Thus, it can be concluded that the trinary force term contributes to only 10 to 30\% of the binary term. The rms ratio of the transverse torque component is observed to be in the range [0.26, 1.13]. The comparatively larger range for torque is mainly due to the binary torque model underperforming for a dataset (see Table \ref{tab:torque_r2}). The corresponding ratio for the streamwise torque component is not computed due to the limitation that the binary term cannot account for this quantity.

\begin{table}
   \caption{Statistical analysis of test sample force predictions obtained using $\bm{F}_{1i}$, $\tilde{\bm{F}}_i[M_2=26,2]$ \& $\tilde{\bm{F}}_i[M_2=26,M_3=10,3]$ models and their comparison with respective particle-resolved (PR) simulation data.}
    \label{tab:force_stats}
   \begin{center}
    \begin{ruledtabular}
    \begin{tabular}{cccccccccccc}
        \rule{0pt}{4ex}$\langle Re \rangle$ & $\langle \phi \rangle$ & \multicolumn{4}{c}{$|\langle \bm{F} \rangle |$} & \multicolumn{3}{c}{$\sigma_{\mathrm{Drag}}$} & \multicolumn{3}{c}{$\sigma_{\mathrm{Lift}}$} \\
        & & PR &  $\bm{F}_{1i}$ & $\tilde{\bm{F}}_i[2]$ & $\tilde{\bm{F}}_i[3] $ & PR & $\tilde{\bm{F}}_i[2]$ & $\tilde{\bm{F}}_i[3] $ & PR & $\tilde{\bm{F}}_i[2]$ & $\tilde{\bm{F}}_i[3] $ \\[2ex]
        \hline
        \rule{0pt}{4ex}9.86 & 0.10 & 2.76 & 3.14 & 2.84 & 2.81 & 0.7823 & 0.7510 & 0.7520 & 0.4661 & 0.4368 & 0.4690 \\
        121.36 & 0.10 & 9.55 & 7.08 & 9.54 & 9.53 & 3.1110 & 2.4820 & 2.5262 & 1.4986 & 1.3442 & 1.4512 \\
        6.95 & 0.21 & 4.03 & 4.84  & 4.02 & 4.00 & 0.9581 & 0.8774 & 0.8798 & 0.7454 & 0.6616 & 0.6803 \\
        73.40 & 0.21 & 9.51 & 8.30 & 9.54 & 9.47 & 2.7987 & 2.4118 & 2.4994 & 1.6178 & 1.4297 & 1.4733 \\
        27.81 & 0.40 & 12.68 & 14.51 & 12.69 & 12.67 & 2.0613 & 1.8689 & 1.9283 & 1.5028 & 1.3872 & 1.4330 \\
        73.42 & 0.40 & 19.14 & 18.35 & 18.99 & 18.90 & 3.6087 & 3.3962 & 3.5509 & 2.4135 & 2.3316 & 2.4221\\
        2.20 & 0.10 & 2.63 & 2.55 & 2.63 & 2.63 & 0.5425 & 0.4941 & 0.4997 & 0.3973 & 0.3474 & 0.3611 \\
        10.92 & 0.10 & 3.28 & 3.15 & 3.19 & 3.19 & 0.7351 & 0.6504 & 0.6597 & 0.4839 & 0.4057 & 0.4374 \\
        165.96 & 0.10 & 9.48 & 8.15 & 9.53 & 9.49 & 2.4191 & 2.0396 & 2.0913 & 1.5662 & 1.2302 & 1.3251\\
        0.25 & 0.20 & 4.42 & 4.00 & 4.42 & 4.41 & 0.7971 & 0.6819 & 0.6848 & 0.6737 & 0.5908 & 0.6005\\
        2.48 & 0.20 & 4.47 & 4.31 & 4.47 & 4.46 & 0.8116 & 0.7243 & 0.7285 & 0.6707 & 0.5798 & 0.5951\\
        49.78 & 0.20 & 7.72 & 7.11 & 7.75 & 7.75 & 1.8194 & 1.5955 & 1.5994 & 1.2608  & 1.0695 & 1.1245\\
        187.25 & 0.20 & 14.24 & 12.16 & 14.21 & 14.19 & 3.6989 & 2.9954 & 3.0572 & 2.6873 & 2.1671 & 2.2788 \\
        3.25 & 0.40 & 11.75 & 11.60 & 11.72 & 11.70 & 2.7535 & 2.2873 & 2.3119 & 2.1329 & 1.5490 & 1.5951 \\
        64.35 & 0.40 & 19.67 & 17.47 & 19.73 & 19.66 & 5.0963 & 3.9178 & 4.0322 & 3.6885 & 2.8854 & 2.9399 \\
        245.65 & 0.40 & 36.62 & 29.69 & 36.47 & 36.24 & 9.5692 & 7.1343 & 7.4556 & 7.1215 & 5.3349 & 5.4929 \\
    \end{tabular}
    \end{ruledtabular}
   \end{center}
\end{table}

\begin{table}
   \caption{Statistical analysis of test sample torque predictions obtained using $\tilde{\bm{T}}_i[M_2=26,2]$ \& $\tilde{\bm{T}}_i[M_2=26,M_3=10,3]$ and their comparison with corresponding particle-resolved (PR) simulation data.}
    \label{tab:torque_stats}
   \begin{center}
   \begin{ruledtabular}
    \begin{tabular}{cccccccc}
        \rule{0pt}{4ex}$\langle Re \rangle$ & $\langle \phi \rangle$ & \multicolumn{3}{c}{$\sigma_{\mathrm{Torque},\parallel}$} & \multicolumn{3}{c}{$\sigma_{\mathrm{Torque},\perp}$}\\
        & & PR & $\tilde{\bm{T}}_i[2]$ & $\tilde{\bm{T}}_i[3]$ & PR & $\tilde{\bm{T}}_i[2]$ & $\tilde{\bm{T}}_i[3]$ \\[2ex] 
        \hline
        \rule{0pt}{4ex}9.86 & 0.10 & 0.2407 & 0.0525 & 0.2090 & 0.9253 & 0.9389 & 0.9848 \\
        121.36 & 0.10 & 0.5043 & 0.0253 & 0.3621 & 1.3689 & 0.8834 &   1.2477 \\
        6.95 & 0.21 & 0.4195 & 0.0991 & 0.3270 & 1.2987 & 1.1596 & 1.2687\\
        73.40 & 0.21 & 0.6441 & 0.0677 & 0.4365 & 1.7103 & 1.2492 & 1.6479 \\
        27.81 & 0.40 & 0.6734 & 0.0808 & 0.5222 & 2.0589 & 1.6960 & 2.0102 \\
        73.42 & 0.40 & 1.0285 & 0.0885 & 0.7665 & 2.5955 & 1.9596 & 2.4156\\
        2.20 & 0.10 & 0.2375 & 0.0493 & 0.1897 & 0.9241 & 0.8779 & 0.9239 \\
        10.92 & 0.10 & 0.2405 & 0.0471 & 0.1654 & 0.9838 & 0.7693 & 0.8692\\
        165.96 & 0.10 & 0.4339 & 0.0208 & 0.2269 & 1.1972 & 0.7821 & 1.0430\\
        0.25 & 0.20 & 0.4295 & 0.0743 & 0.3514 & 1.4509 & 1.3778 & 1.4472\\
        2.48 & 0.20 & 0.4217 & 0.0726 & 0.3324 & 1.4465 & 1.3860 & 1.4377\\
        49.78 & 0.20 & 0.5078 & 0.0575 & 0.3520 & 1.6549 & 1.3675 & 1.5794\\
        187.25 & 0.20 & 0.8446 & 0.0390 & 0.4093 & 2.0219 & 1.4511 & 1.7918\\
        3.25 & 0.40 & 1.0256 & 0.0826 & 0.8007 & 2.4789 & 2.2178 & 2.3772 \\
        64.35 & 0.40 & 1.3063 & 0.1911 & 0.7421 & 2.9899 & 2.0702 & 2.3461 \\
        245.65 & 0.40 & 2.0385 & 0.1246 & 1.0671 & 3.9776 & 3.1240 & 3.2752\\
    \end{tabular}
    \end{ruledtabular}
   \end{center}
\end{table}

\section{Interpretation of the machine learned models}\label{interpret}
This section will explore physical interpretations of the machine learned binary and trinary models of force and torque. The advantage of the hierarchical approach over a direct approach to force/torque modeling based on $M$ nearest neighbors is that we can clearly interpret the binary influence of a neighbor, and the trinary influence of each pairs of neighbors, and so on in a systematic manner.

\subsection{Binary force and torque maps}
\begin{figure}
    \centering
    \begin{subfigure}[b]{\textwidth}
       \centering
        \includegraphics[width=0.32\textwidth,keepaspectratio=true]{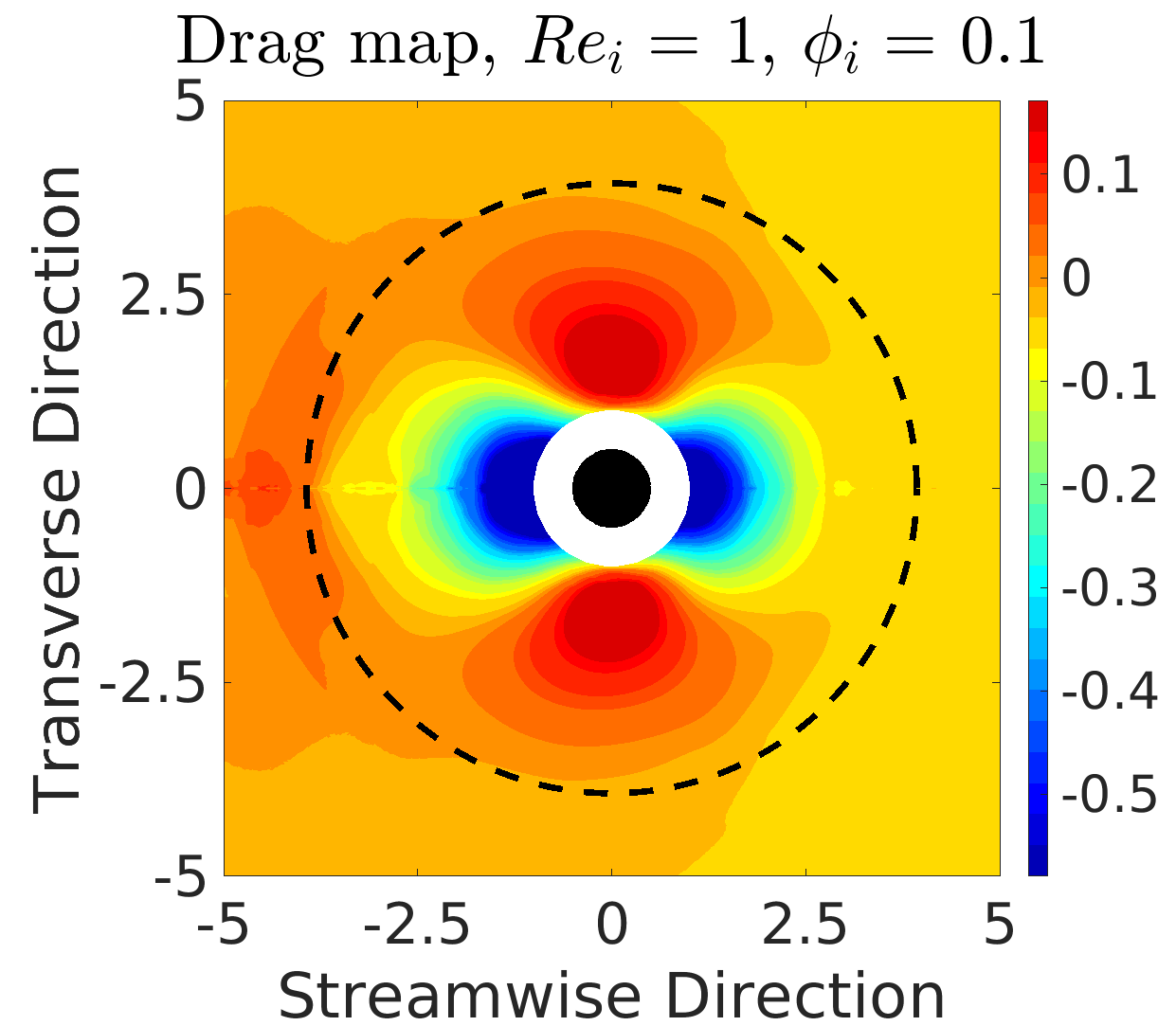}
         \includegraphics[width=0.32\textwidth,keepaspectratio=true]{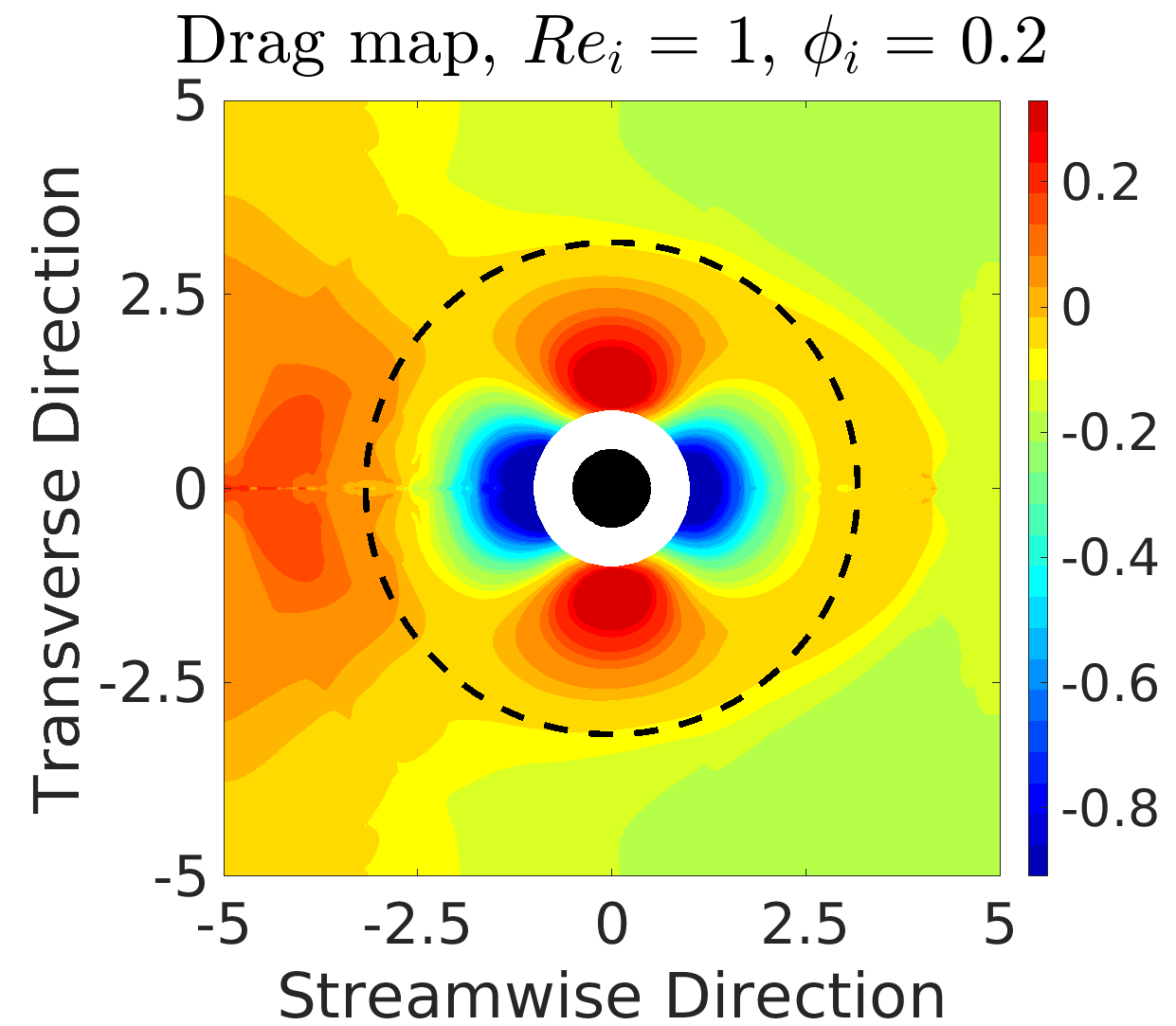}
          \includegraphics[width=0.32\textwidth,keepaspectratio=true]{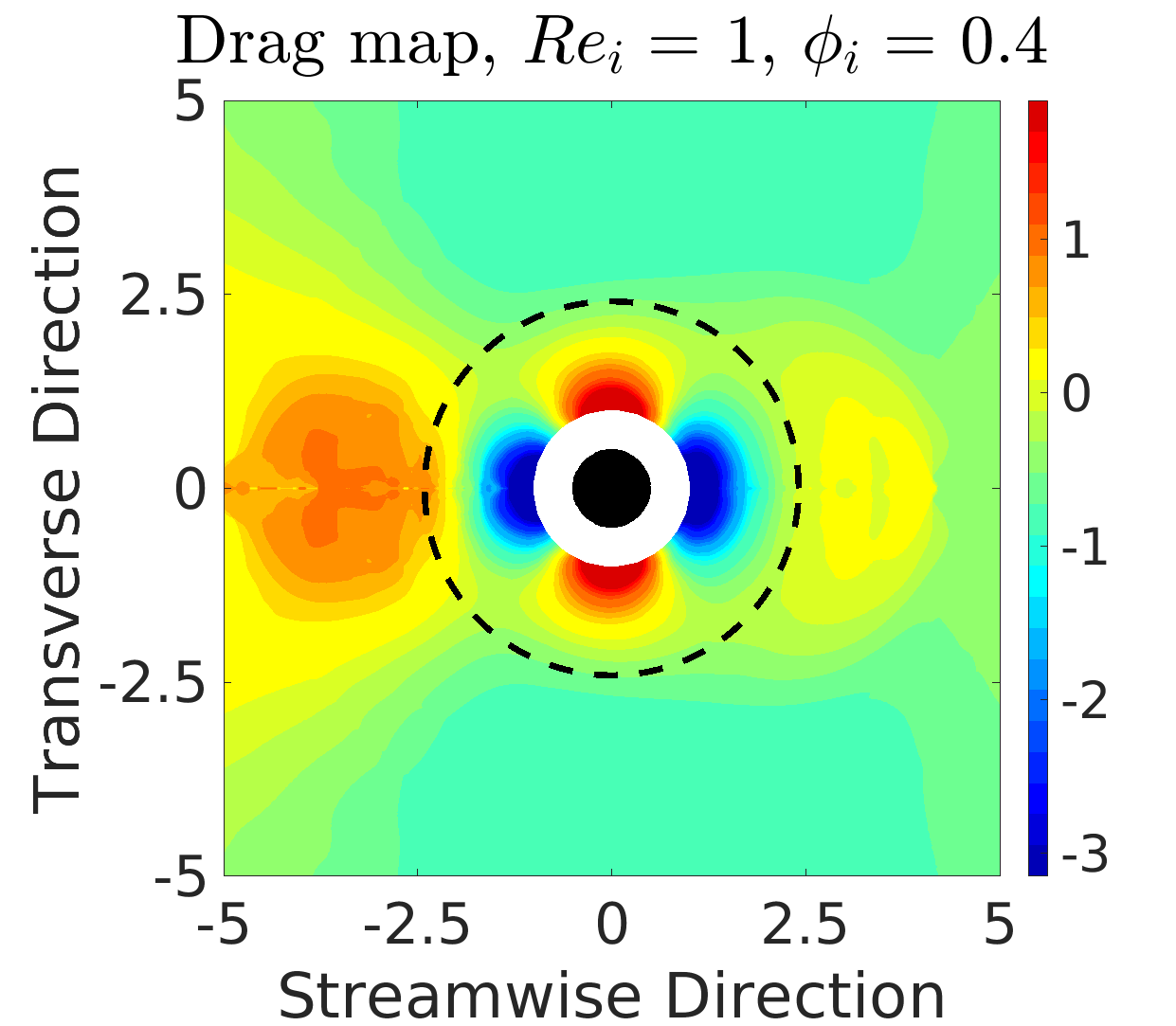}
    \end{subfigure}
    \begin{subfigure}[b]{\textwidth}
       \centering
        \includegraphics[width=0.32\textwidth,keepaspectratio=true]{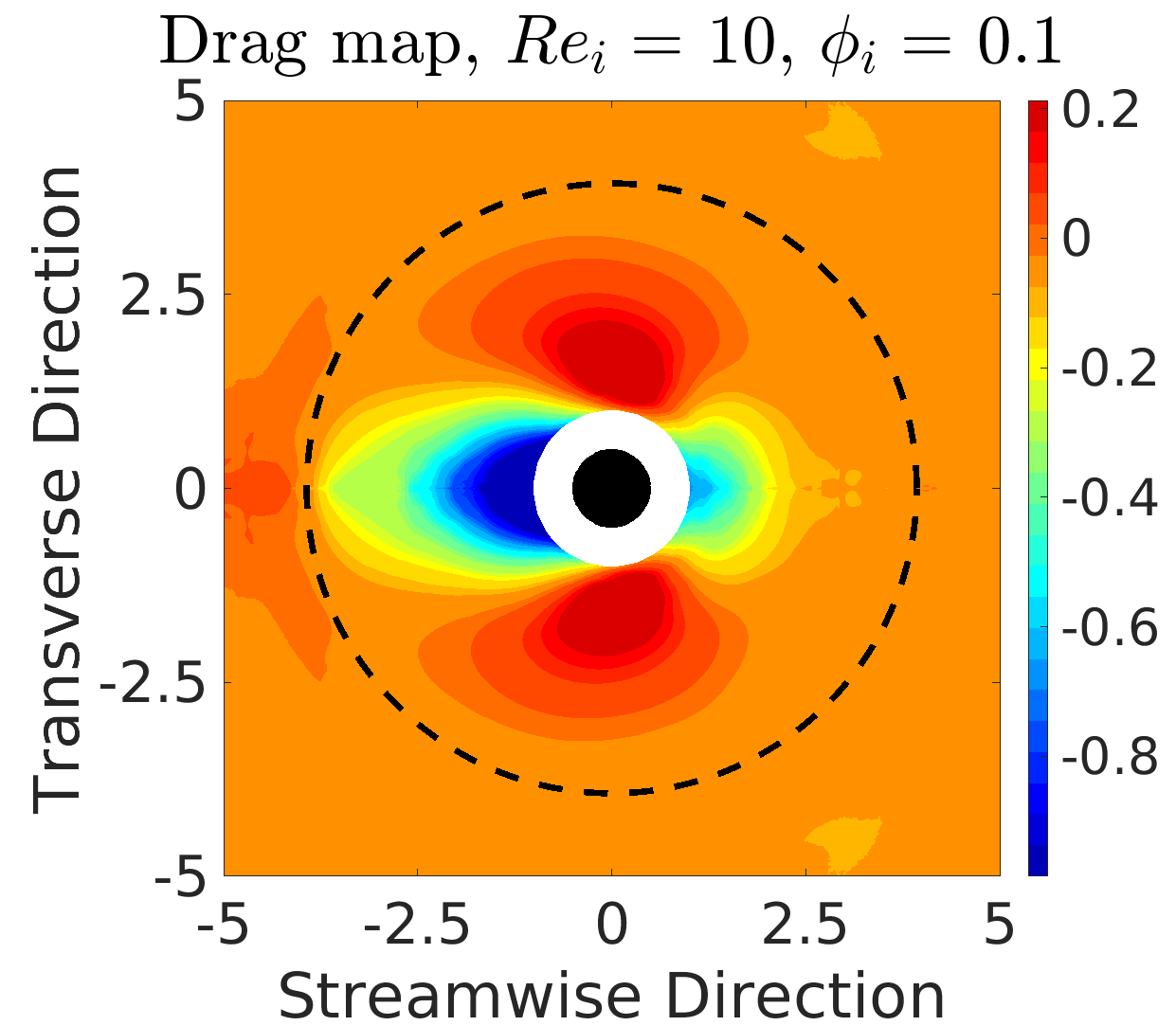}
         \includegraphics[width=0.32\textwidth,keepaspectratio=true]{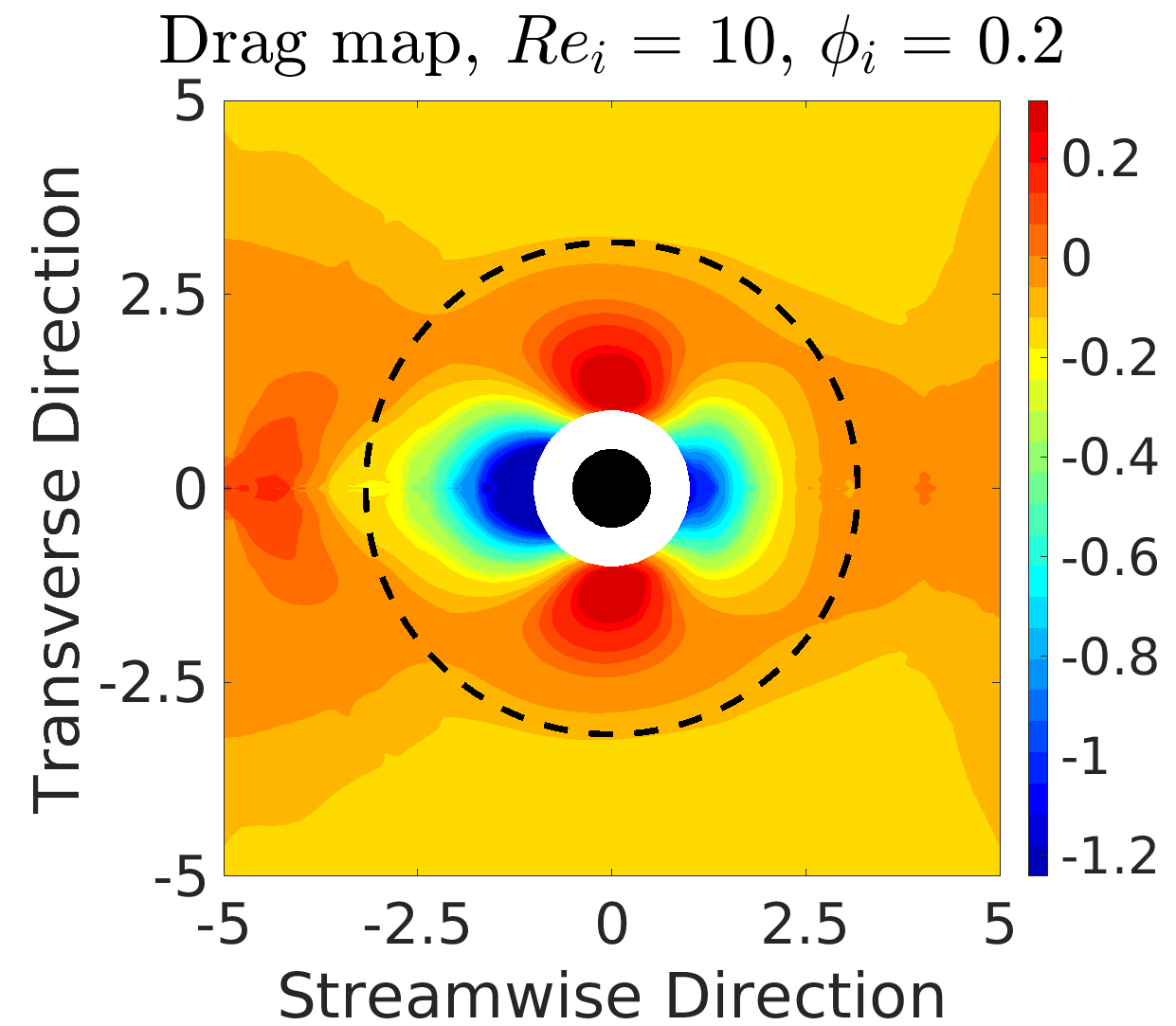}
          \includegraphics[width=0.32\textwidth,keepaspectratio=true]{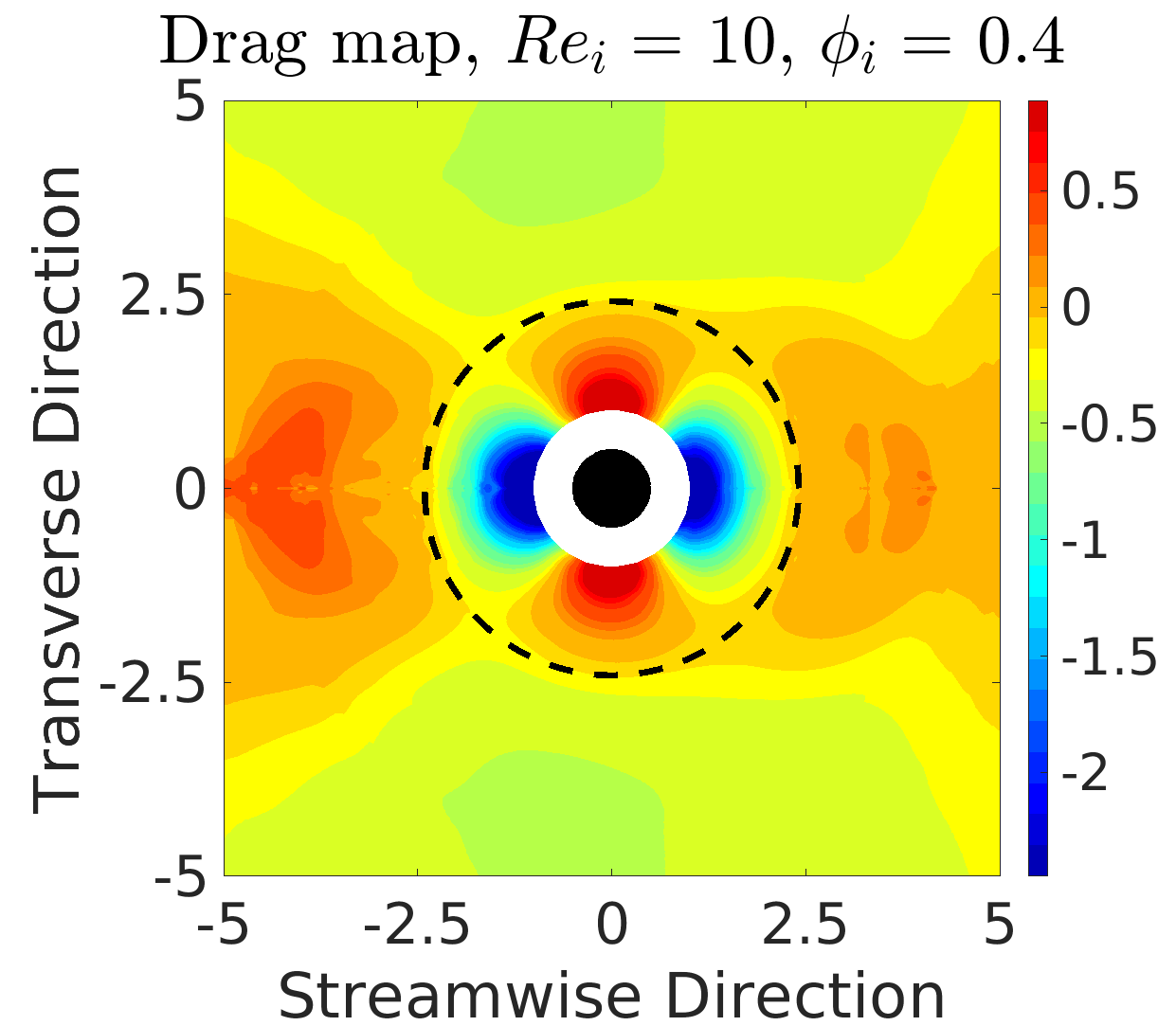}
    \end{subfigure}
    \begin{subfigure}[b]{\textwidth}
       \centering
        \includegraphics[width=0.32\textwidth,keepaspectratio=true]{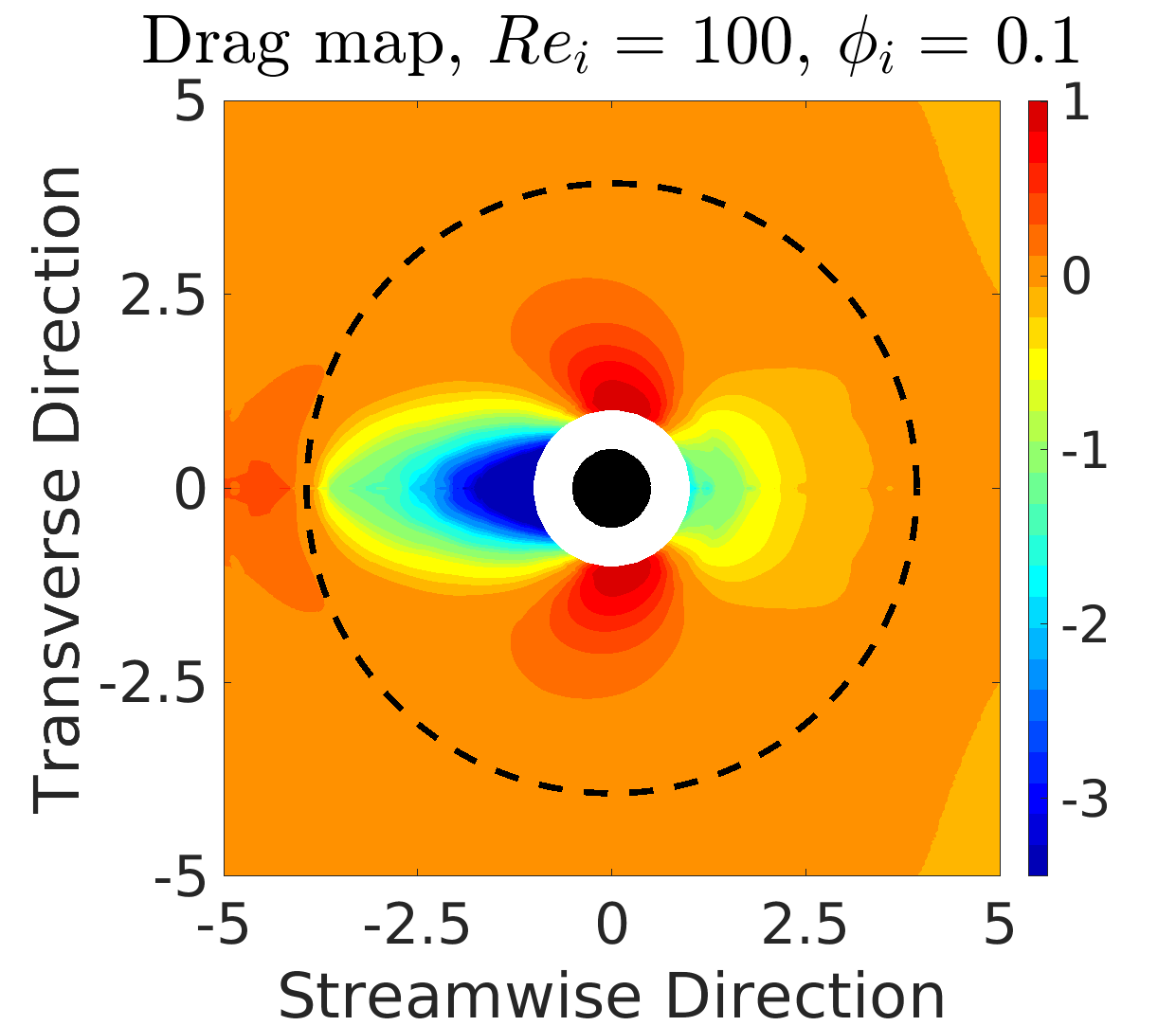}
         \includegraphics[width=0.32\textwidth,keepaspectratio=true]{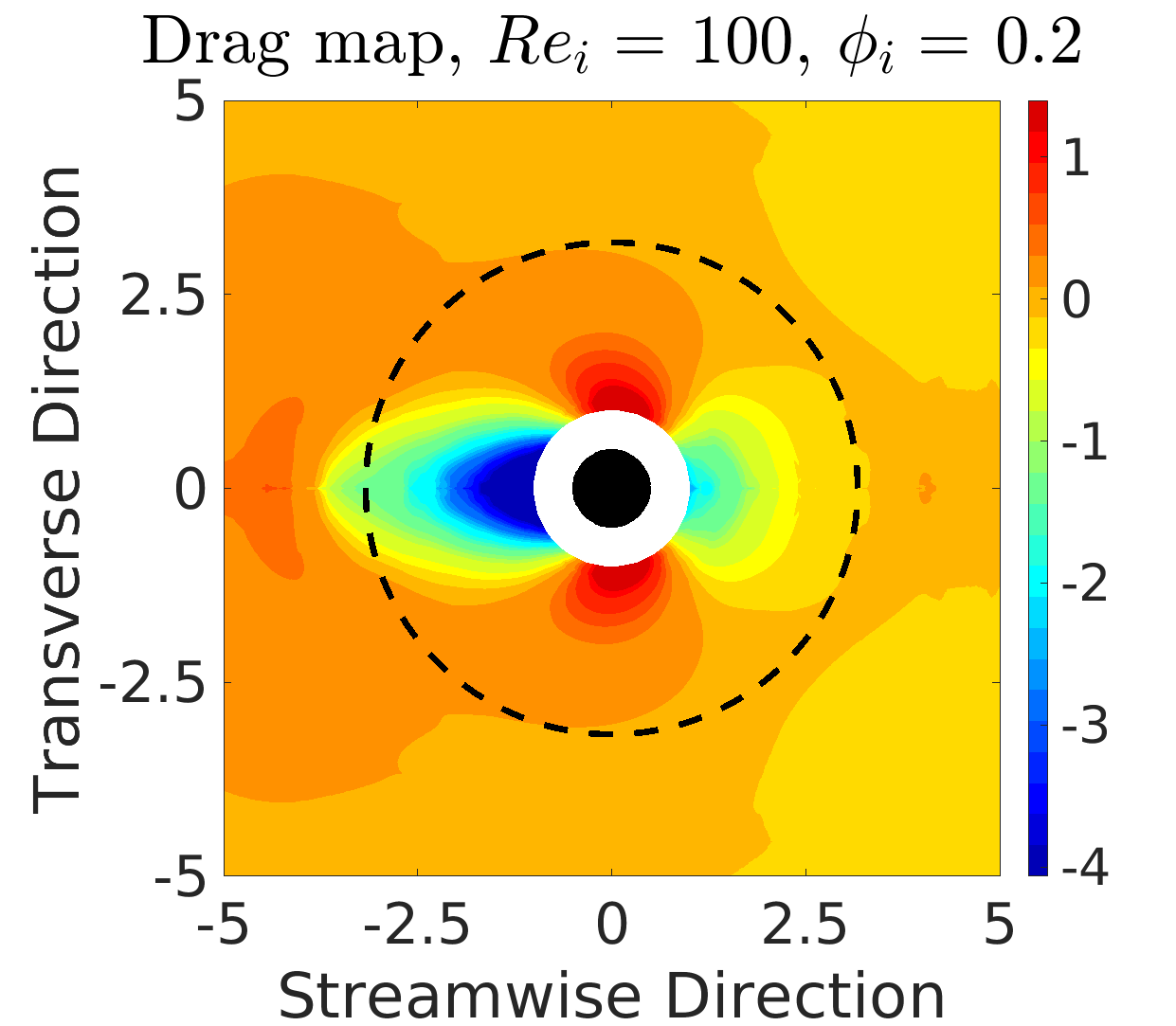}
          \includegraphics[width=0.32\textwidth,keepaspectratio=true]{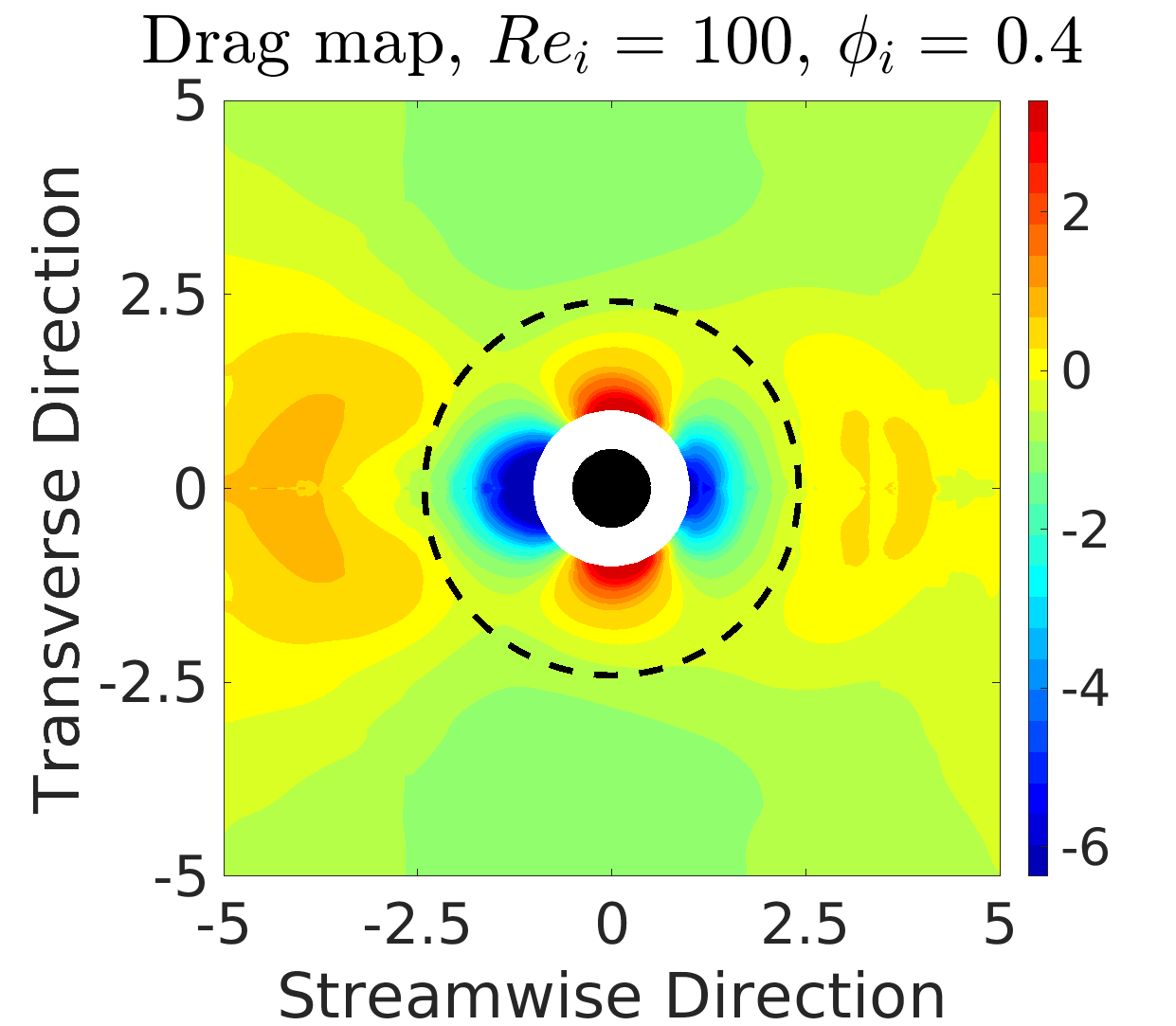}
    \end{subfigure}
    \begin{subfigure}[b]{\textwidth}
       \centering
        \includegraphics[width=0.32\textwidth,keepaspectratio=true]{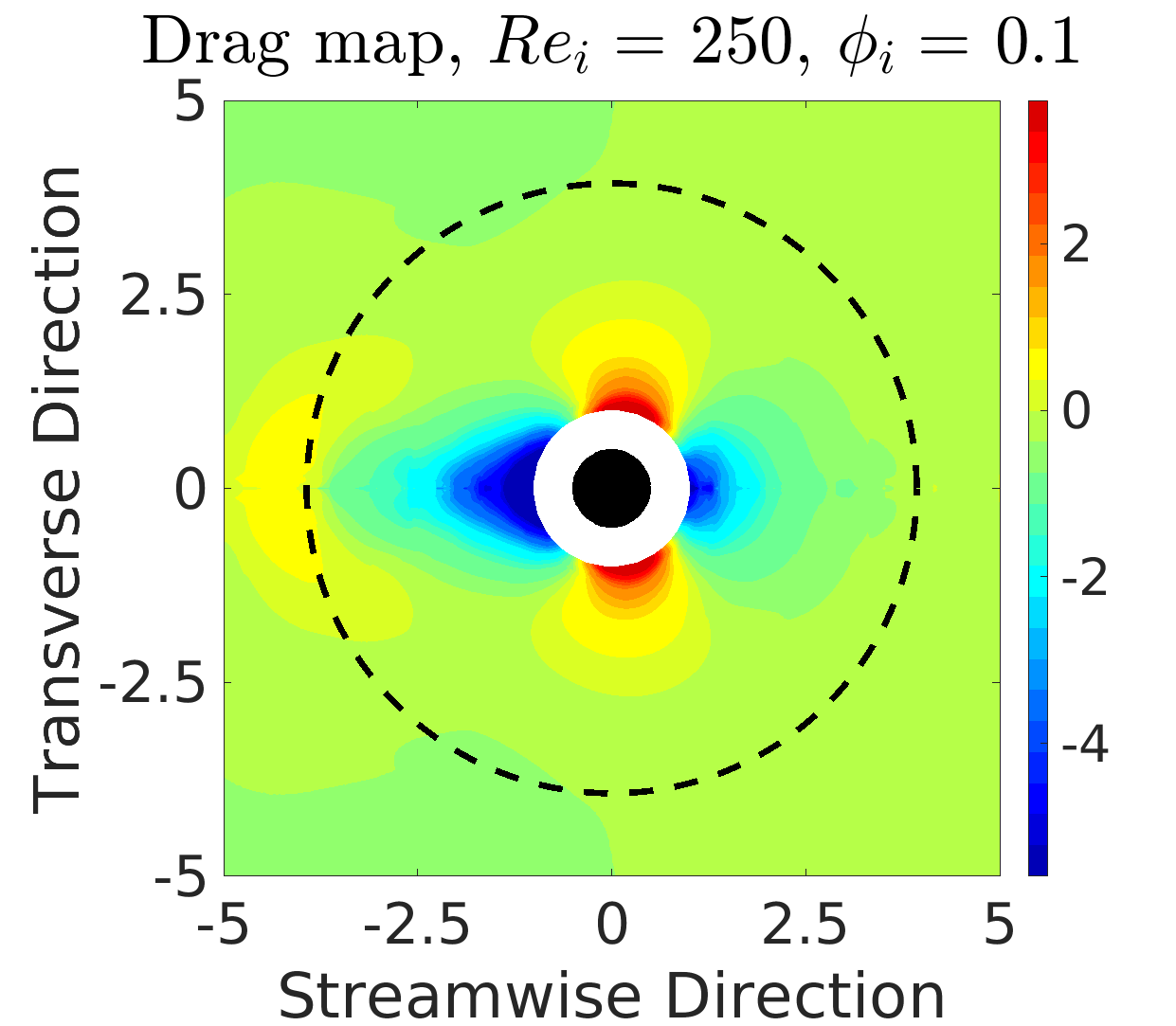}
         \includegraphics[width=0.32\textwidth,keepaspectratio=true]{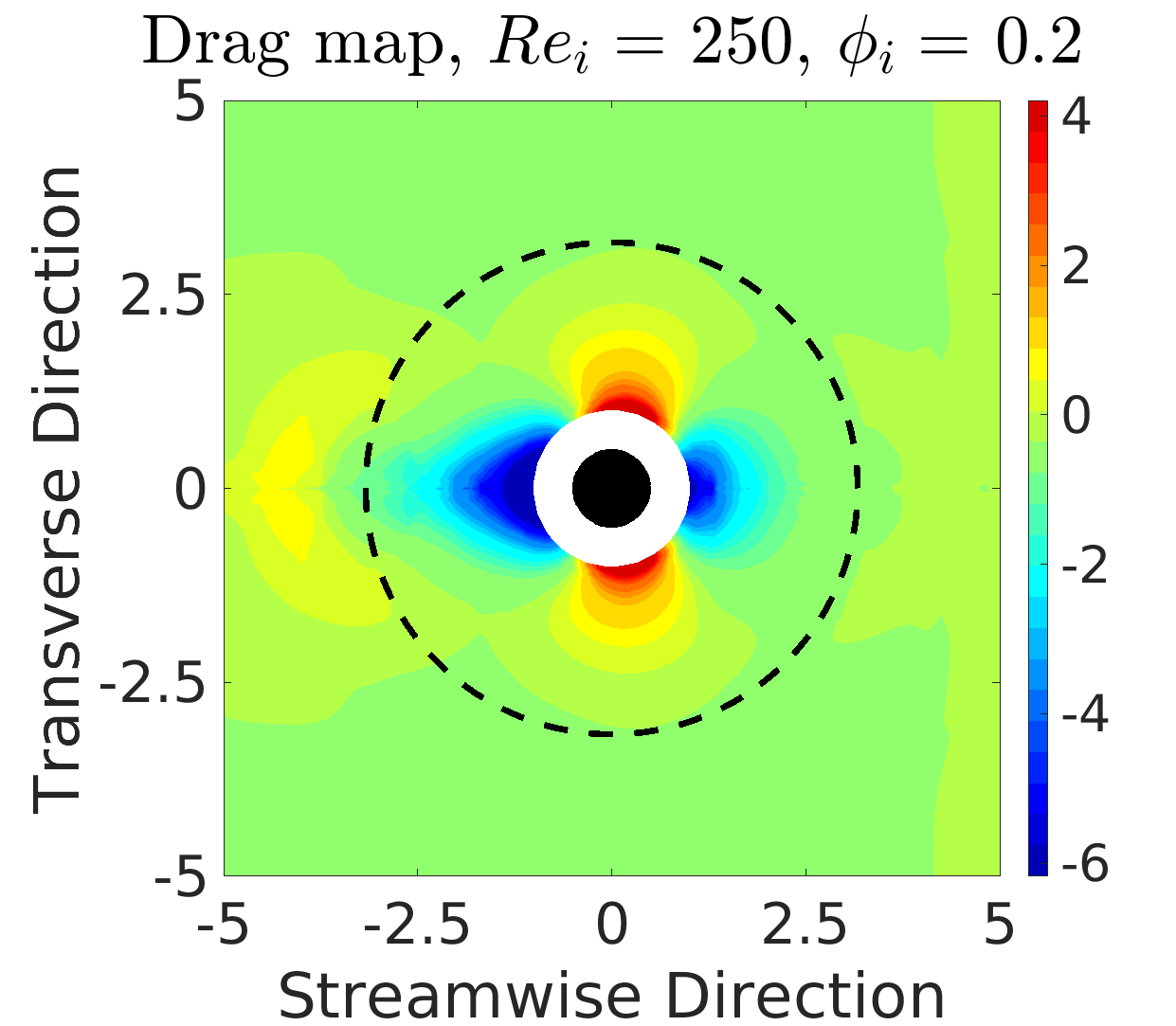}
          \includegraphics[width=0.32\textwidth,keepaspectratio=true]{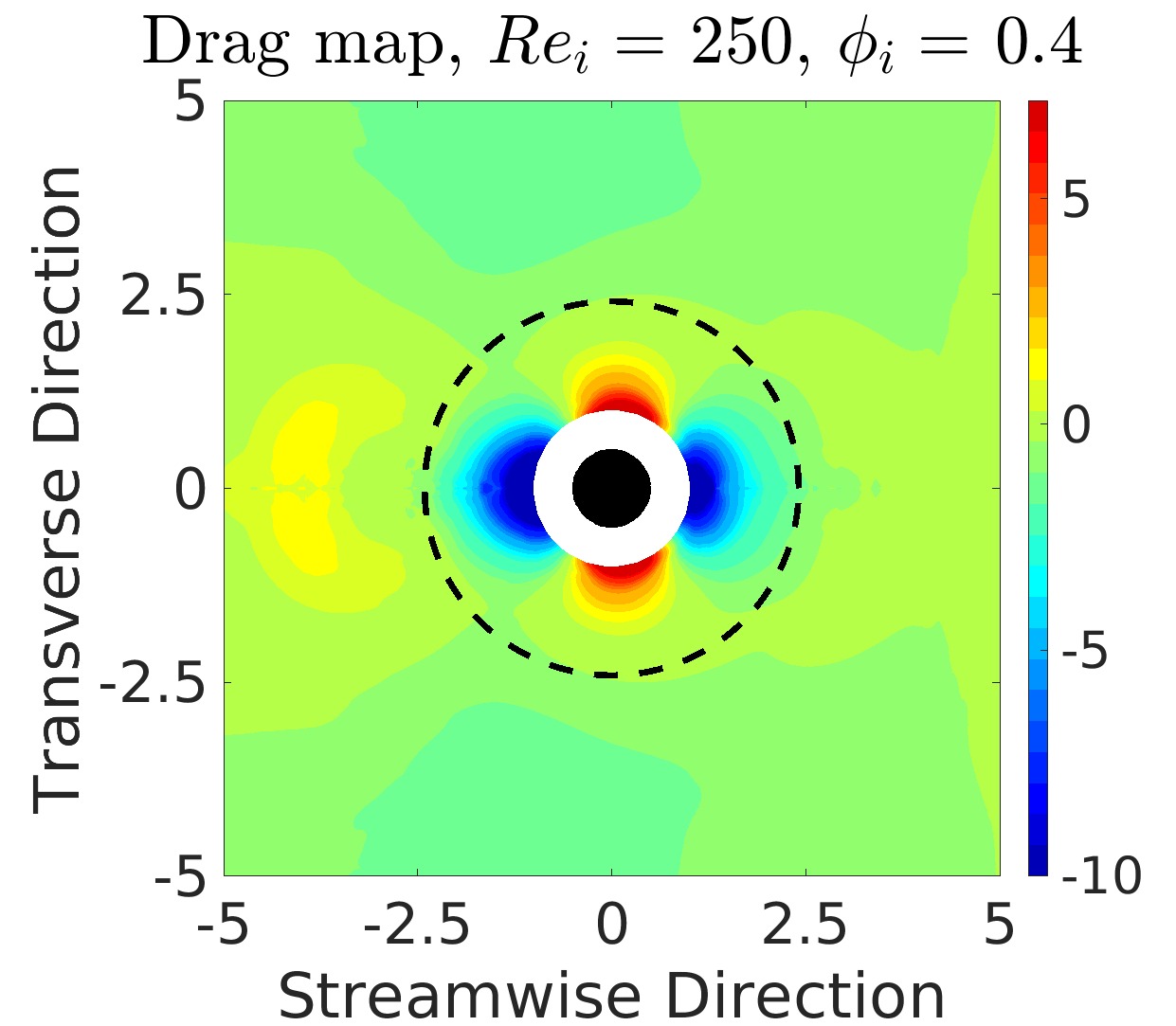}
    \end{subfigure}
    \caption{Binary-interaction drag (streamwise force component) maps obtained using $\tilde{\bm{F}}_{2i}$ with $M_2 = 26$.}
    \label{fig:binary_drag_maps}
\end{figure}

\begin{figure}
    \centering
    \begin{subfigure}[b]{\textwidth}
       \centering
        \includegraphics[width=0.32\textwidth,keepaspectratio=true]{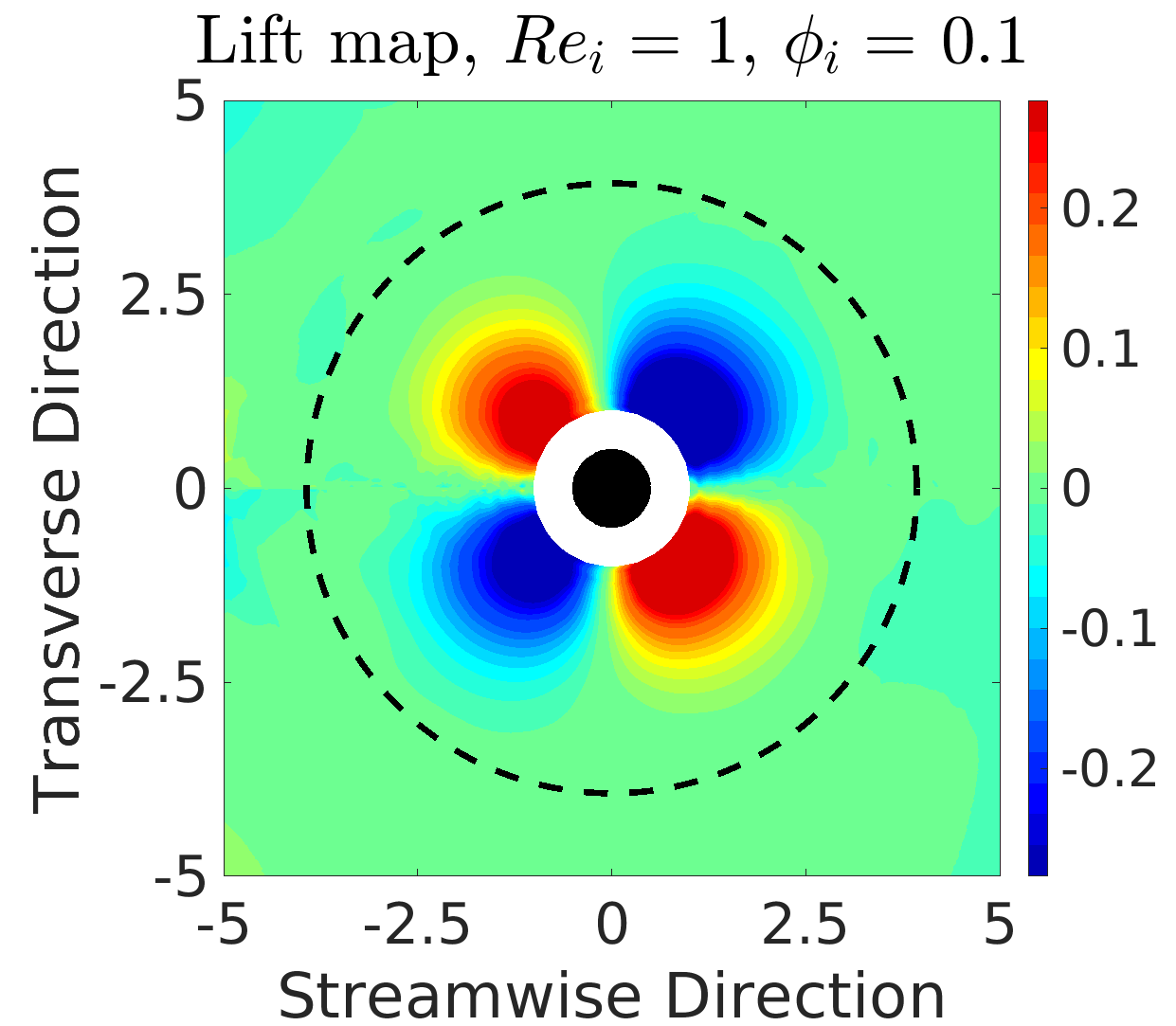}
         \includegraphics[width=0.32\textwidth,keepaspectratio=true]{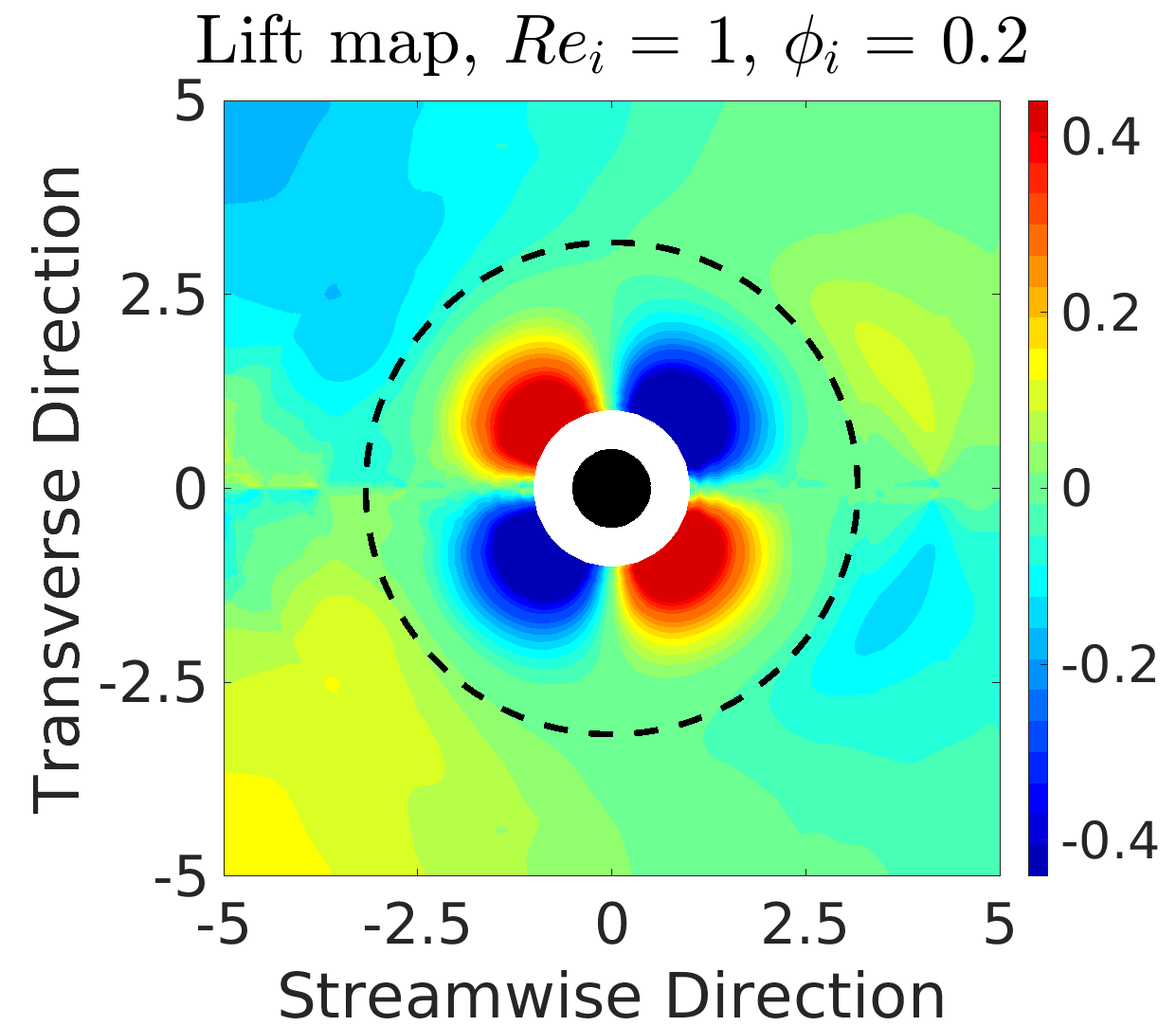}
          \includegraphics[width=0.32\textwidth,keepaspectratio=true]{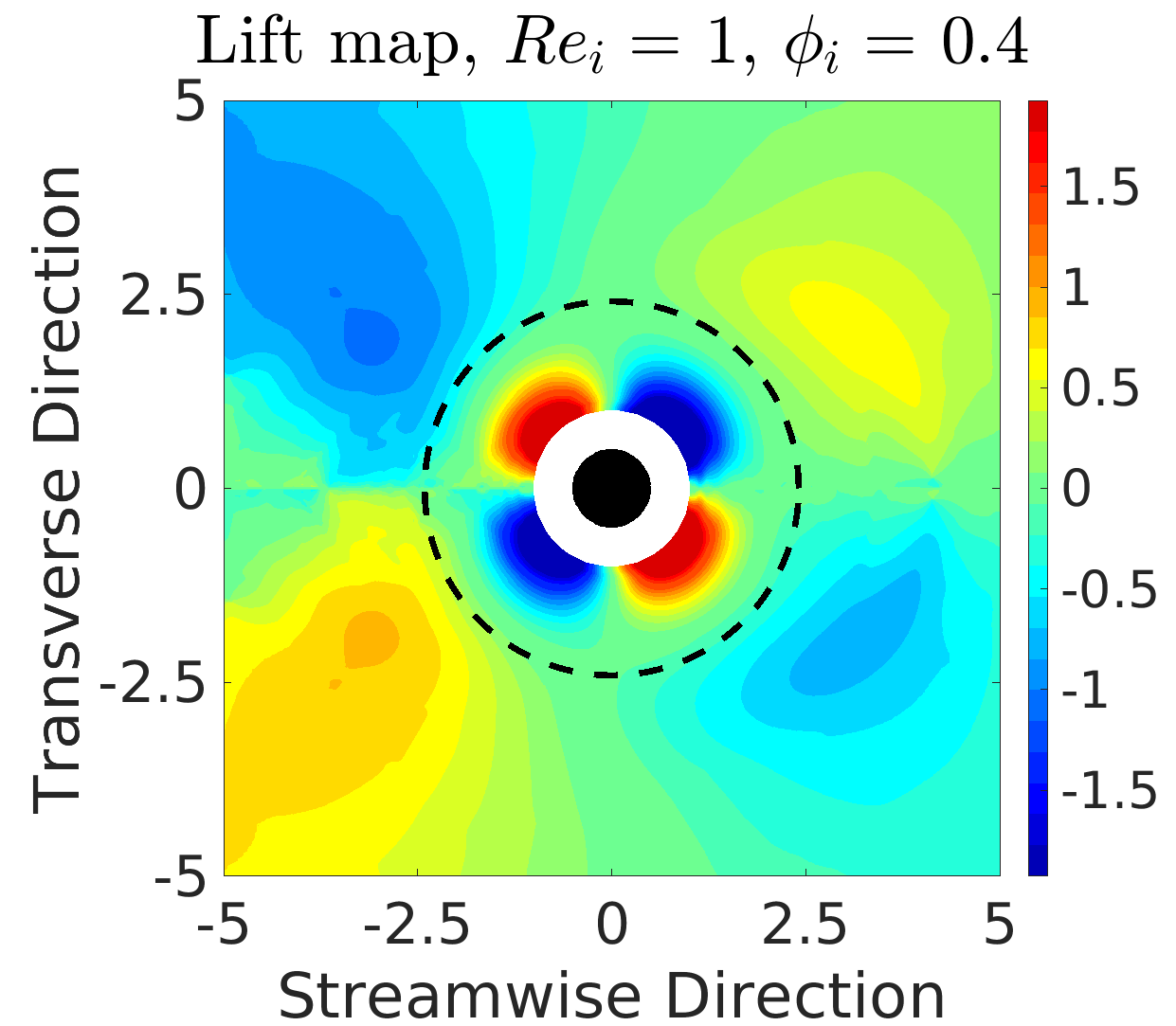}
    \end{subfigure}
    \begin{subfigure}[b]{\textwidth}
       \centering
        \includegraphics[width=0.32\textwidth,keepaspectratio=true]{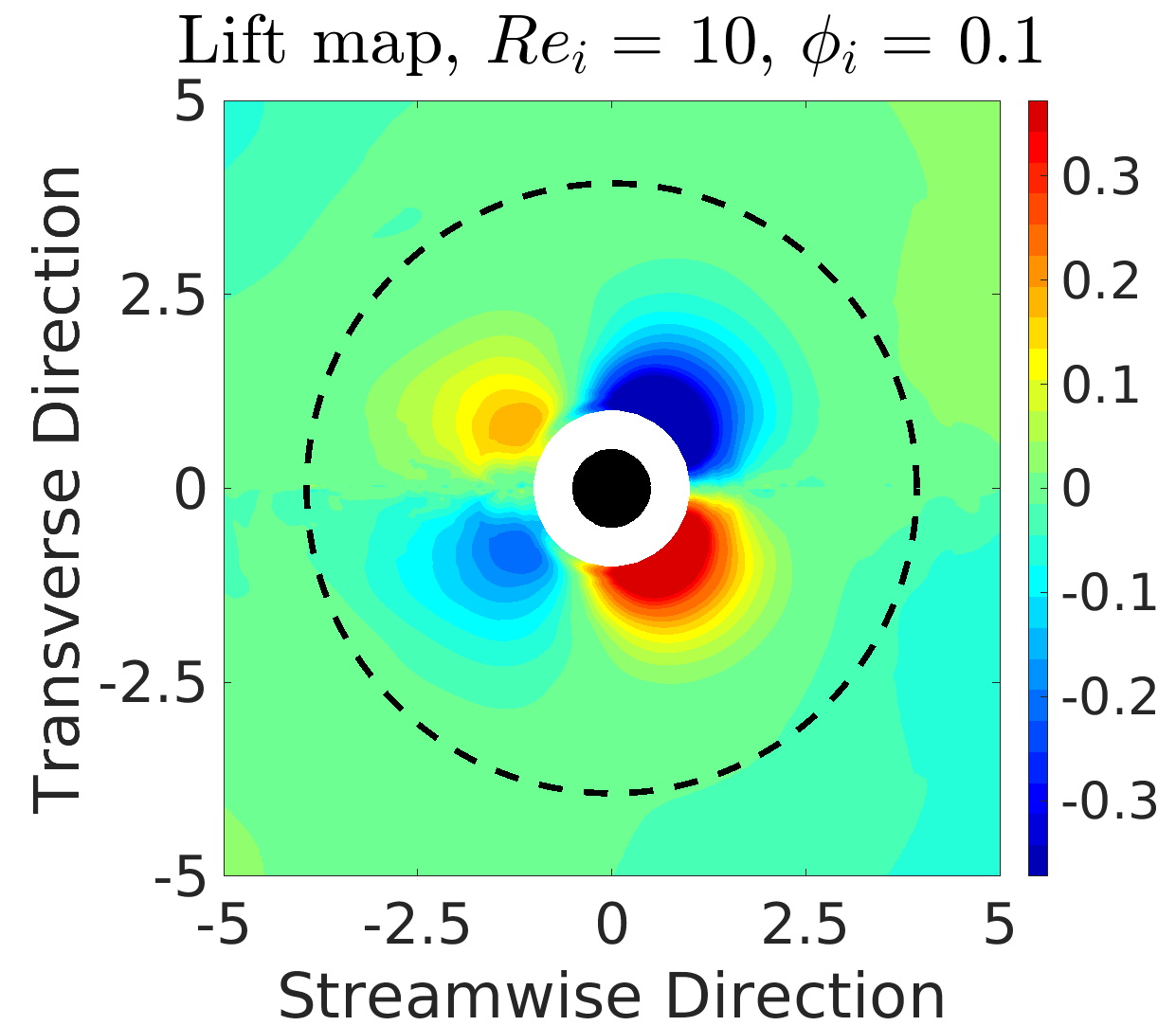}
         \includegraphics[width=0.32\textwidth,keepaspectratio=true]{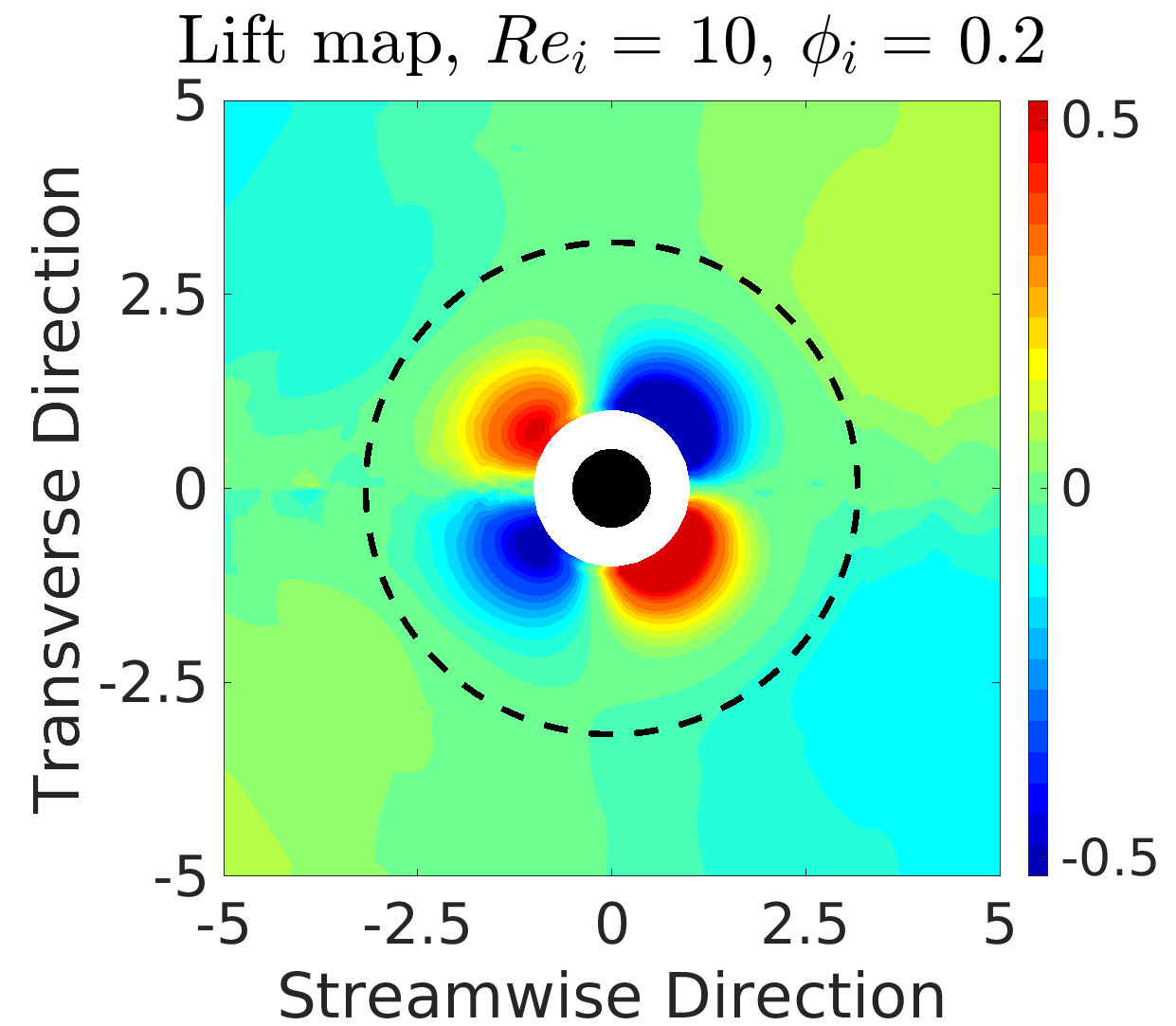}
          \includegraphics[width=0.32\textwidth,keepaspectratio=true]{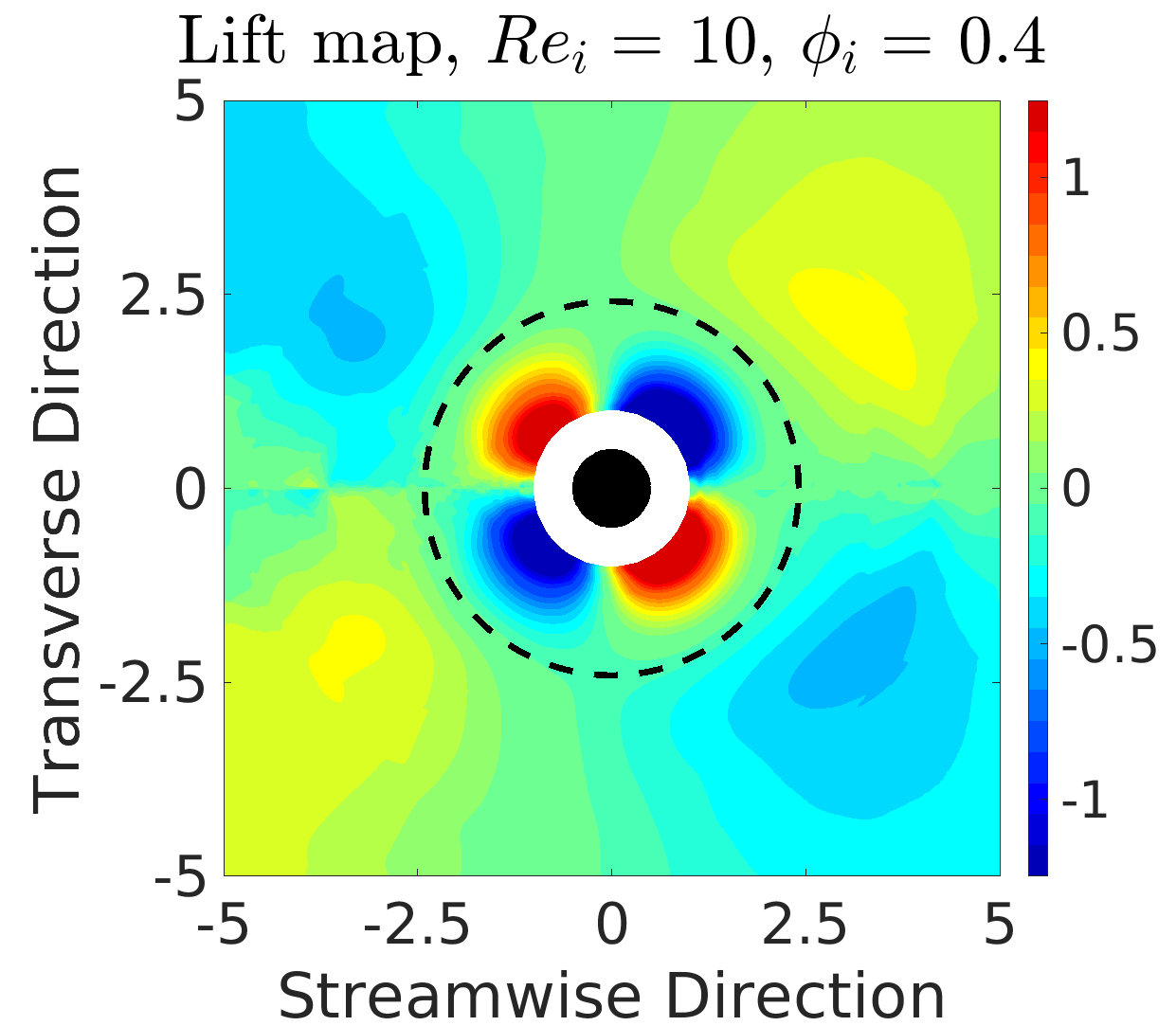}
    \end{subfigure}
    \begin{subfigure}[b]{\textwidth}
       \centering
        \includegraphics[width=0.32\textwidth,keepaspectratio=true]{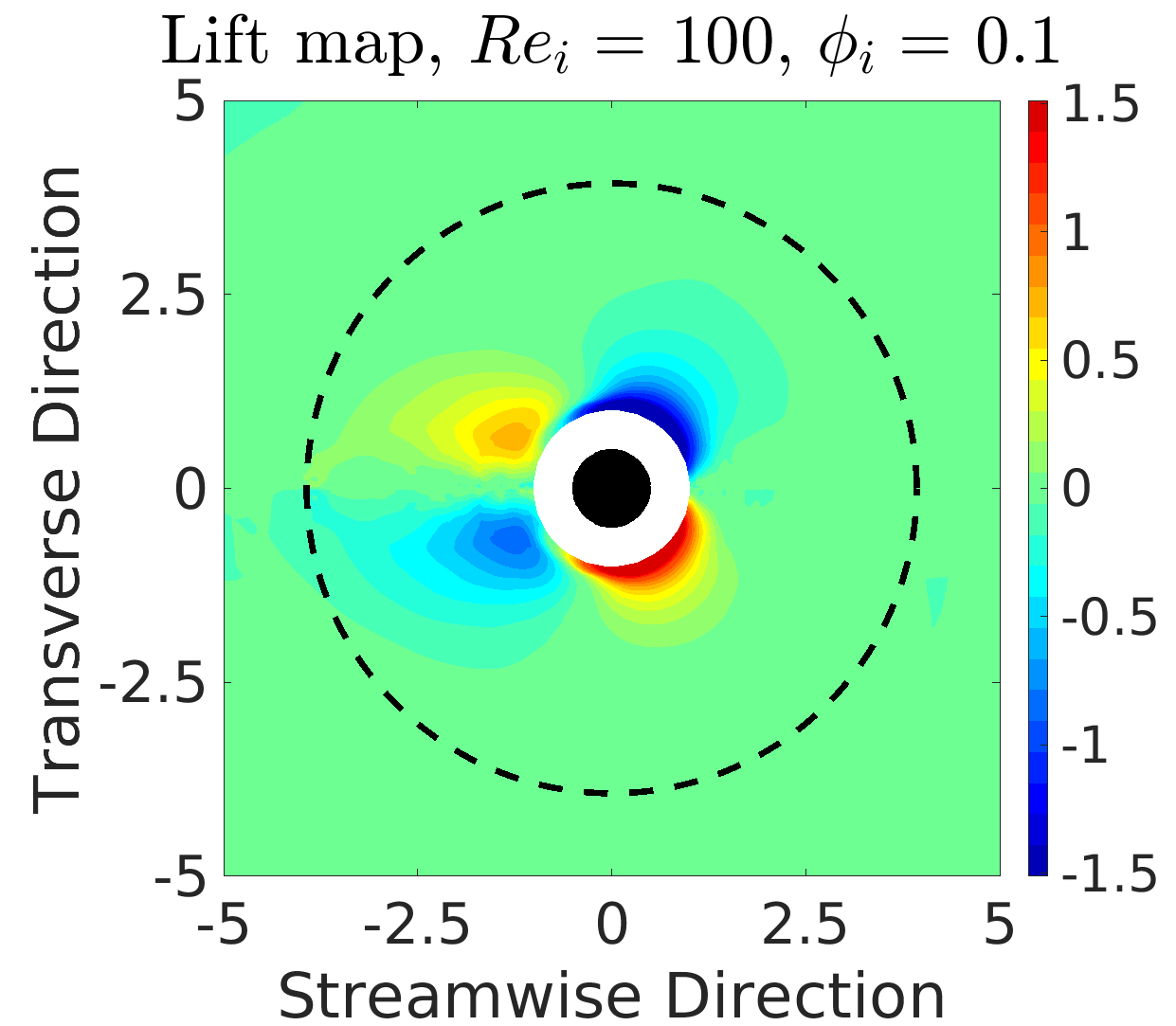}
         \includegraphics[width=0.32\textwidth,keepaspectratio=true]{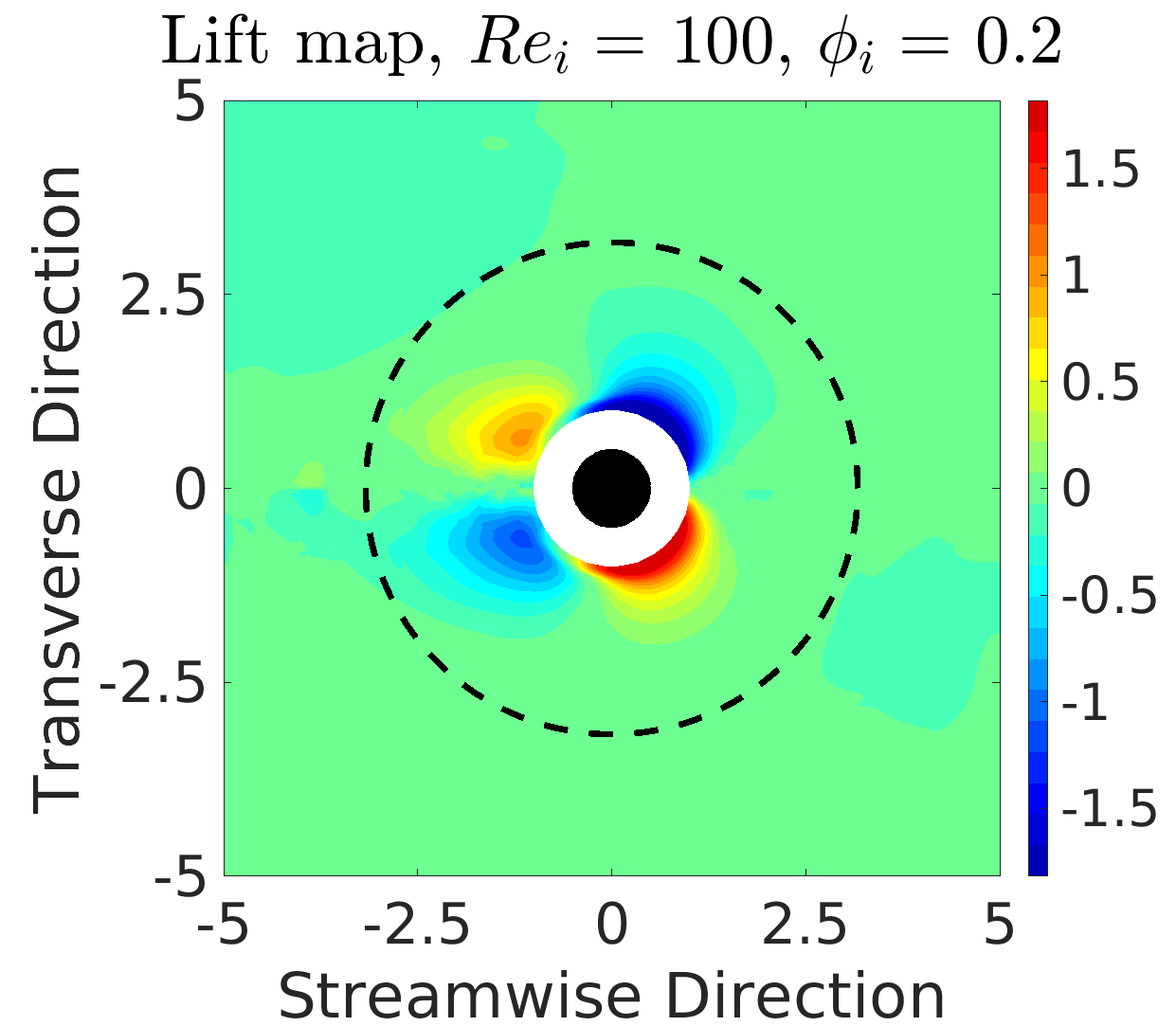}
          \includegraphics[width=0.32\textwidth,keepaspectratio=true]{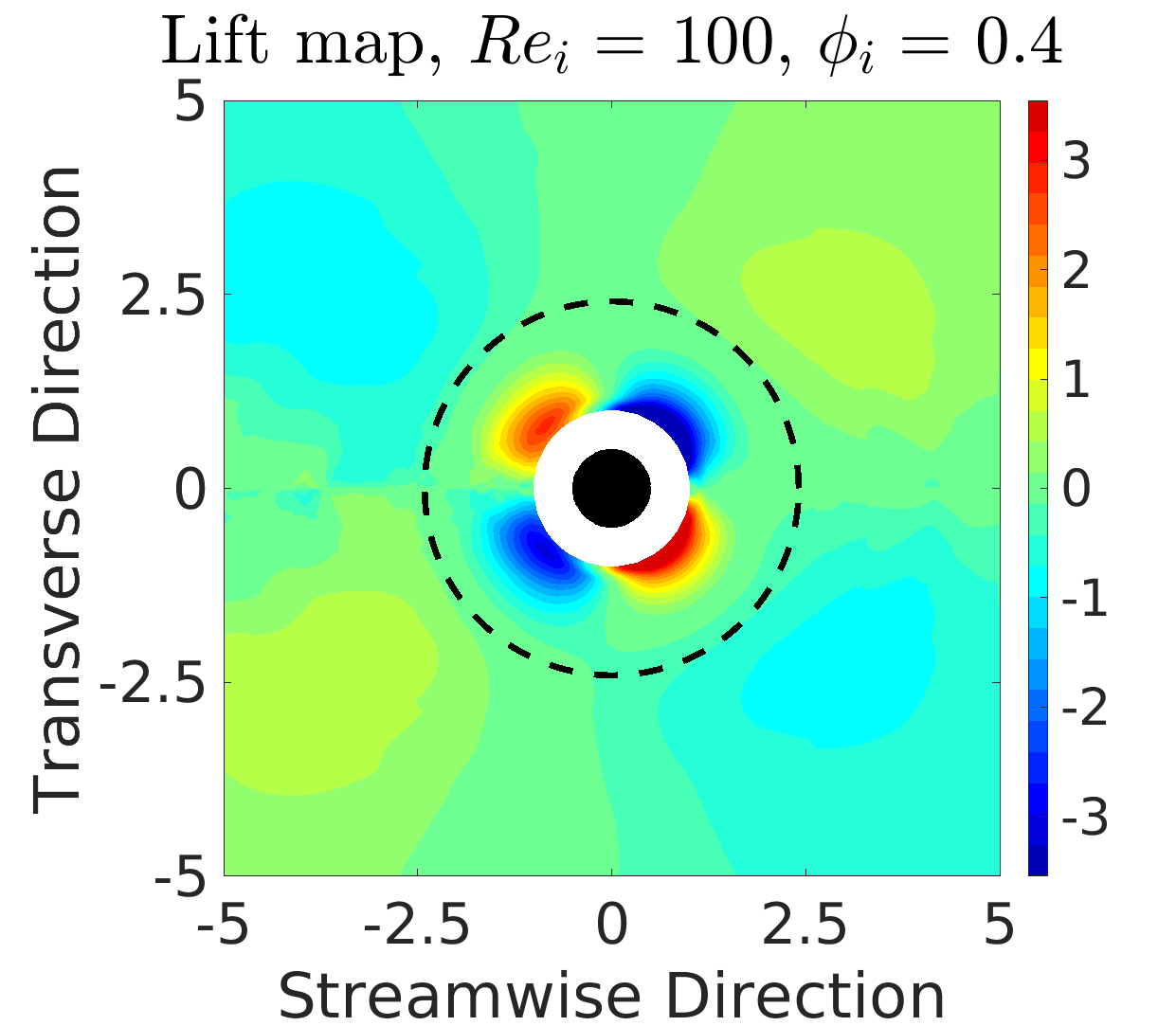}
    \end{subfigure}
    \begin{subfigure}[b]{\textwidth}
       \centering
        \includegraphics[width=0.32\textwidth,keepaspectratio=true]{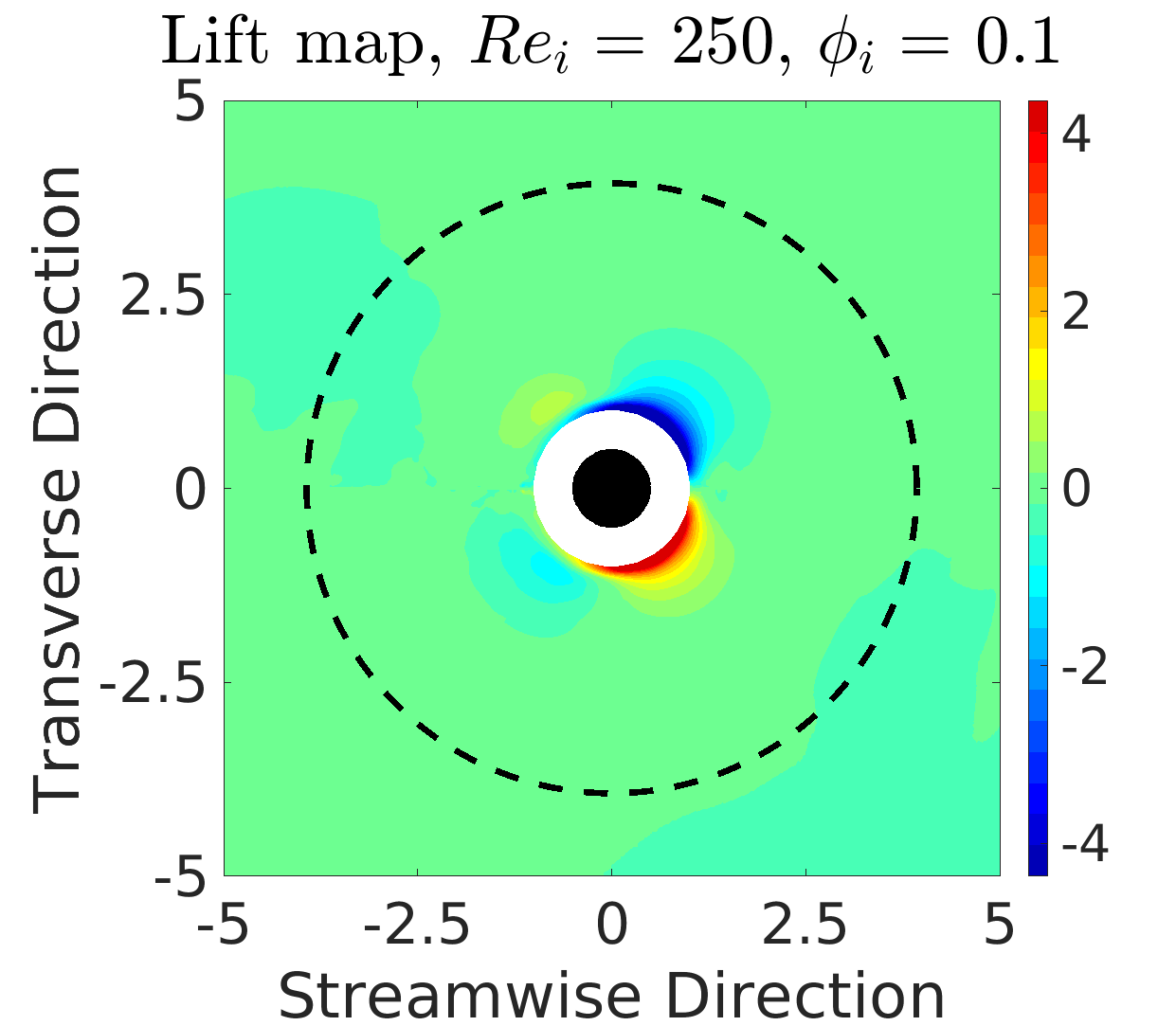}
         \includegraphics[width=0.32\textwidth,keepaspectratio=true]{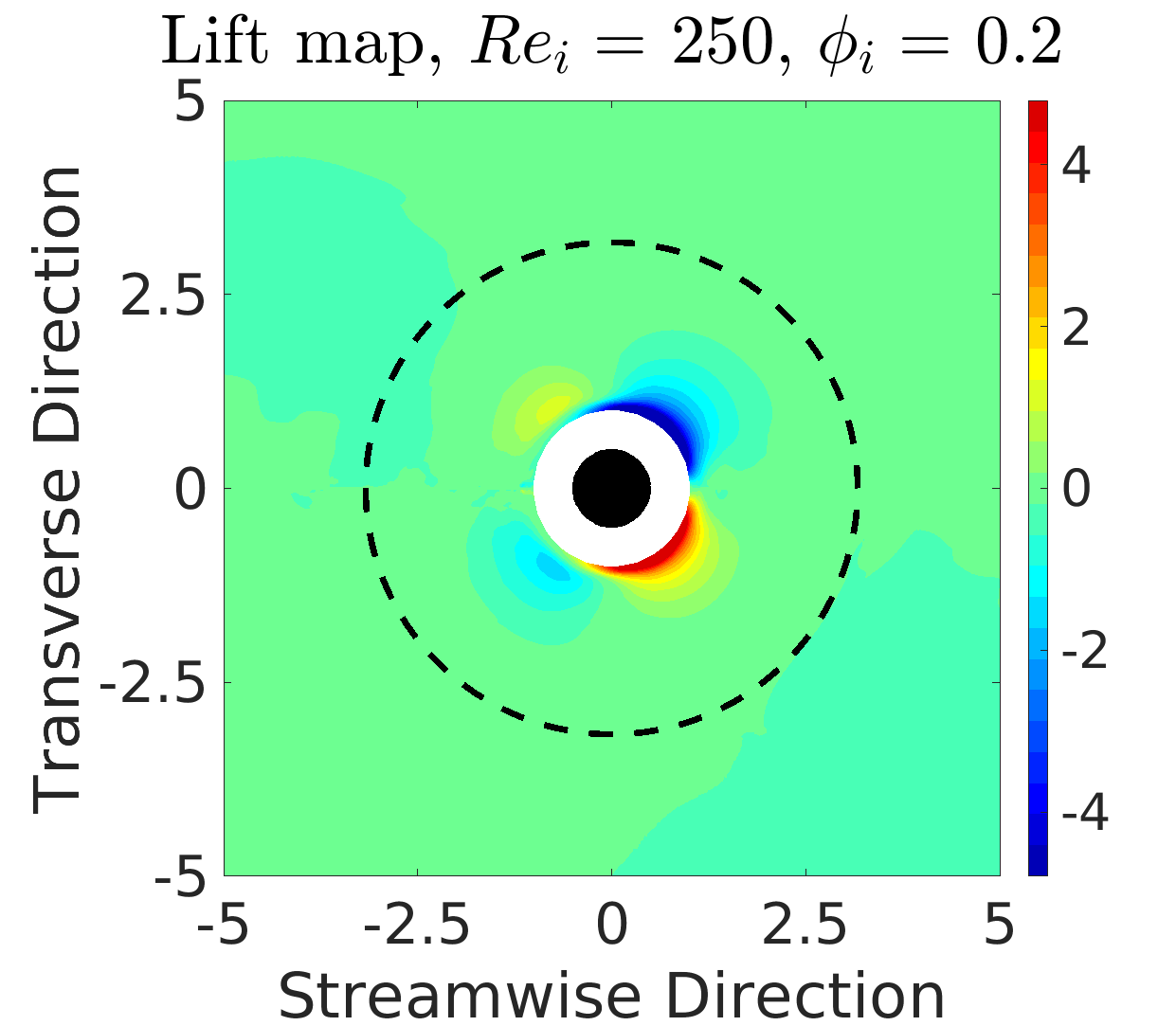}
          \includegraphics[width=0.32\textwidth,keepaspectratio=true]{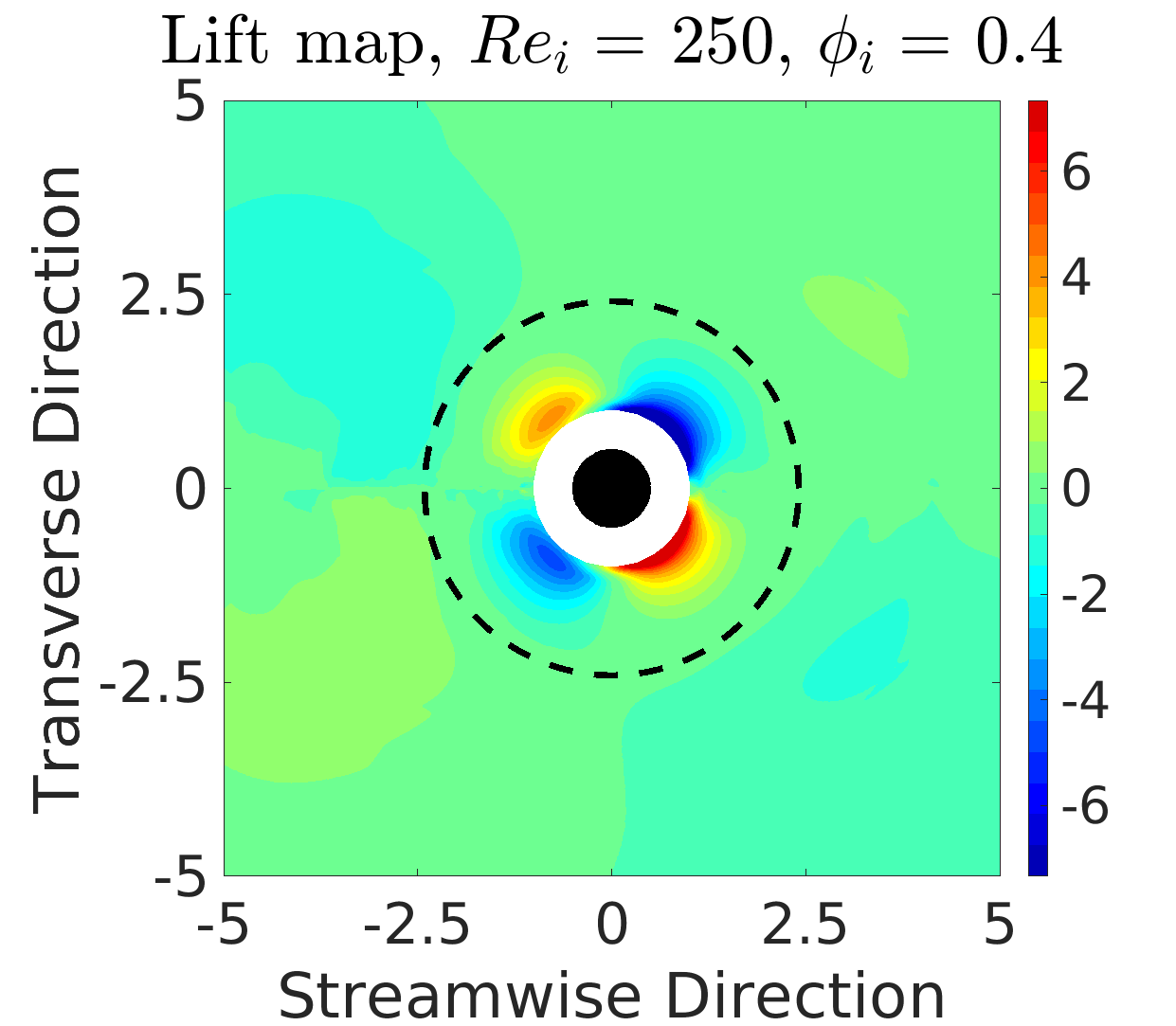}
    \end{subfigure}
    \caption{Binary-interaction lift (transverse force component) maps obtained using $\tilde{\bm{F}}_{2i}$ with $M_2=26$.}
    \label{fig:binary_lift_maps}
\end{figure}

\begin{figure}
    \centering
    \begin{subfigure}[b]{\textwidth}
       \centering
        \includegraphics[width=0.32\textwidth,keepaspectratio=true]{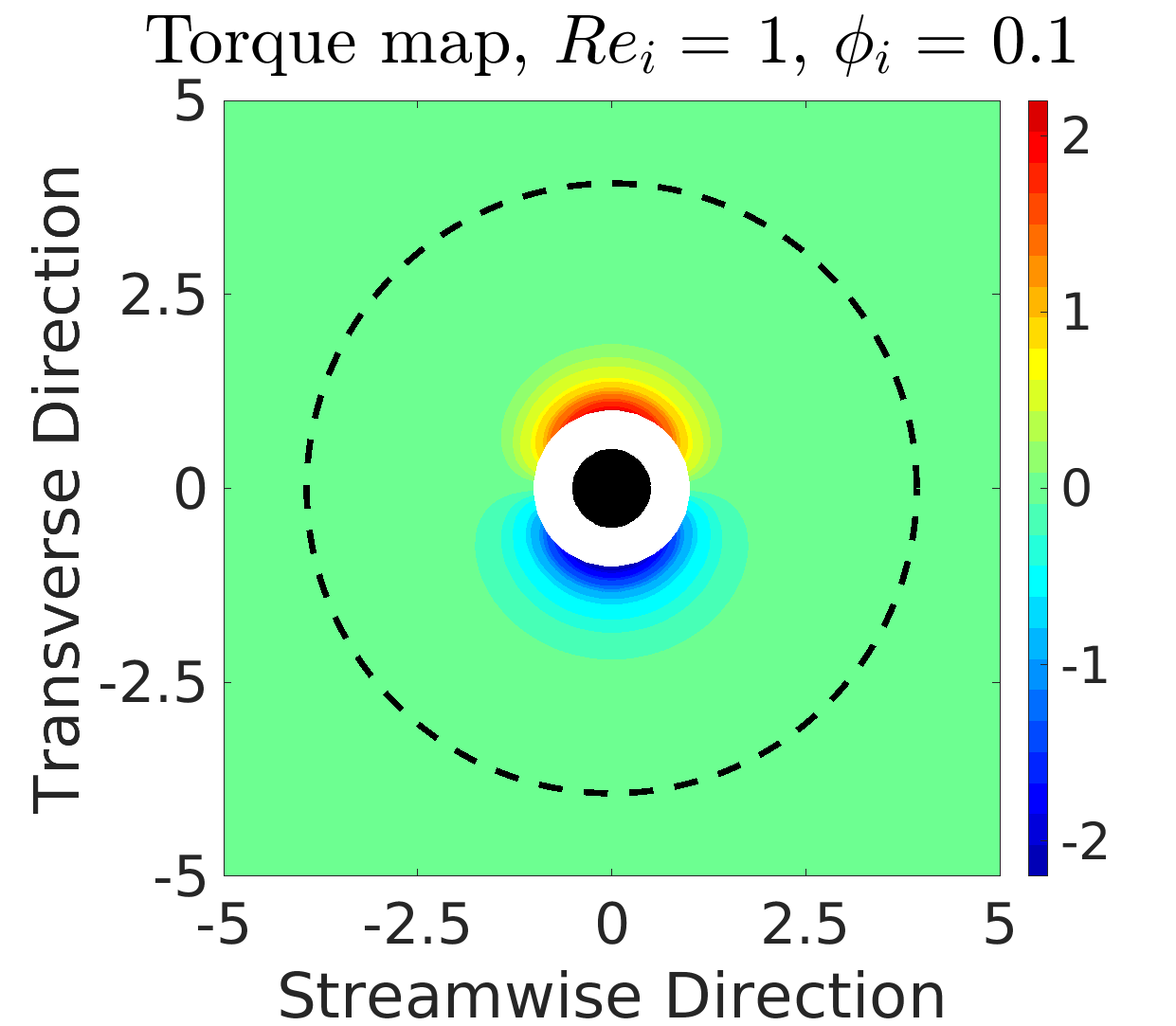}
         \includegraphics[width=0.32\textwidth,keepaspectratio=true]{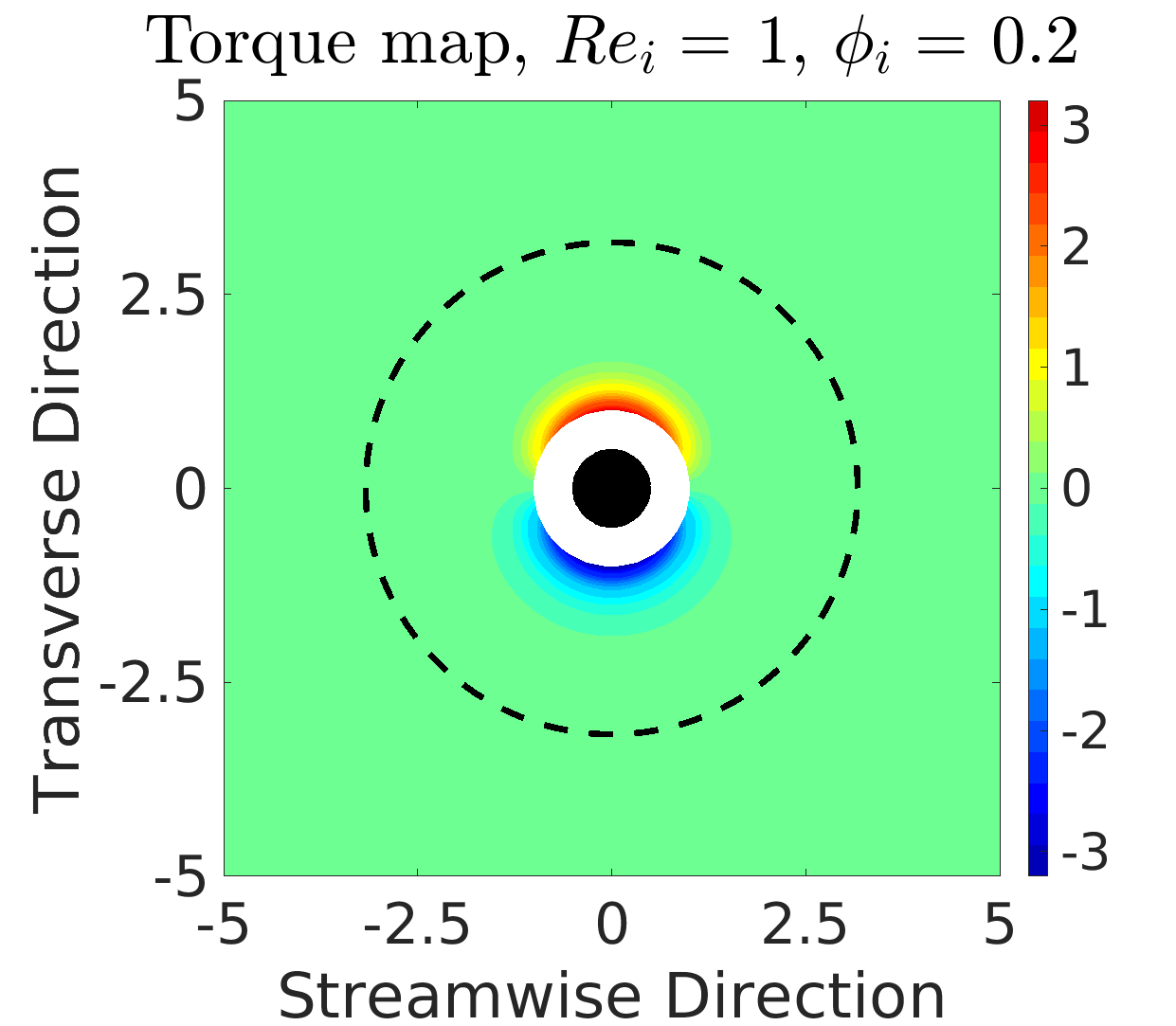}
          \includegraphics[width=0.32\textwidth,keepaspectratio=true]{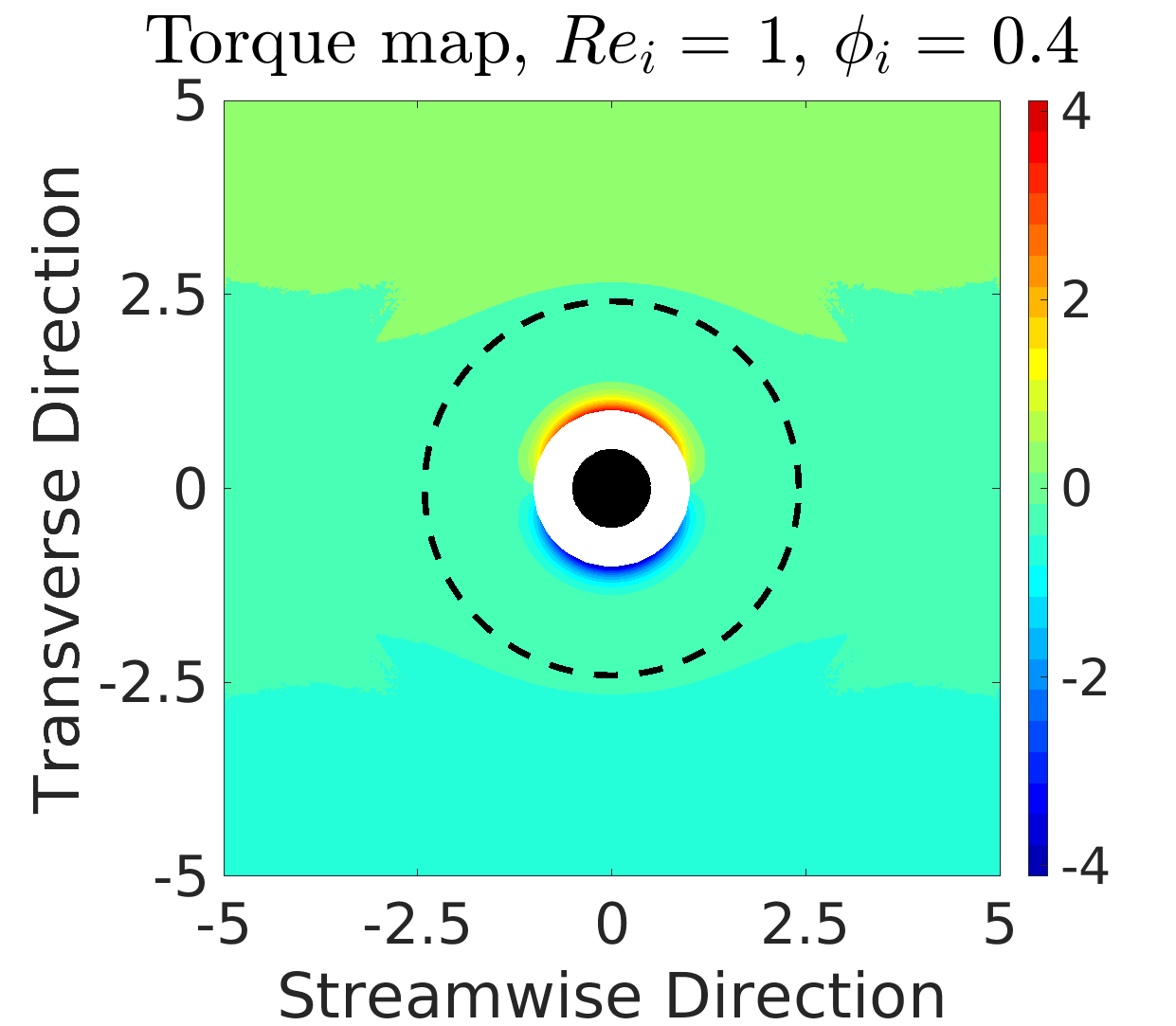}
    \end{subfigure}
    \begin{subfigure}[b]{\textwidth}
       \centering
        \includegraphics[width=0.32\textwidth,keepaspectratio=true]{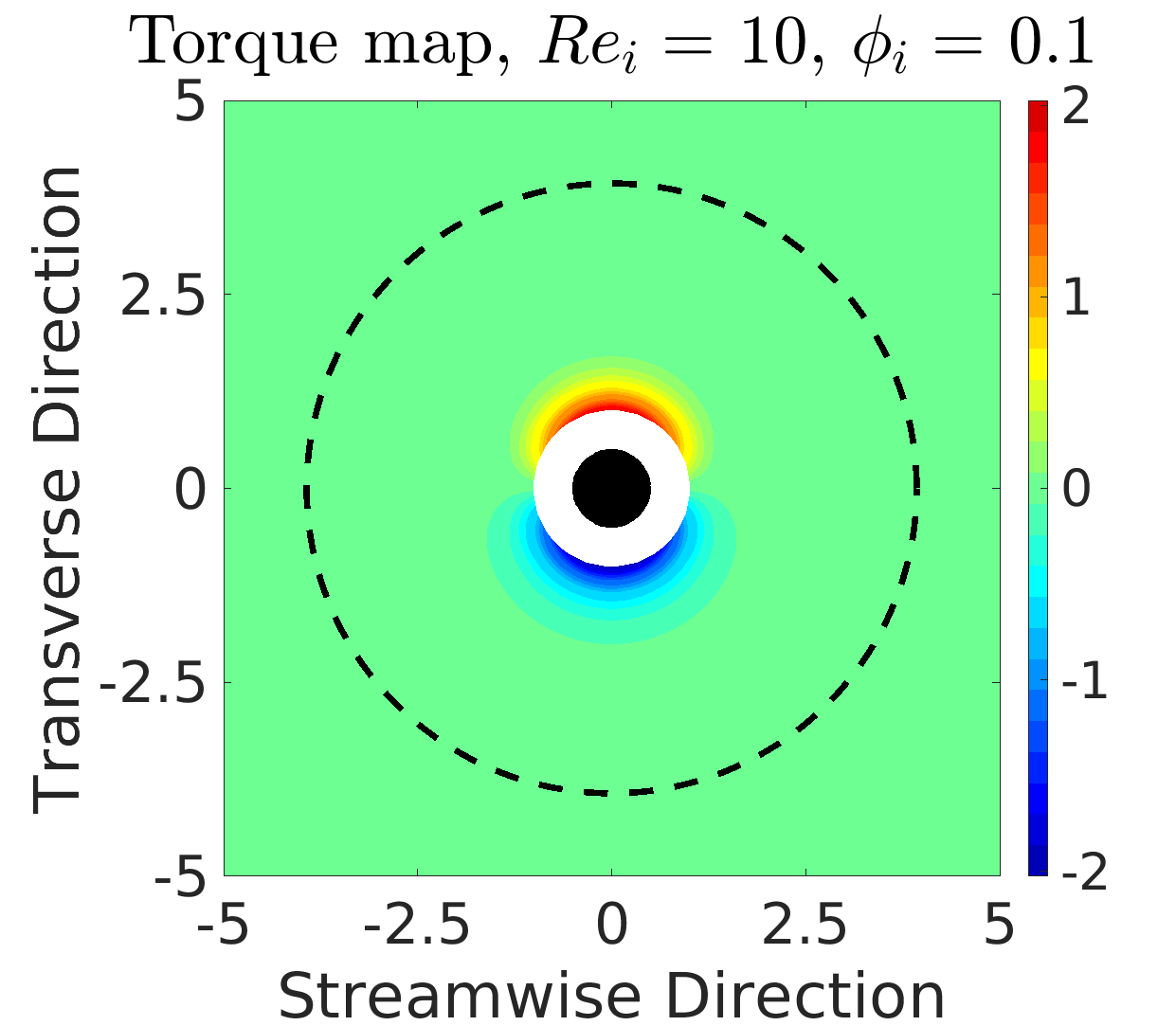}
         \includegraphics[width=0.32\textwidth,keepaspectratio=true]{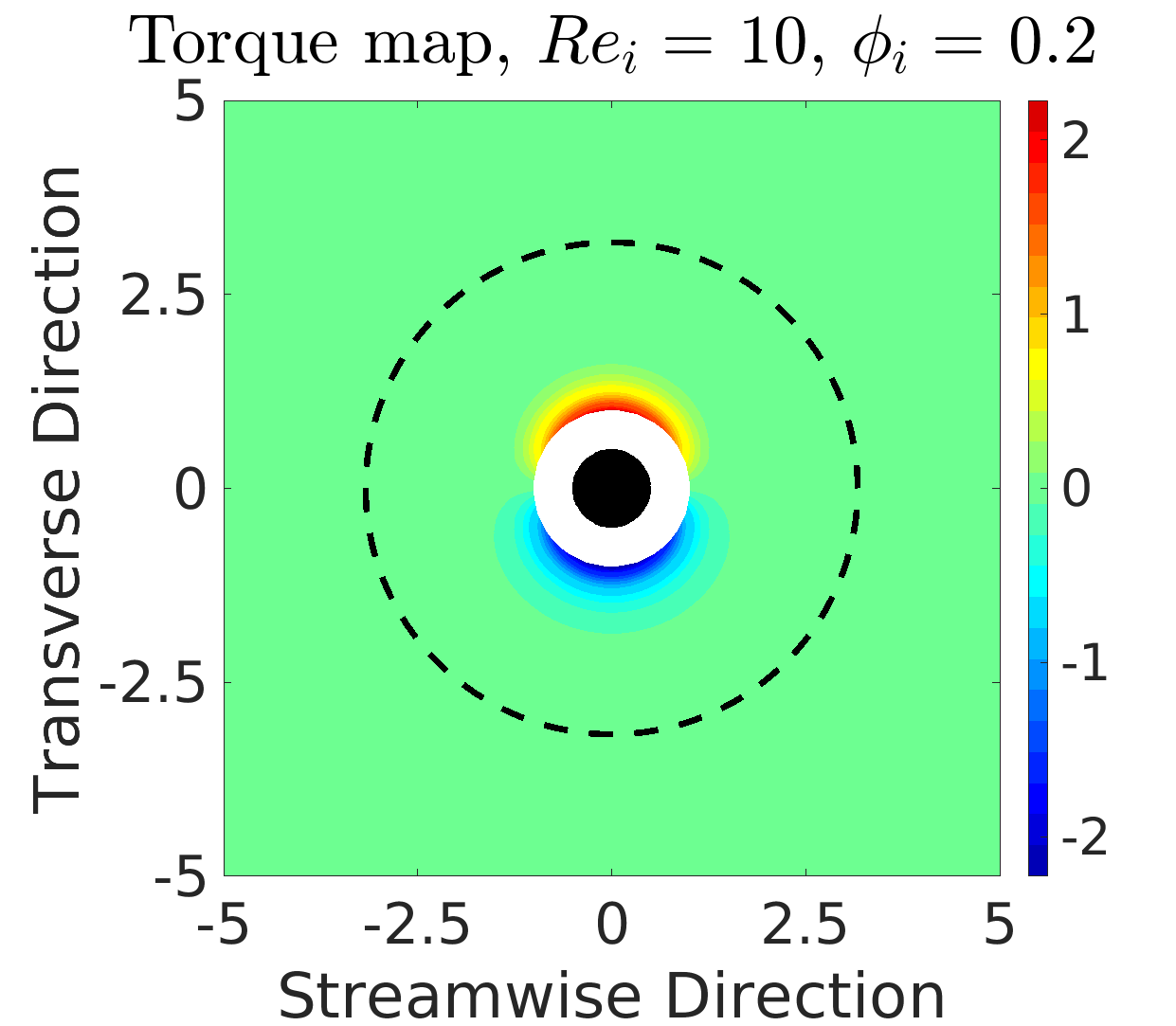}
          \includegraphics[width=0.32\textwidth,keepaspectratio=true]{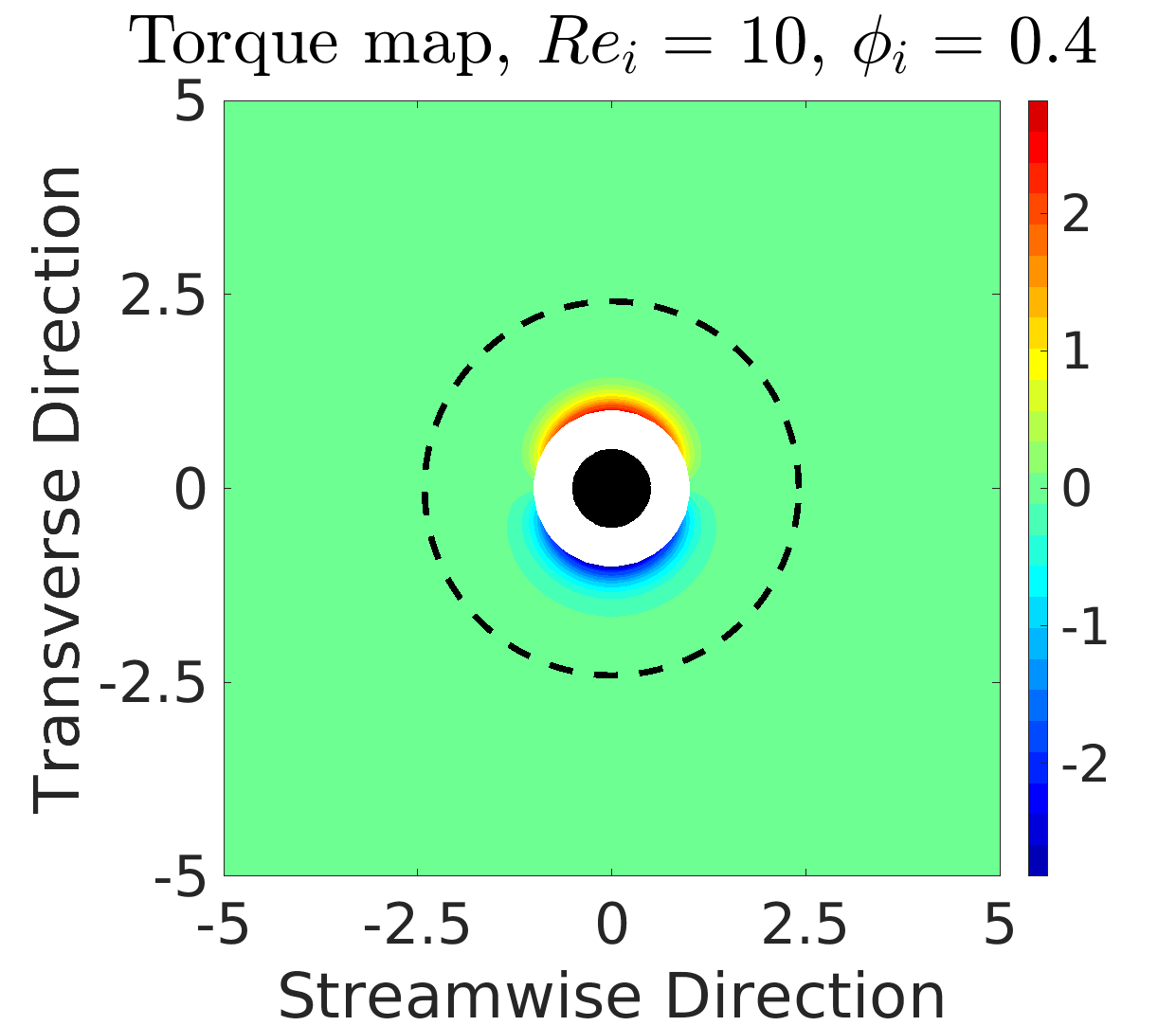}
    \end{subfigure}
    \begin{subfigure}[b]{\textwidth}
       \centering
        \includegraphics[width=0.32\textwidth,keepaspectratio=true]{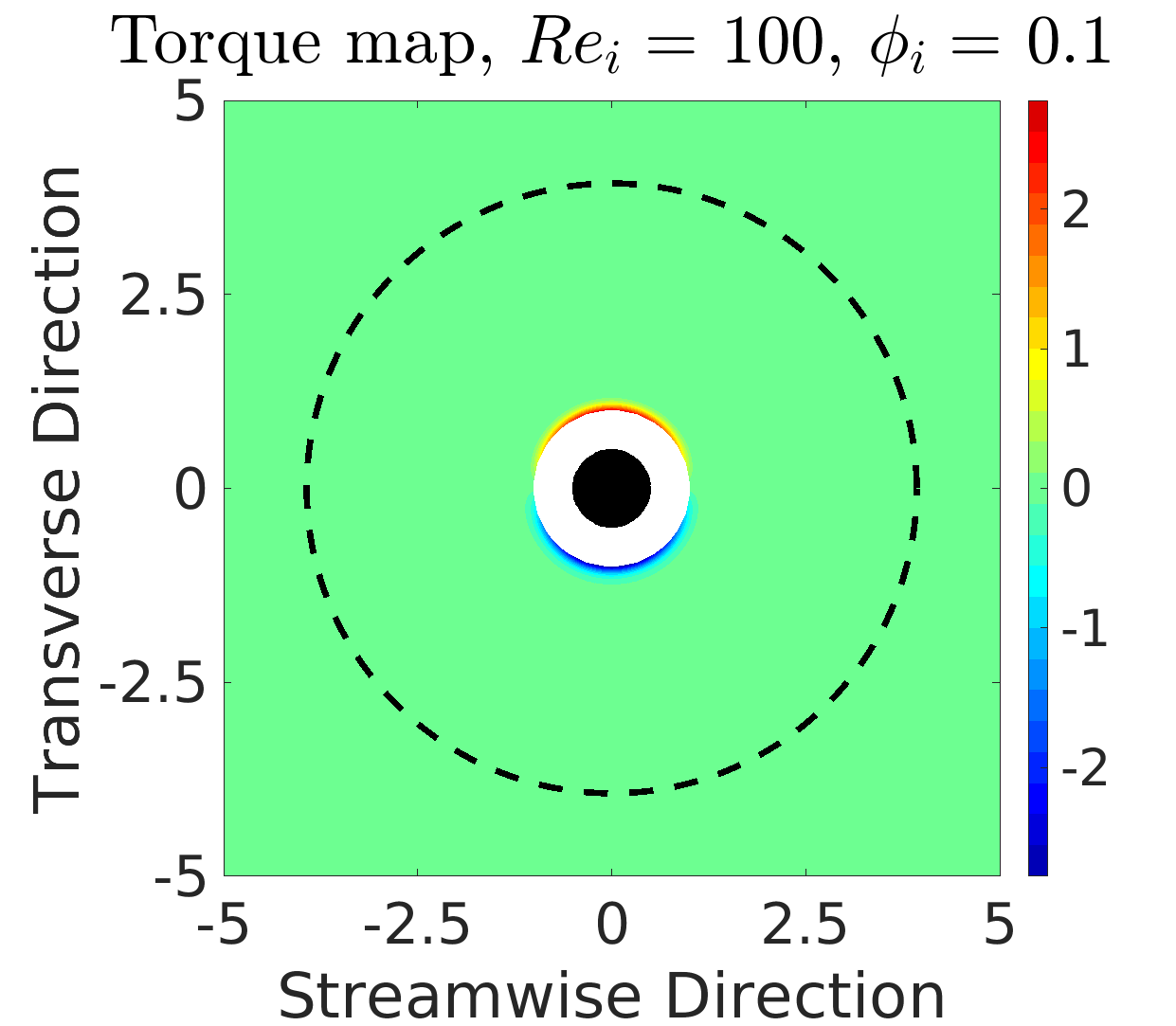}
         \includegraphics[width=0.32\textwidth,keepaspectratio=true]{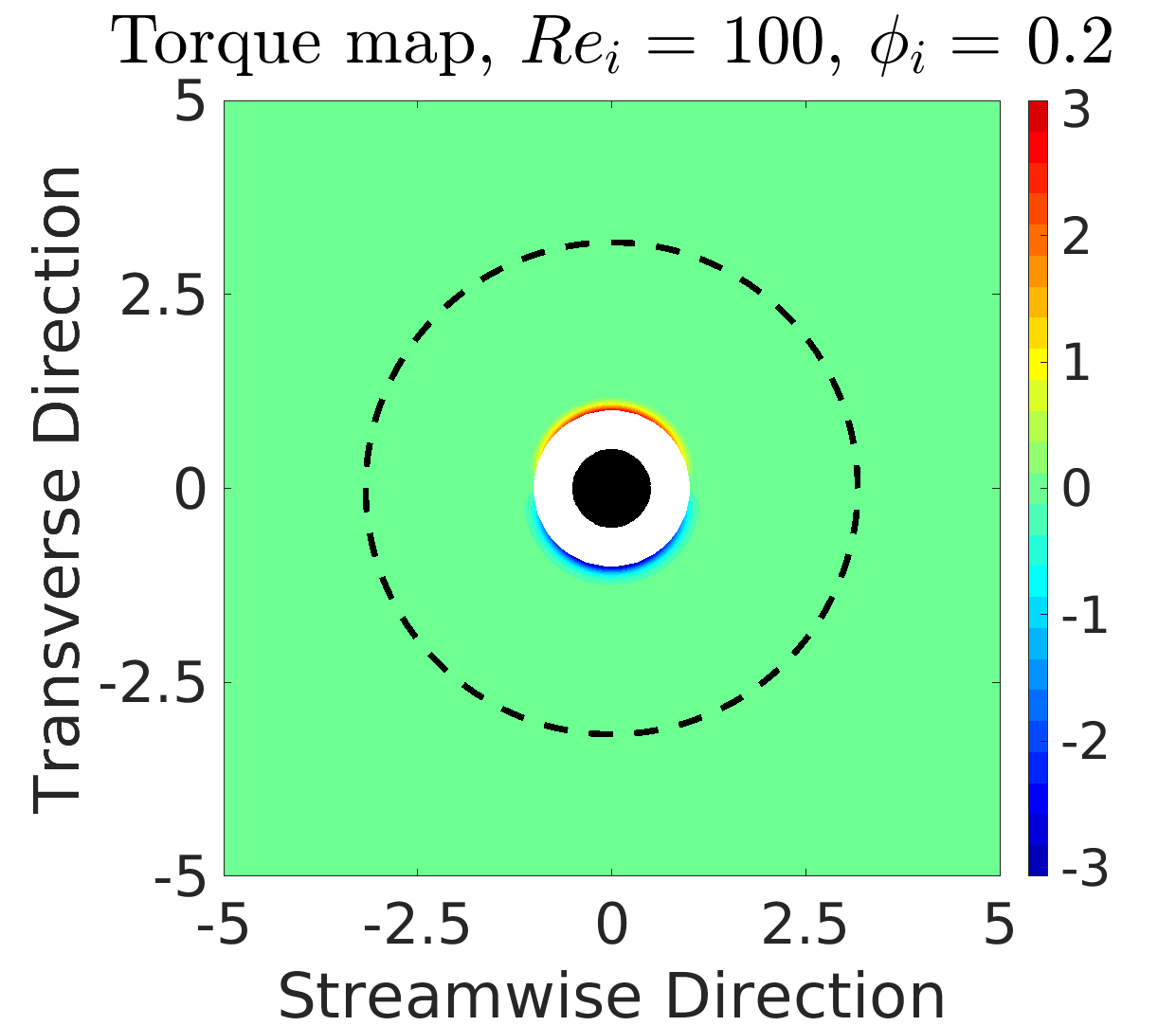}
          \includegraphics[width=0.32\textwidth,keepaspectratio=true]{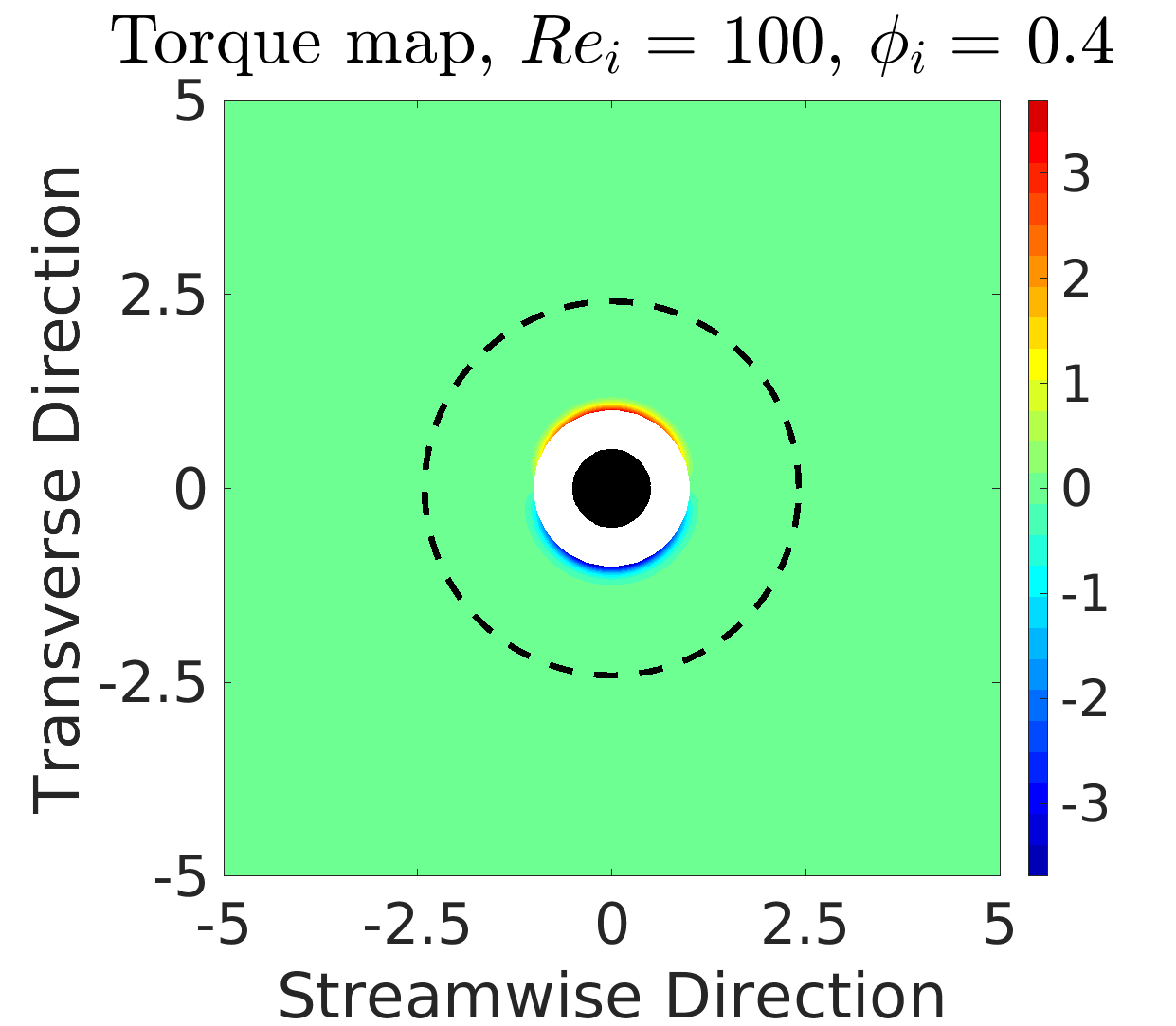}
    \end{subfigure}
    \begin{subfigure}[b]{\textwidth}
       \centering
        \includegraphics[width=0.32\textwidth,keepaspectratio=true]{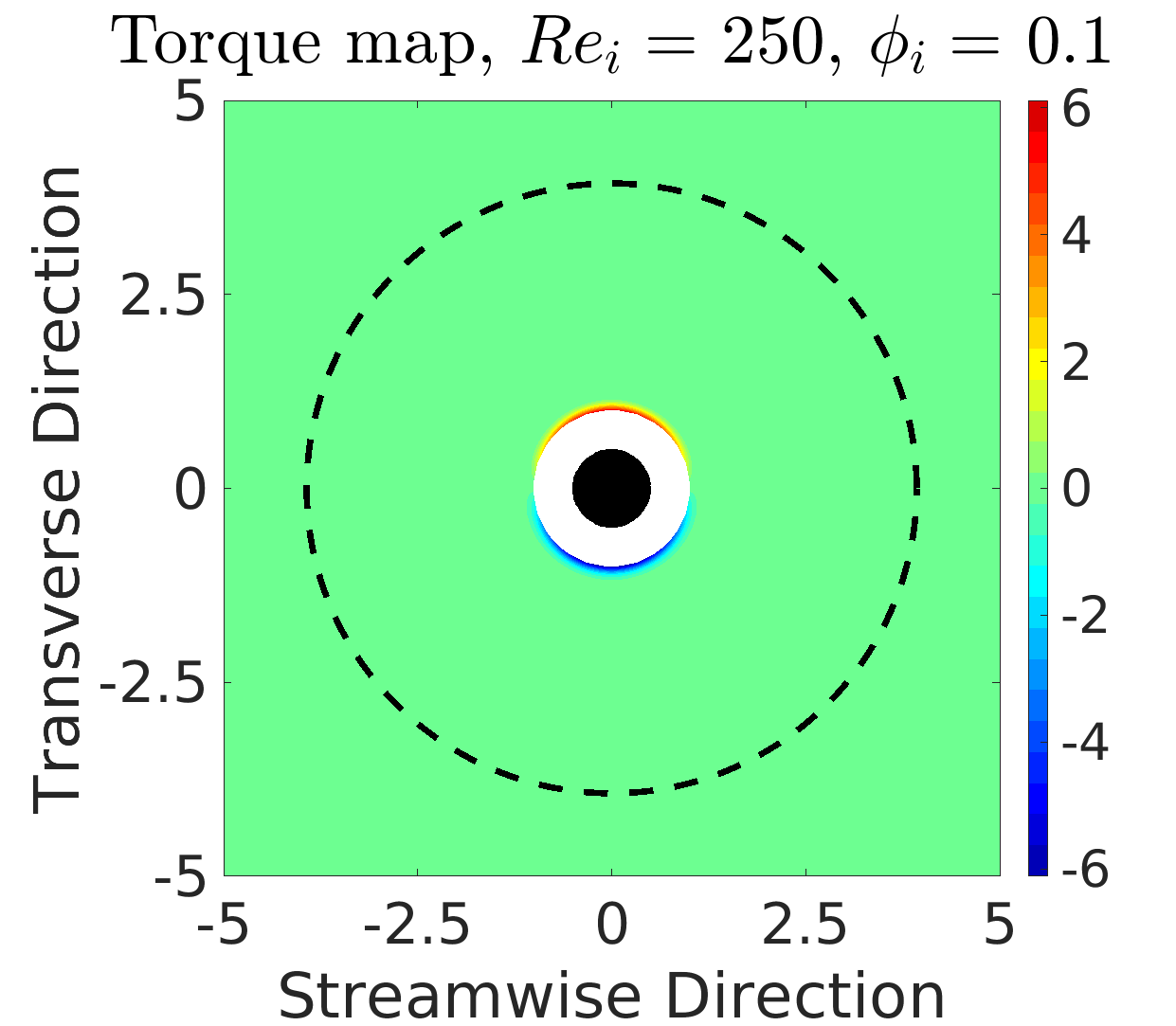}
         \includegraphics[width=0.32\textwidth,keepaspectratio=true]{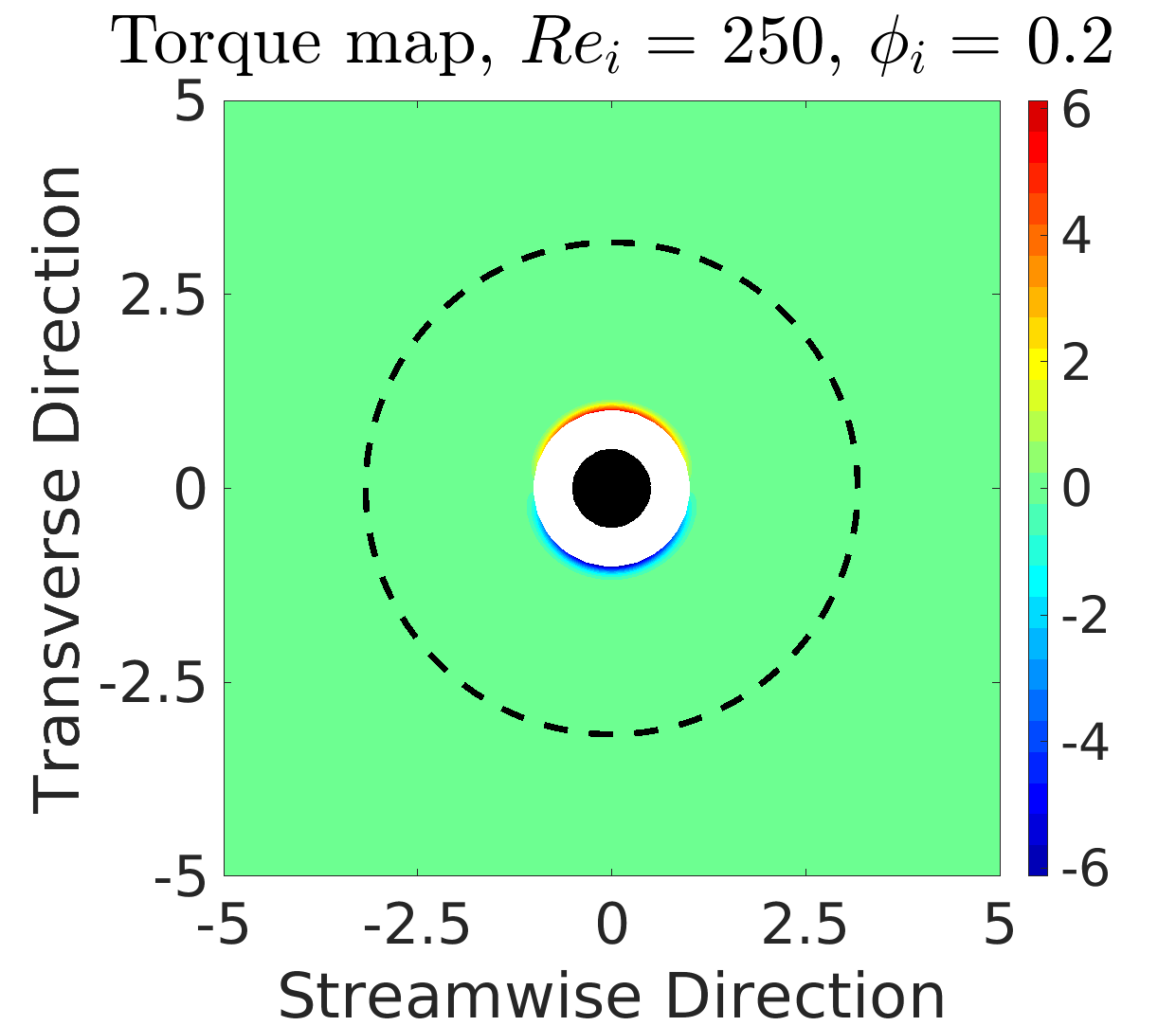}
          \includegraphics[width=0.32\textwidth,keepaspectratio=true]{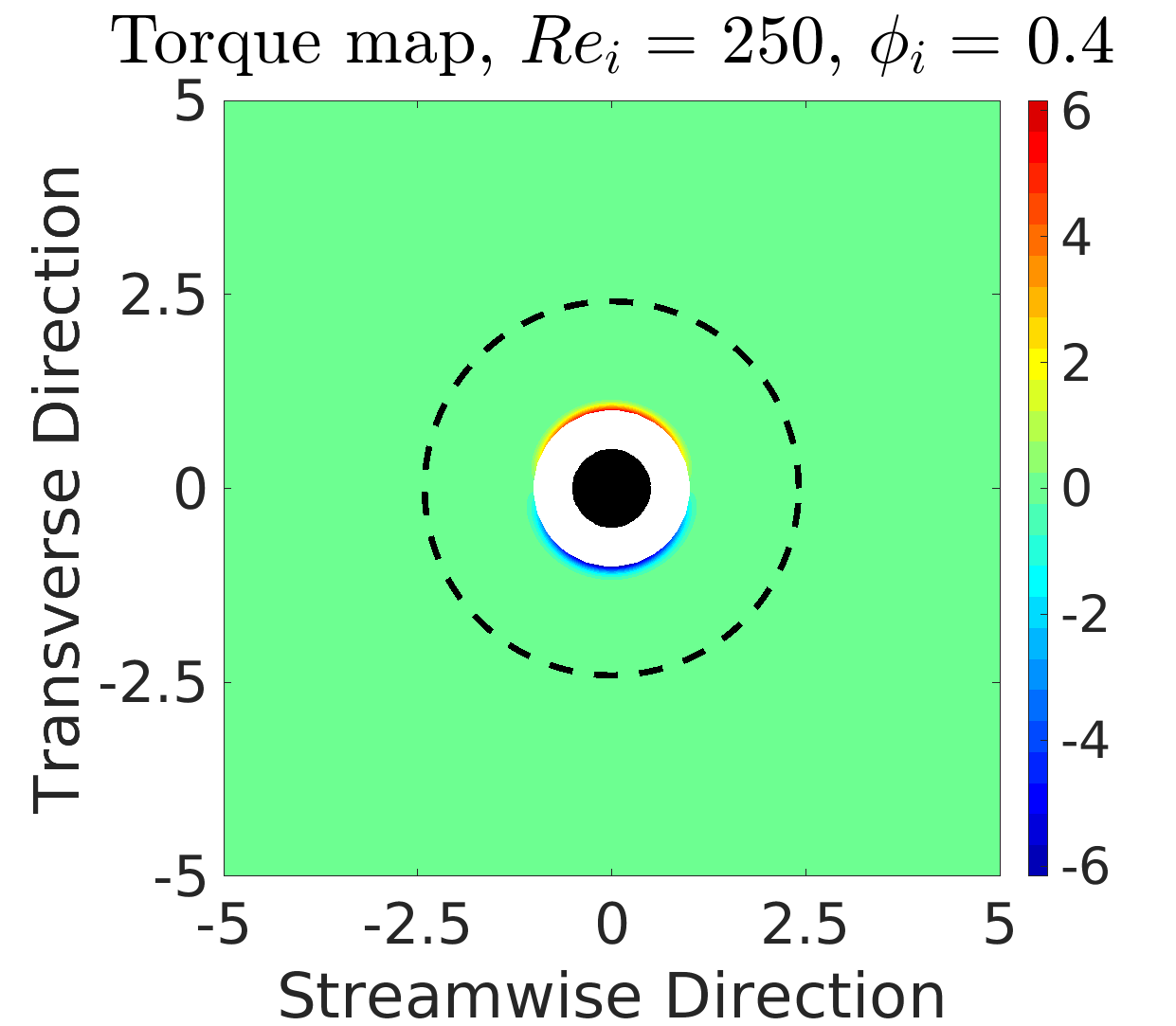}
    \end{subfigure}
    \caption{Binary-interaction torque maps obtained using $\tilde{\bm{T}}_{2i}$ with $M_2=26$.}
    \label{fig:binary_torque_maps}
\end{figure}

The low dimensionality of binary interaction models means that they can be easily visualized and interpreted. The physics of flow over a two-body system can be essentially visualized using a plane that is formed by the mean-flow direction (streamwise direction) and the line connecting the centers of both particles. Binary-interaction maps for streamwise force (drag), transverse force (lift) and transverse torque components are plotted in Figures \ref{fig:binary_drag_maps}, \ref{fig:binary_lift_maps} \& \ref{fig:binary_torque_maps} respectively. The maps shown in these figures are averaged over 5 different versions of the respective models that are a product of $(k=5)$-fold cross-validation. These maps are to be interpreted in the following manner. The black filled circle of unit non-dimensional diameter at the center represents the reference particle on the streamwise-transverse plane. The white wider circle indicates the region where a neighboring particle's center cannot exist due to nonoverlapping condition. The value at any other location in a map denotes the influence a neighbor, which has its center at that position, would have on the reference particle for the quantity of interest. Perturbation maps for $Re_i =$ \{1, 10, 100, 250\} at $\phi_i =$ \{0.10, 0.20, 0.40\} with $Re_i$ increasing along rows and $\phi_i$ along columns are shown in these figures. Among the considered datasets the farthest $M_2=26^{th}$ neighbor for $\langle \phi \rangle=$\{0.10, 0.20, 0.40\} is at \{3.93, 3.17, 2.41\} respectively. Dashed circles with the above mentioned radius at respective volume fractions are drawn on the maps. The binary maps have been properly learnt only in regions within these dashed circles.

The force perturbation induced by a neighbor in a two-body system exists only in the plane formed by the streamwise direction and the line joining particle centers. The universal nature of the models presented in this work enables one to easily plot maps for $Re_i$ and $\phi_i$ combinations that do not precisely exist in training datasets. The force maps depict smooth contours closer to the reference particle where there is a higher probability for the considered $M_2=26$ neighbors to appear. However, the region further away from the reference particle, where the influence is traditionally expected to become negligible, has unphysical values due to either rare or no occurrence of neighbors in the training data. This provides a clear idea on the range of applicability of the binary maps. It can also be observed that the drag and lift maps at a fixed Reynolds number have a similar shape for different volume fractions. The blue regions upstream and downstream of the reference particle (i.e., to the left and right of the black circle) represent negative binary influence of the neighbor on the drag of the reference particle. Thus, the binary influence of an upstream or downstream neighbor contributes to drag reduction. Whereas, the binary influence of a neighbor to the side contributes to drag increase. In the lift map, only off-axis neighbors have an influence on the lift force. An off-axis neighbor located upstream of the reference particle results in a binary influence, whose transverse force is directed towards the neighbor. Whereas, an off-axis neighbor located downstream results in a transverse force that is directed away from the neighbor. It can be observed that at low Reynolds numbers, upstream and downstream neighbors have nearly similar drag and lift influence, indicating fore-aft symmetry. This symmetry is broken at higher Reynolds numbers, where an upstream neighbor has a stronger effect on drag, while a downstream neighbor has a stronger lift influence. These binary maps are similar to those obtained with the PIEP model \citet{akiki_jfm} as well to those obtained by \citet{wachs_piep}.

Torque perturbation induced by a neighbor in the two-body setup is directed perpendicular to the considered plane. Interestingly the range of influence of torque on the reference particle at a given ($Re_i$, $\phi_i$) is considerably smaller than the corresponding influence range for force. Therefore, satisfactory torque predictions can be obtained using fewer neighbors compared to force predictions. A detailed quantitative analysis of neighbor-truncation error for force and torque predictions is presented in Appendix \ref{app_A}.

It can be observed that binary maps for conditions outside of training data are also included in Figures \ref{fig:binary_drag_maps}, \ref{fig:binary_lift_maps}, \& \ref{fig:binary_torque_maps}. For example, the maps at $(Re_i,\phi_i)=$ (250,0.10), and (250,0.20) can be perceived as an extrapolation from the case (250,0.40) that was included in the training dataset. The maps appear to follow the respective characteristic trends observed in the other maps. As the volume fraction decreases at a fixed Reynolds number, the force maps display an increase in the radius of influence and a decrease in perturbation magnitude. On the other hand, it is evident from $(\phi_i=0.40)$ maps that the models produce unphysical results even slightly outside of the dashed circle, i.e. for $|\bm{r}_j| > 2.41$. 
This suggests that the machine learned models have difficulty in extrapolating trends from seen to unseen (lower to higher) volume fractions at intermediate particle separations that are just beyond the training region. Based on this discussion we believe that the models have a higher generalizability for unseen Reynolds numbers than unseen particle volume fractions.

\subsection{Trinary force and torque maps}
As an example, the trinary interaction maps are shown for $(Re_i,\phi_i) = (50,0.20)$. The three-body interaction problem must be visualized in 3D by fixing the position of one neighbor (called the {\it{first neighbor}} along with the reference particle, while varying the position of the {\it{second neighbor}}. The reference particle will be represented using red and the first neighbor will be shown using white. We illustrate for three different locations of the first neighbor. The chosen positions of the first neighbor relative to the reference particle are (-1.1, 0, 0), (1.1, 0, 0), and (0, 1.1, 0), where the mean-flow (streamwise) direction is along $+x$. The reference and first neighbor particles are visualized at twice their actual size, to hide regions where the third particle's center cannot exist (non-overlapping condition). This explains the intersection of the red and white spheres in the figure. Similar to the binary maps, the trinary maps are also averaged over five independent versions of the model as a result of $(k=5)$-fold cross-validation. Furthermore, the maps only depict spatial regions where the models have been trained, i.e., interior of the farthest location of neighbor from $M_3 = 10$ at $\langle \phi \rangle = 0.20$. The value at a given spatial location indicates the influence of the second neighbor with its center at that point along with the first neighbor would have on the reference particle.

\begin{figure}
    \centering
    \begin{subfigure}[b]{\textwidth}
       \centering
       \includegraphics[width=0.49\textwidth,keepaspectratio=true]{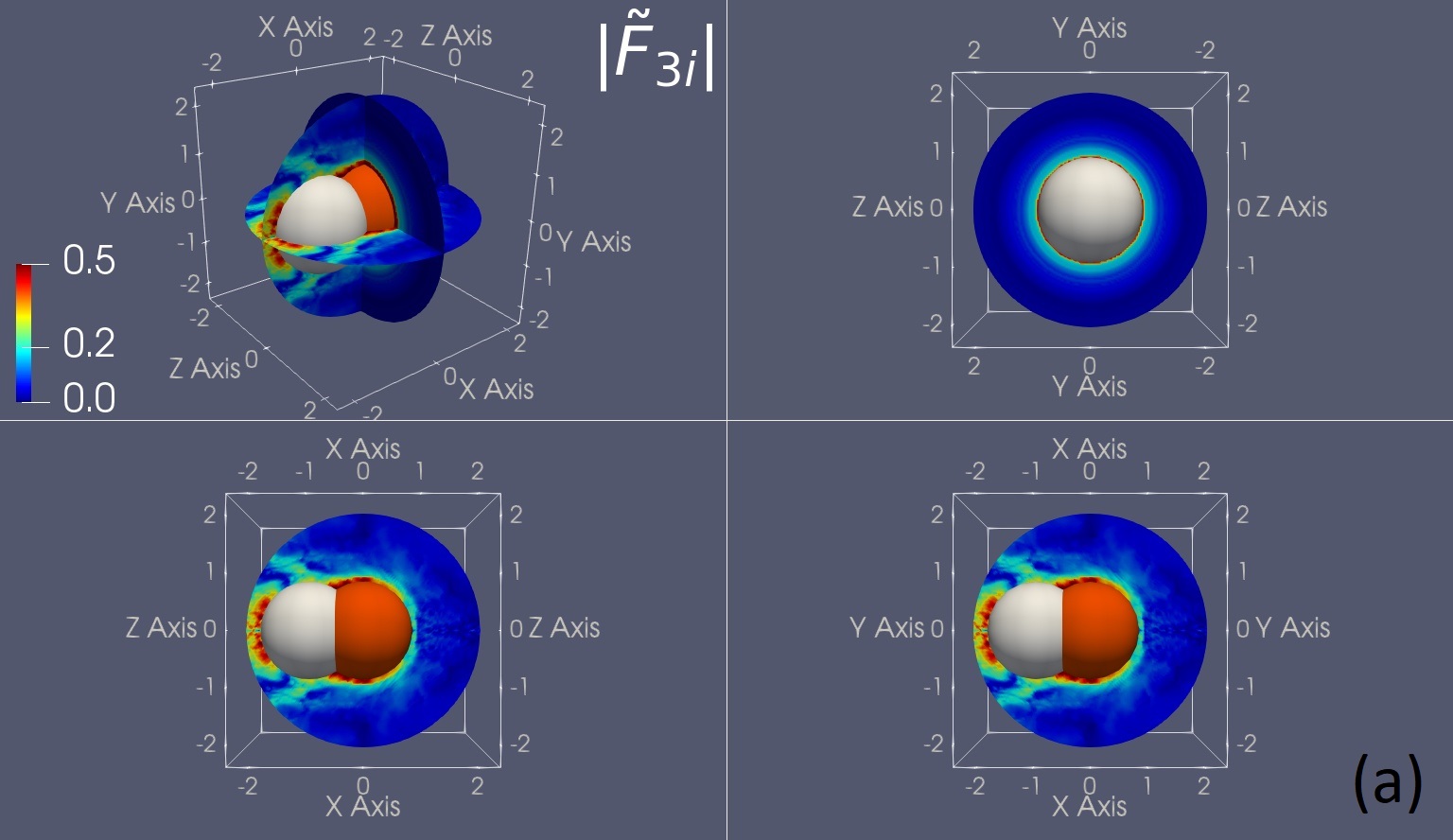}
       \includegraphics[width=0.49\textwidth,keepaspectratio=true]{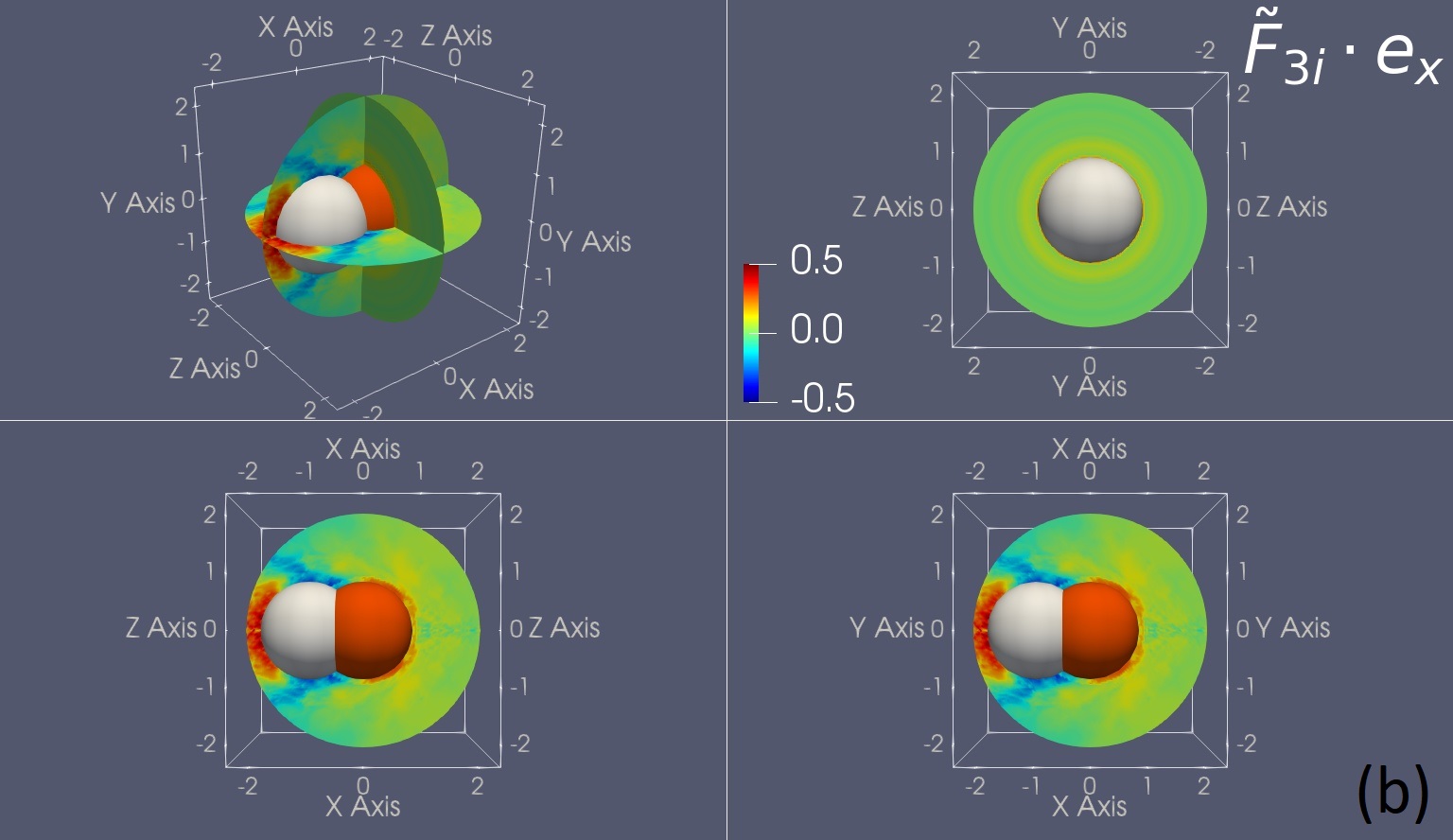}\vspace{0.5mm}
       \includegraphics[width=0.49\textwidth,keepaspectratio=true]{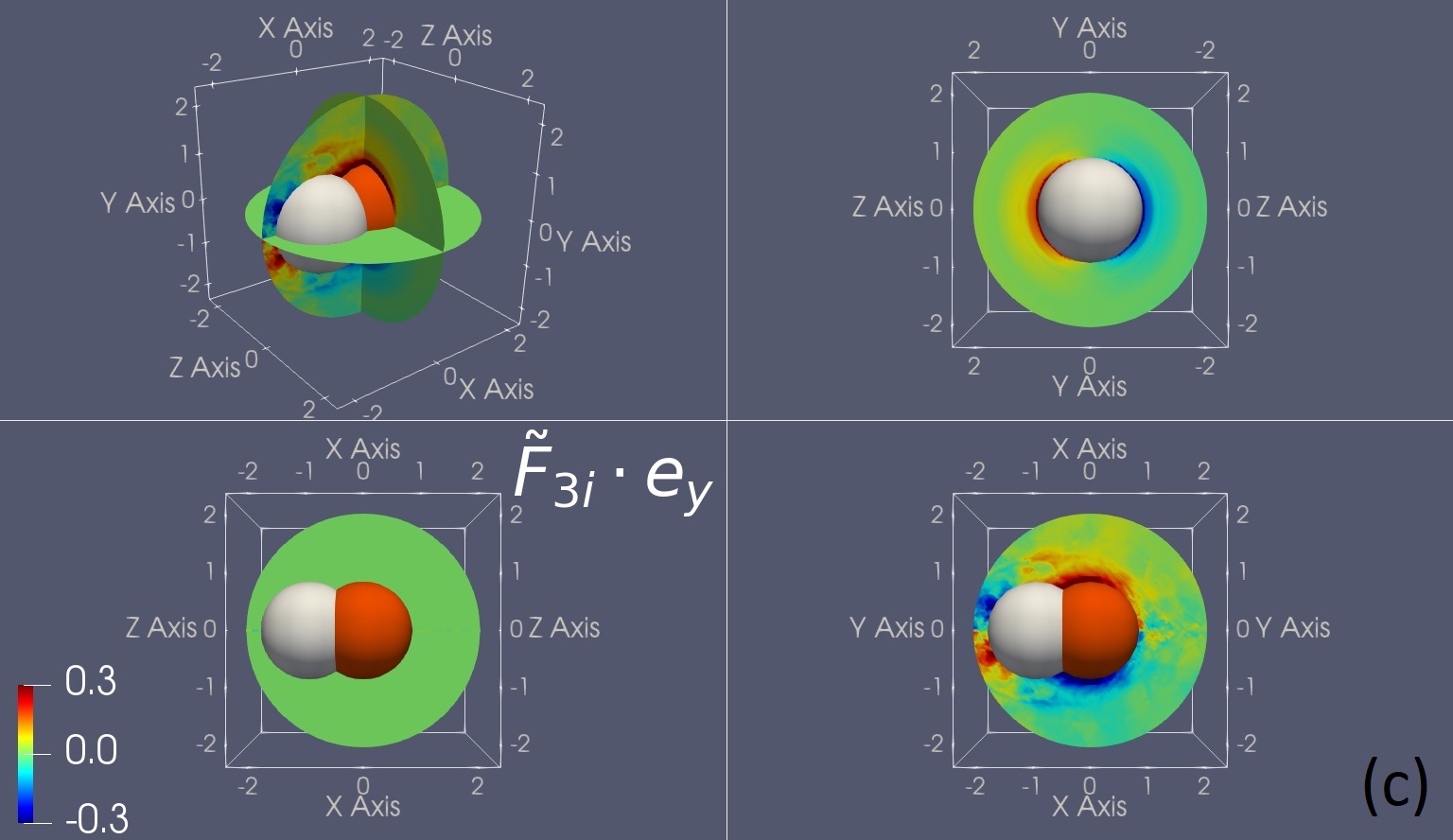}
       \includegraphics[width=0.49\textwidth,keepaspectratio=true]{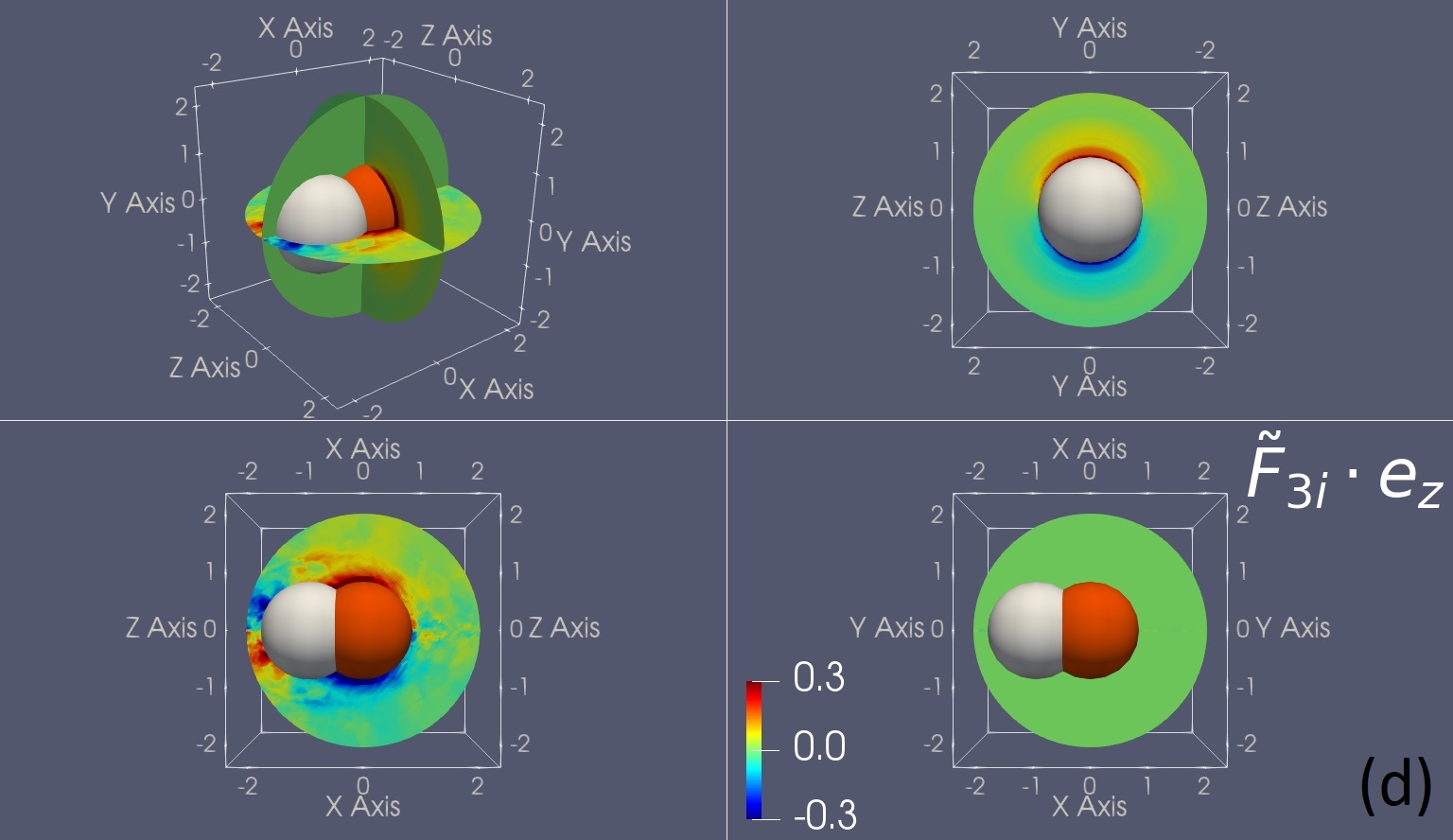}
    \end{subfigure}
    \caption{Trinary-interaction force influence with upstream first neighbor (denoted with white) at (-1.1, 0, 0) relative to reference particle (represented with red). (a) Magnitude of flucuations ($|\tilde{\bm{F}}_{3i}|$), (b) Force perturbation along $x$ ($\tilde{\bm{F}}_{3i} \cdot \bm{e}_x$), (c) Force perturbation along $y$ ($\tilde{\bm{F}}_{3i} \cdot \bm{e}_y$), and (d) Force perturbation along $z$ ($\tilde{\bm{F}}_{3i} \cdot \bm{e}_z$).}
    \label{fig:force_config_1}
\end{figure}

\begin{figure}
    \centering
    \begin{subfigure}[b]{\textwidth}
       \centering
       \includegraphics[width=0.49\textwidth,keepaspectratio=true]{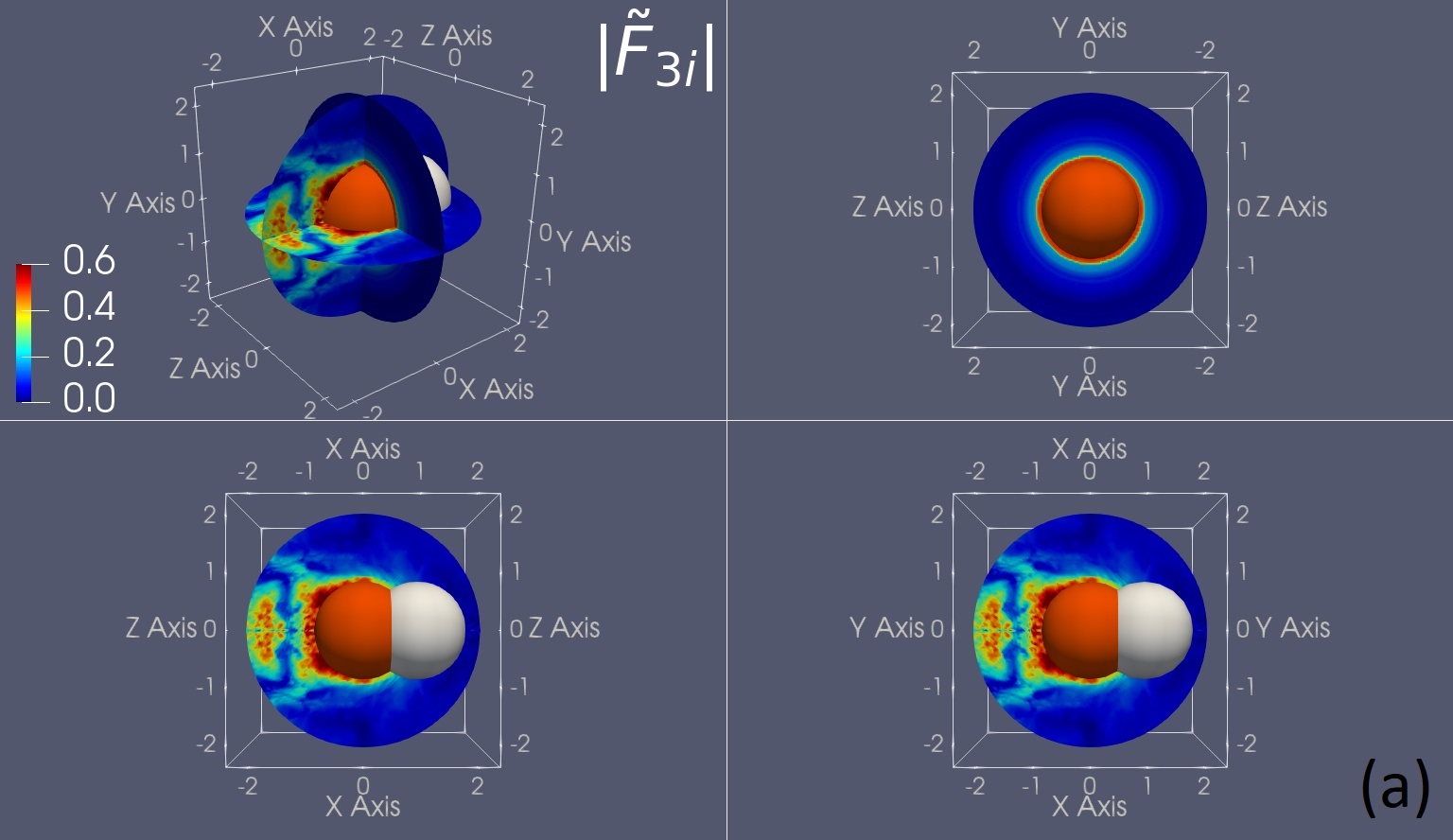}
       \includegraphics[width=0.49\textwidth,keepaspectratio=true]{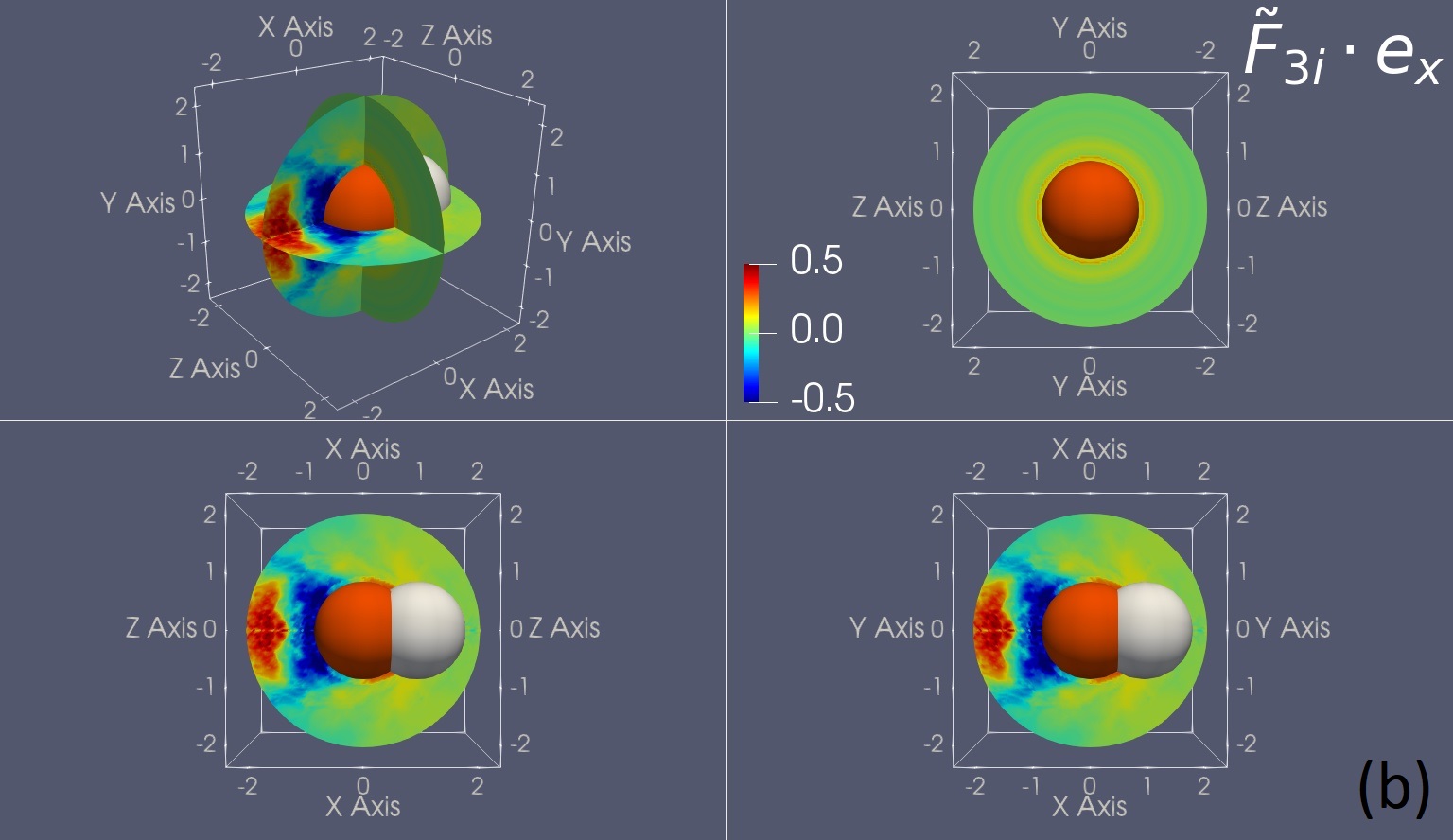}\vspace{0.5mm}
       \includegraphics[width=0.49\textwidth,keepaspectratio=true]{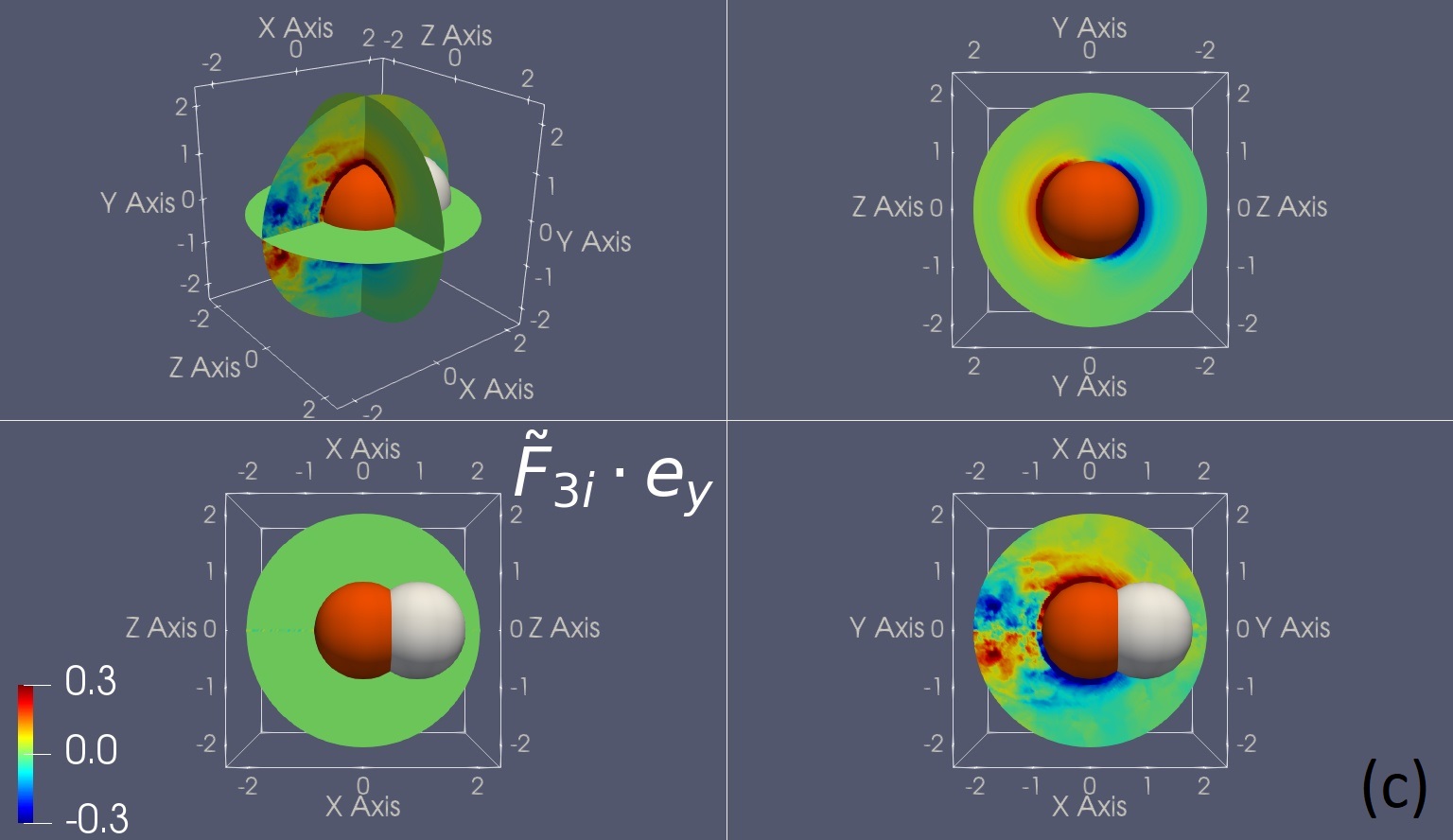}
       \includegraphics[width=0.49\textwidth,keepaspectratio=true]{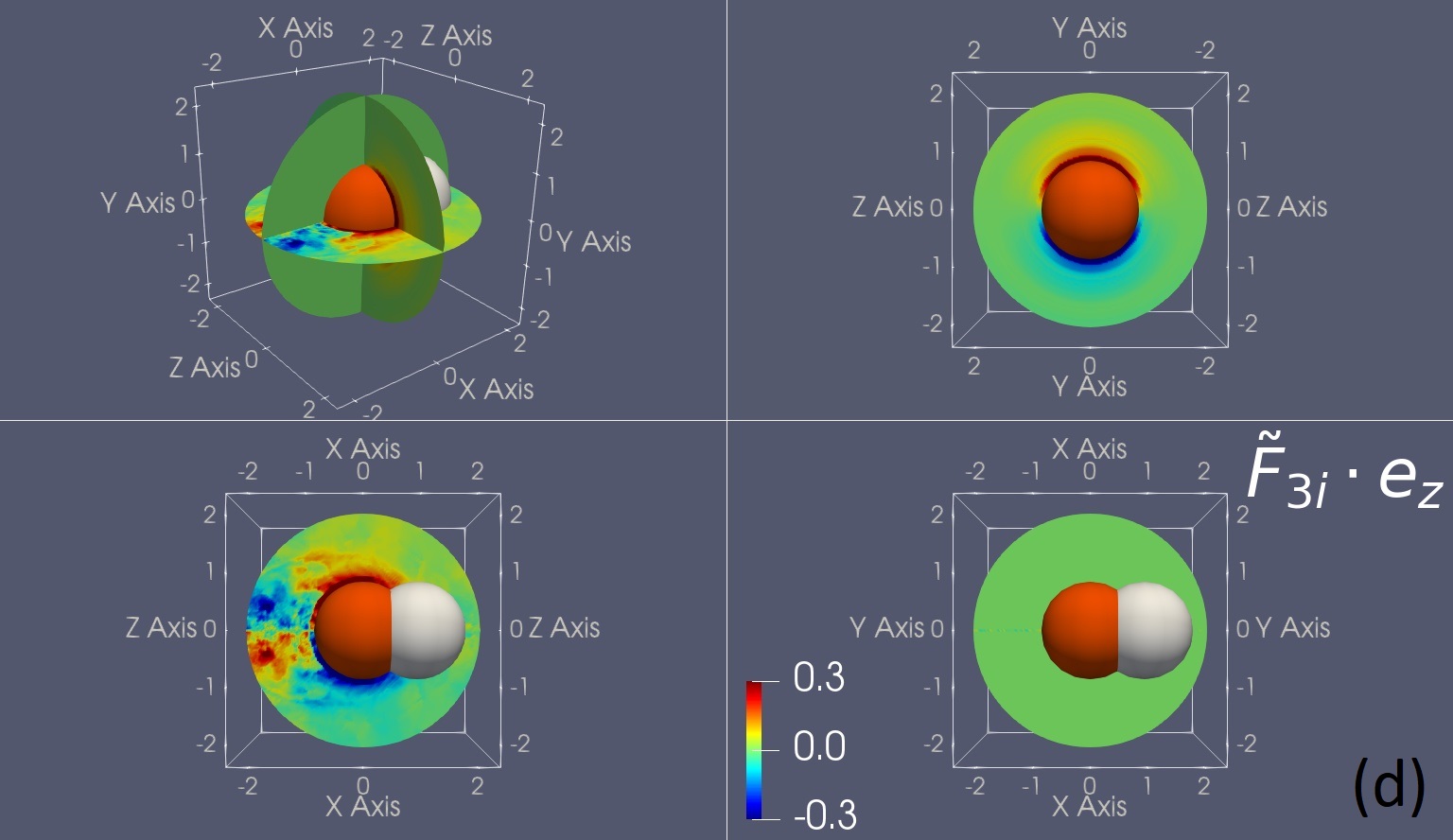}
    \end{subfigure}
    \caption{Trinary-interaction force influence with downstream first neighbor (denoted with white) at (1.1, 0, 0) relative to reference particle (represented with red). (a) Magnitude of flucuations ($|\tilde{\bm{F}}_{3i}|$), (b) Force perturbation along $x$ ($\tilde{\bm{F}}_{3i} \cdot \bm{e}_x$), (c) Force perturbation along $y$ ($\tilde{\bm{F}}_{3i} \cdot \bm{e}_y$), and (d) Force perturbation along $z$ ($\tilde{\bm{F}}_{3i} \cdot \bm{e}_z$).}
    \label{fig:force_config_2}
\end{figure}

\begin{figure}
    \centering
    \begin{subfigure}[b]{\textwidth}
       \centering
       \includegraphics[width=0.49\textwidth,keepaspectratio=true]{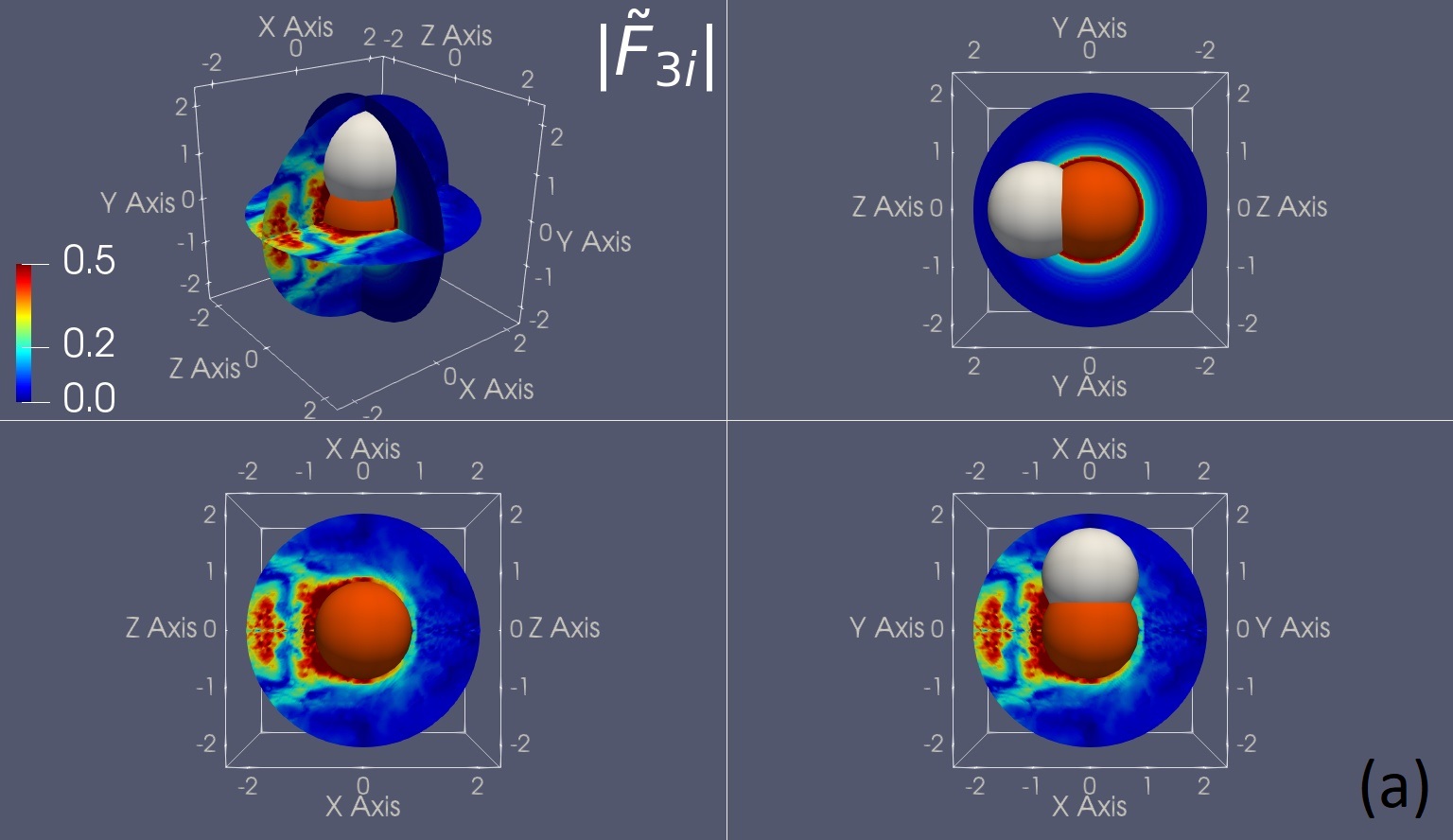}
       \includegraphics[width=0.49\textwidth,keepaspectratio=true]{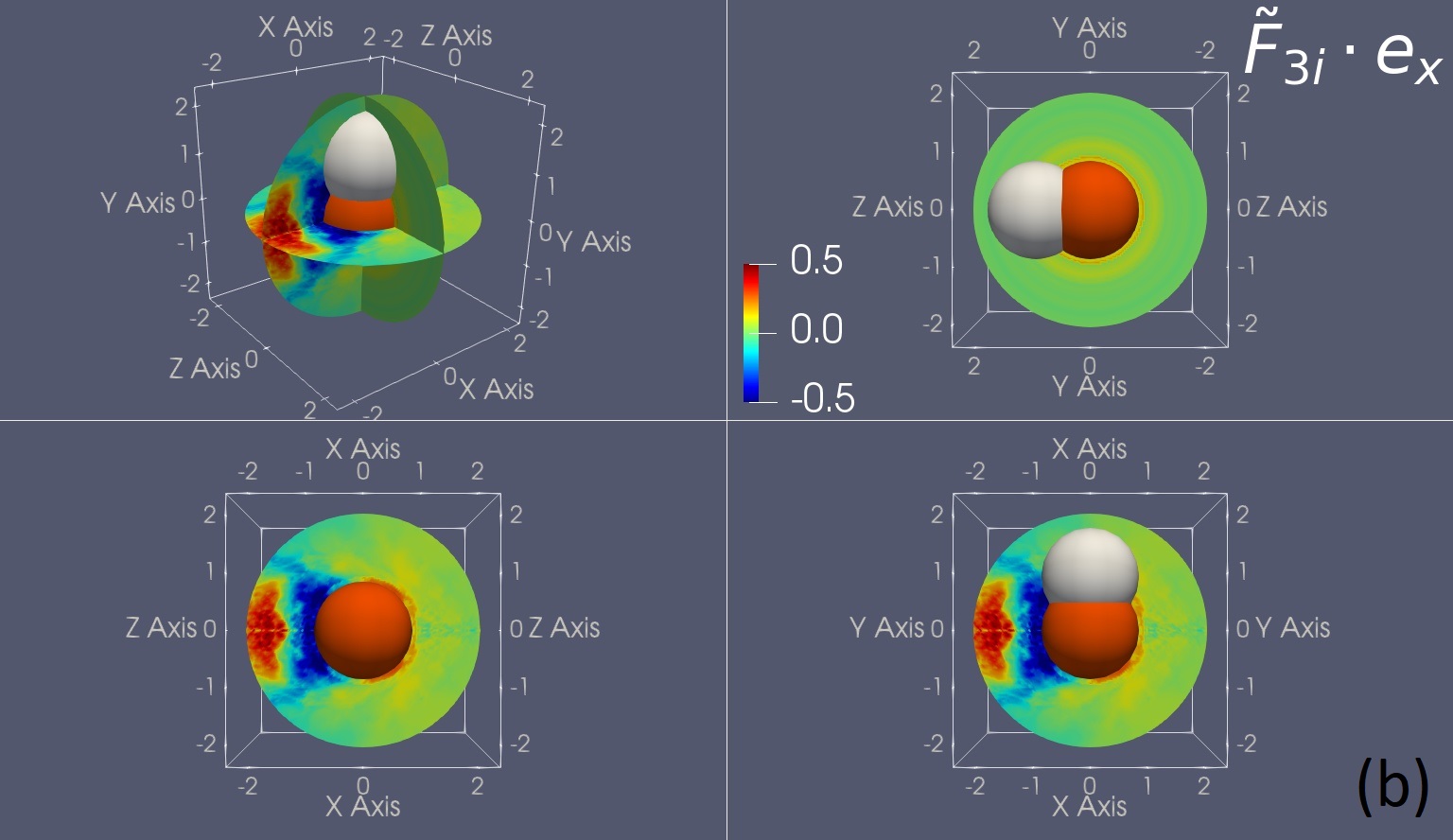}\vspace{0.5mm}
       \includegraphics[width=0.49\textwidth,keepaspectratio=true]{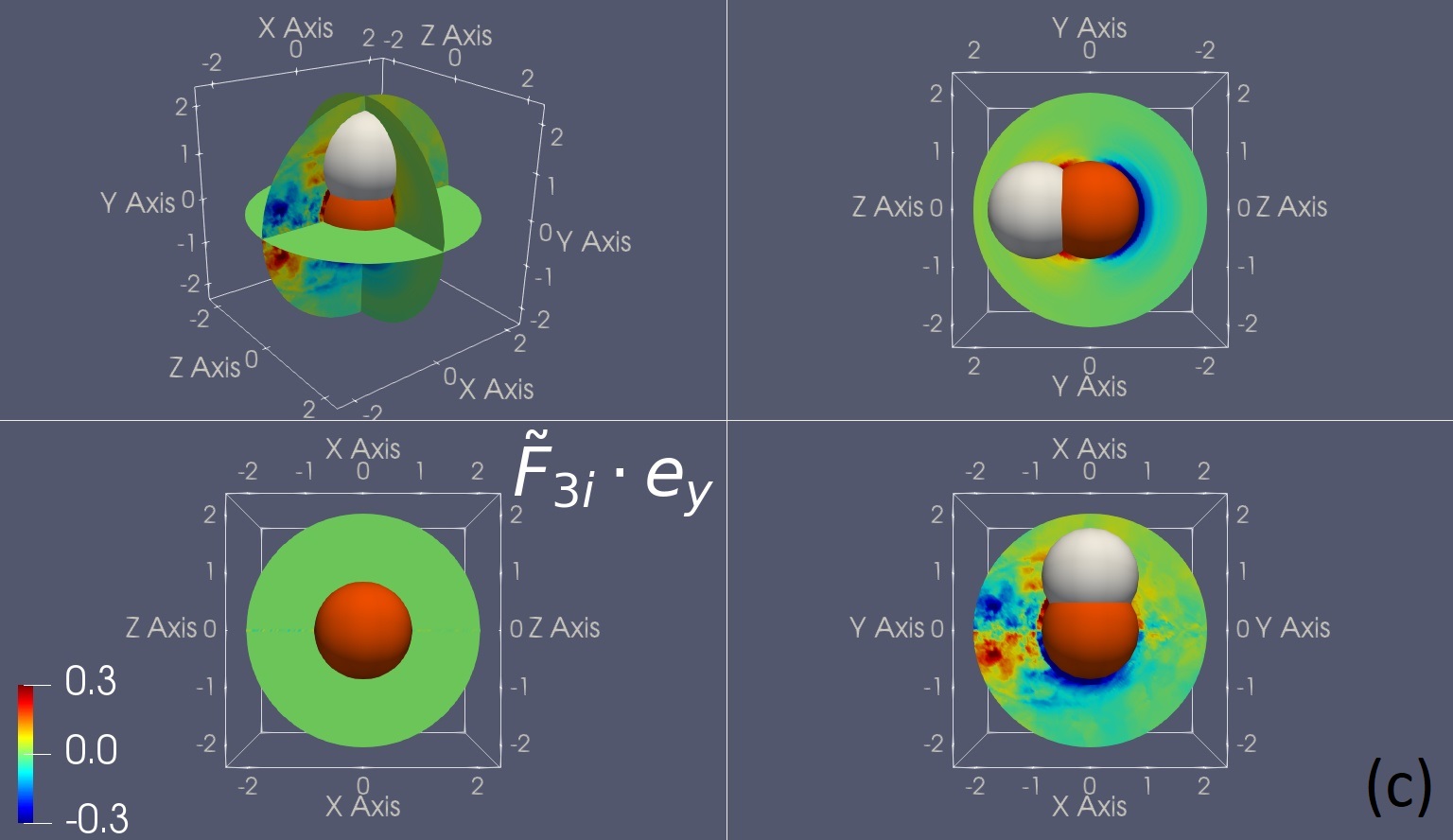}
       \includegraphics[width=0.49\textwidth,keepaspectratio=true]{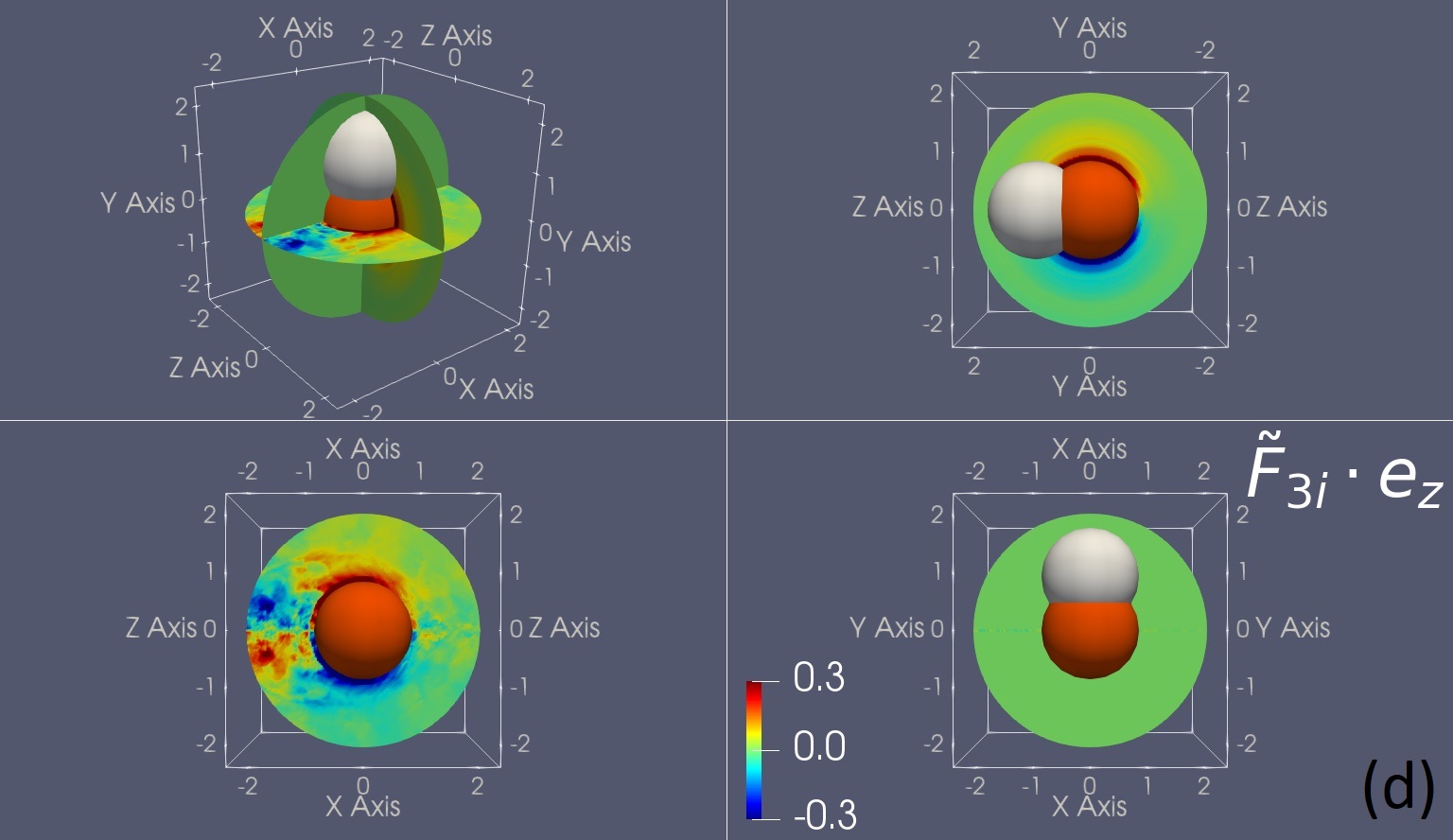}
    \end{subfigure}
    \caption{Trinary-interaction force influence with transverse first neighbor (denoted with white) at (0, 1.1, 0) relative to reference particle (represented with red). (a) Magnitude of flucuations ($|\tilde{\bm{F}}_{3i}|$), (b) Force perturbation along $x$ ($\tilde{\bm{F}}_{3i} \cdot \bm{e}_x$), (c) Force perturbation along $y$ ($\tilde{\bm{F}}_{3i} \cdot \bm{e}_y$), and (d) Force perturbation along $z$ ($\tilde{\bm{F}}_{3i} \cdot \bm{e}_z$).}
    \label{fig:force_config_3}
\end{figure}

The force perturbations induced by the trinary-interaction term $\tilde{\bm{F}}_{3i}$ with $M_3=10$ for the three configurations are shown in Figures \ref{fig:force_config_1}, \ref{fig:force_config_2}, \& \ref{fig:force_config_3}. It can be clearly seen that the presented maps preserve symmetries perfectly. It can also be observed that the physics learned by the trinary-interaction model for all three configurations is mainly close to the reference particle on the upstream side. Interestingly this is true even when the first neighbor is located downstream. Interpretation of the trinary maps requires some care. The total perturbing influence of the first and the second neighbors (whose positions relative to the reference particle are known) can be obtained by adding the binary influence of the two neighbors taken one at a time and the trinary influence of them acting together. Thus, the trinary map must be properly interpreted as a correction to the sum of perturbations induced by the two neighbors as predicted by the binary maps. The trinary maps are noisier than the binary maps even in the region of training. This clearly indicates the fact that the data that was used is not the completely adequate for a fully converged trinary term. The optimal trinary model obtained in the present work clearly attempts to capture key aspects of the underlying physics, as indicated by the consistent improvement seen in the $R^2$ values with the inclusion of the trinary term. Nevertheless, we believe further improvement of the trinary term is possible with the availability of additional training data.

\begin{figure}
    \centering
    \begin{subfigure}[b]{\textwidth}
       \centering
       \includegraphics[width=0.49\textwidth,keepaspectratio=true]{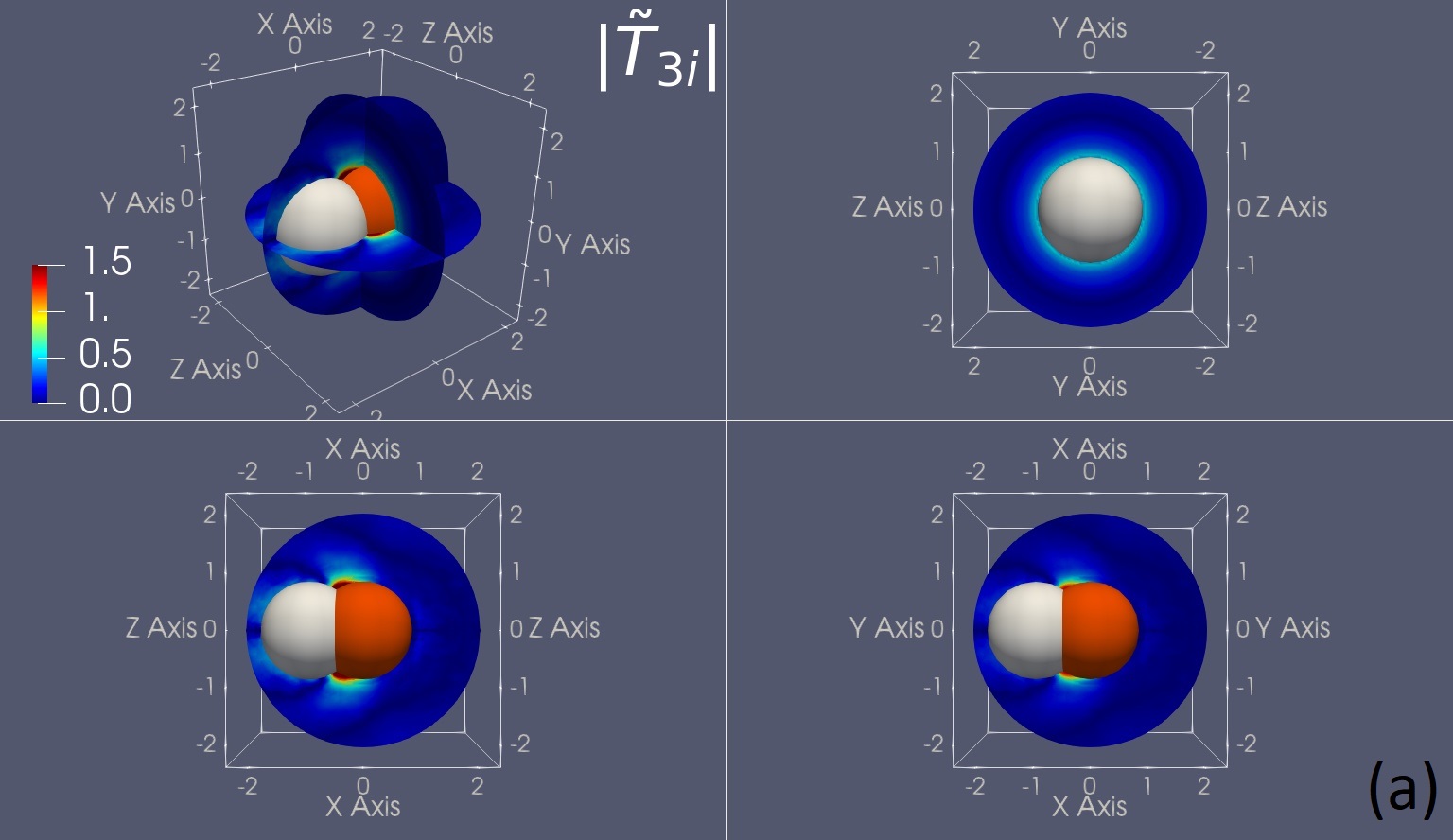}
       \includegraphics[width=0.49\textwidth,keepaspectratio=true]{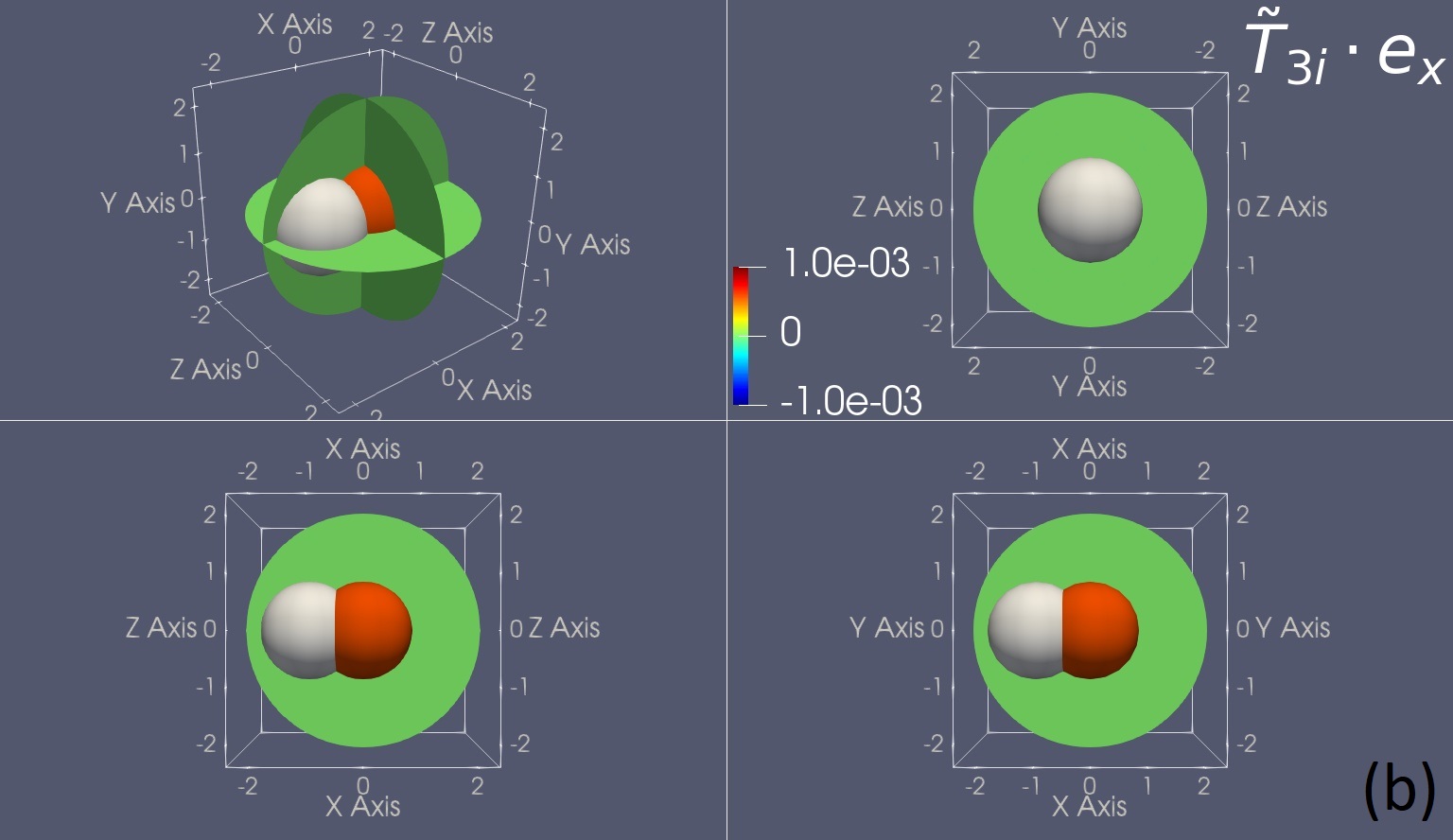}\vspace{0.5mm}
       \includegraphics[width=0.49\textwidth,keepaspectratio=true]{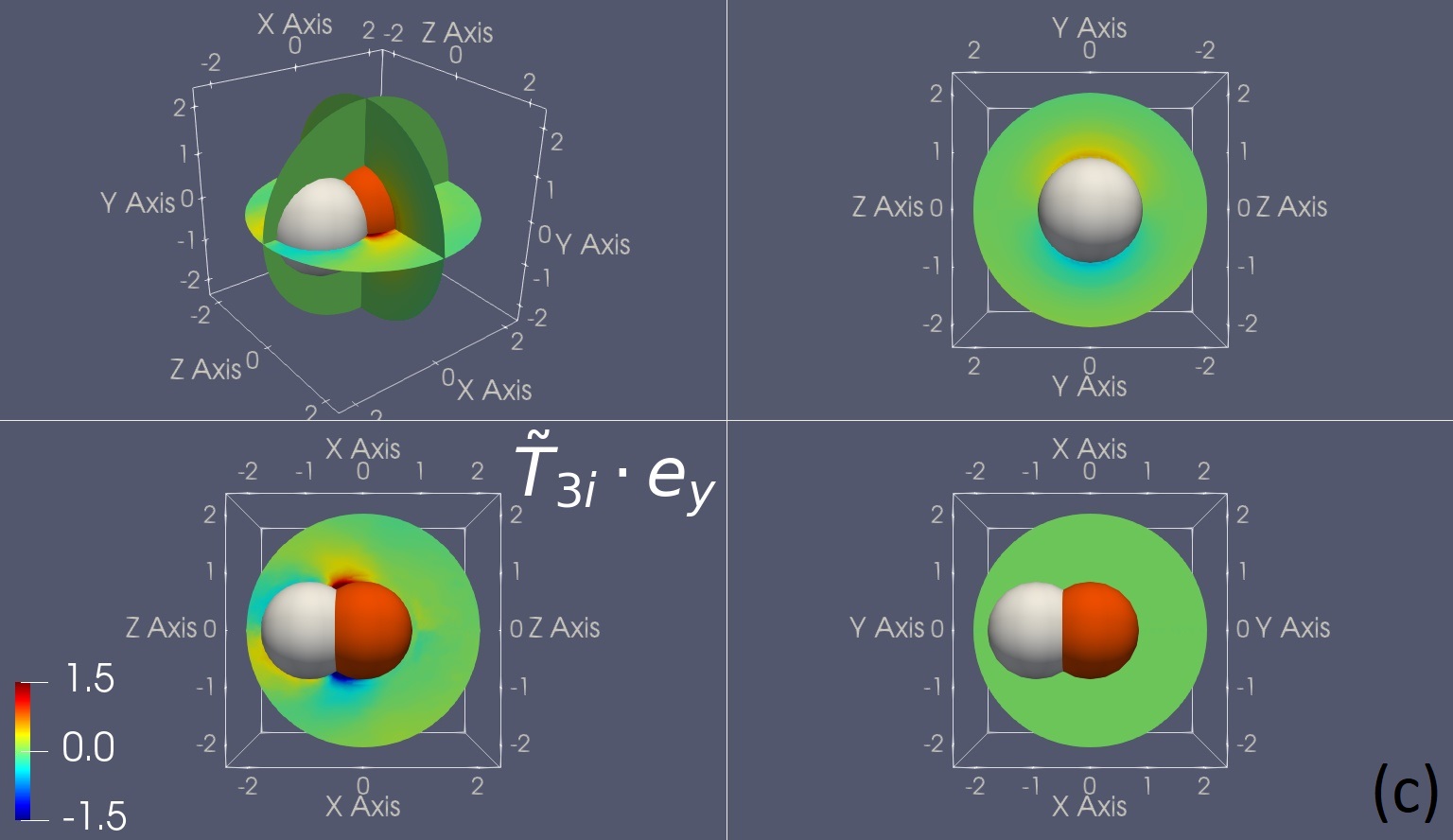}
       \includegraphics[width=0.49\textwidth,keepaspectratio=true]{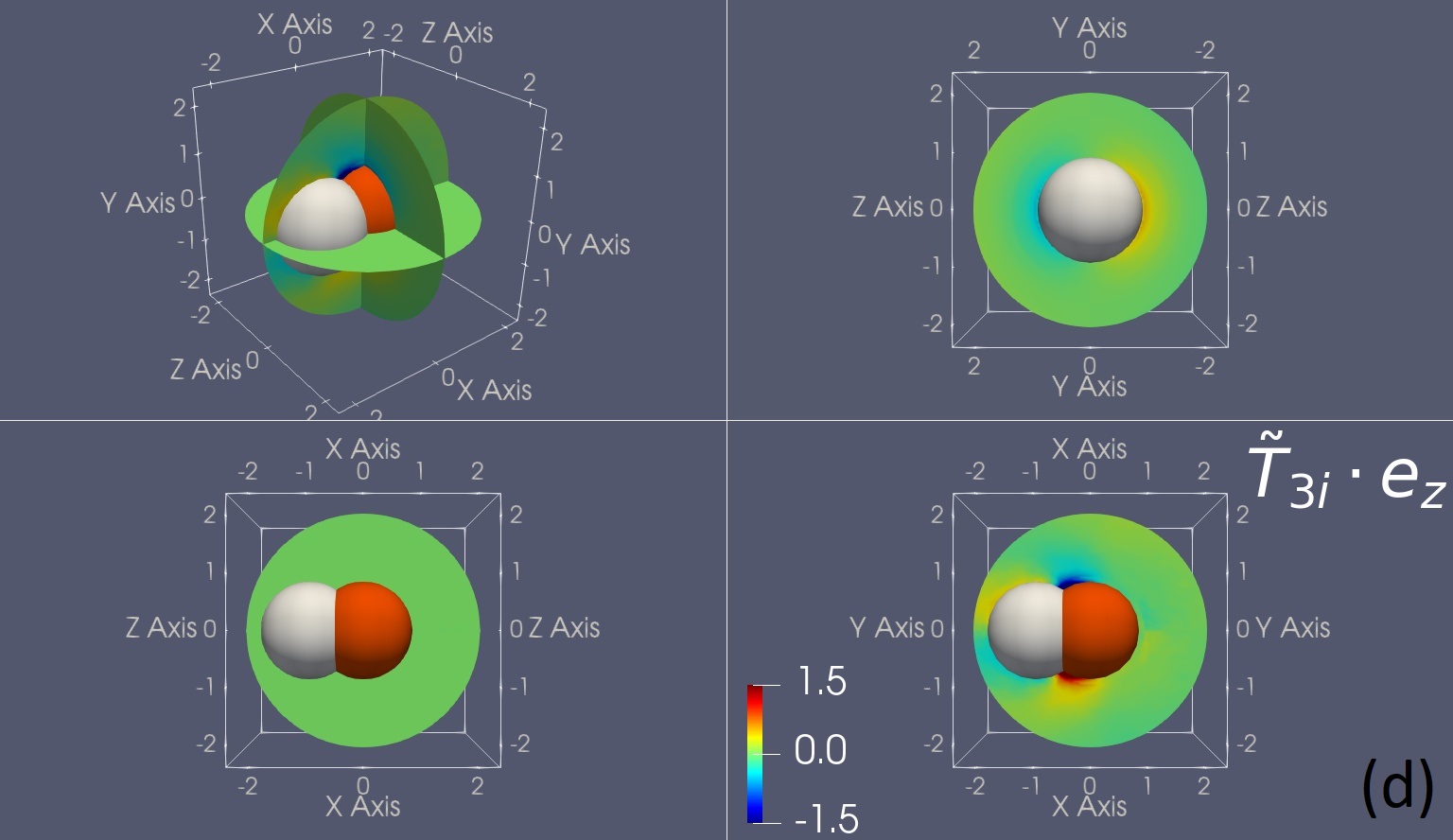}
    \end{subfigure}
    \caption{Trinary-interaction torque influence with upstream first neighbor (denoted with white) at (-1.1, 0, 0) relative to reference particle (represented with red). (a) Magnitude of flucuations ($|\tilde{\bm{T}}_{3i}|$), (b) Torque perturbation along $x$ ($\tilde{\bm{T}}_{3i} \cdot \bm{e}_x$), (c) Torque perturbation along $y$ ($\tilde{\bm{T}}_{3i} \cdot \bm{e}_y$), and (d) Torque perturbation along $z$ ($\tilde{\bm{T}}_{3i} \cdot \bm{e}_z$).}
    \label{fig:torque_config_1}
\end{figure}

\begin{figure}
    \centering
    \begin{subfigure}[b]{\textwidth}
       \centering
       \includegraphics[width=0.49\textwidth,keepaspectratio=true]{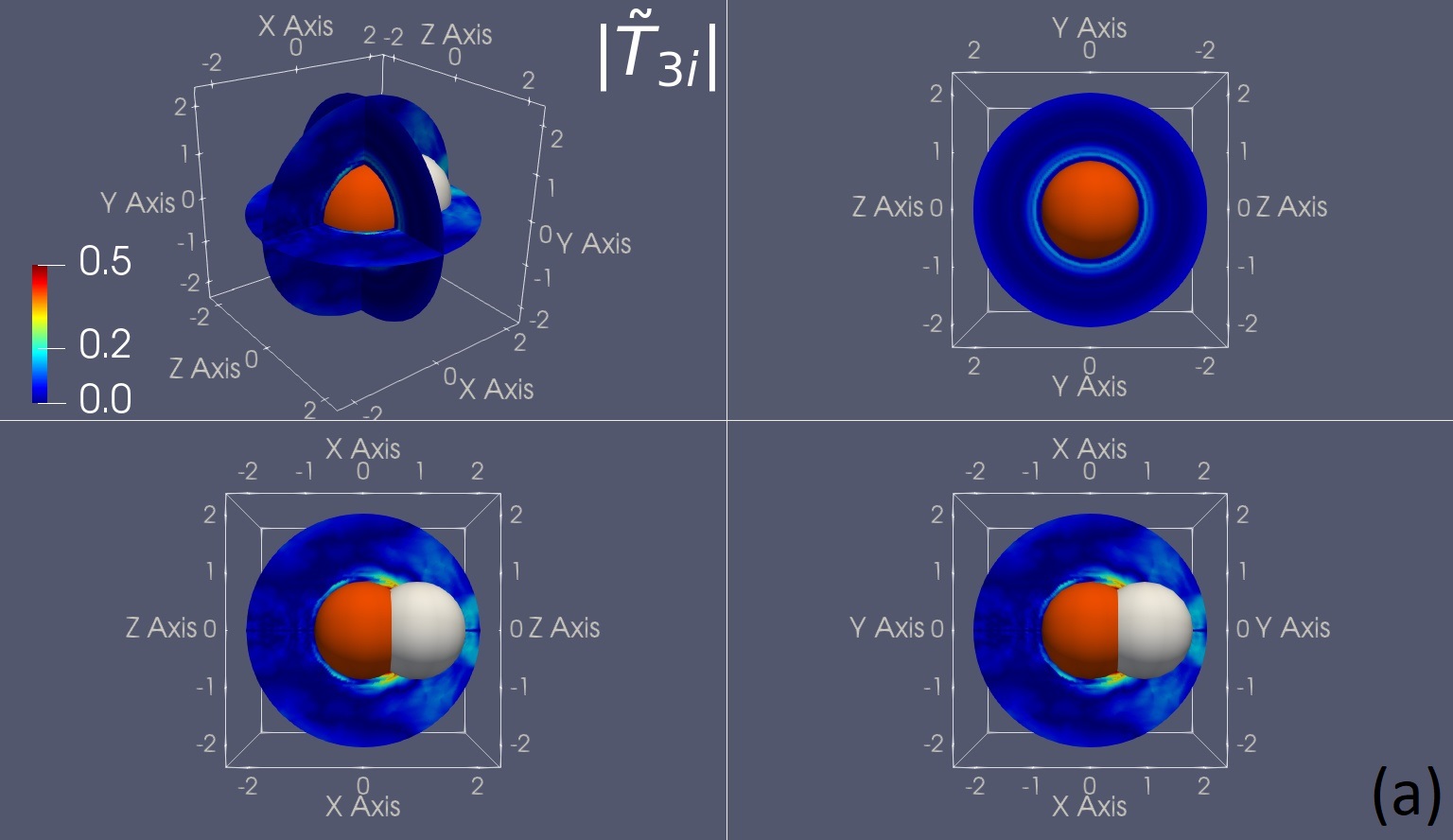}
       \includegraphics[width=0.49\textwidth,keepaspectratio=true]{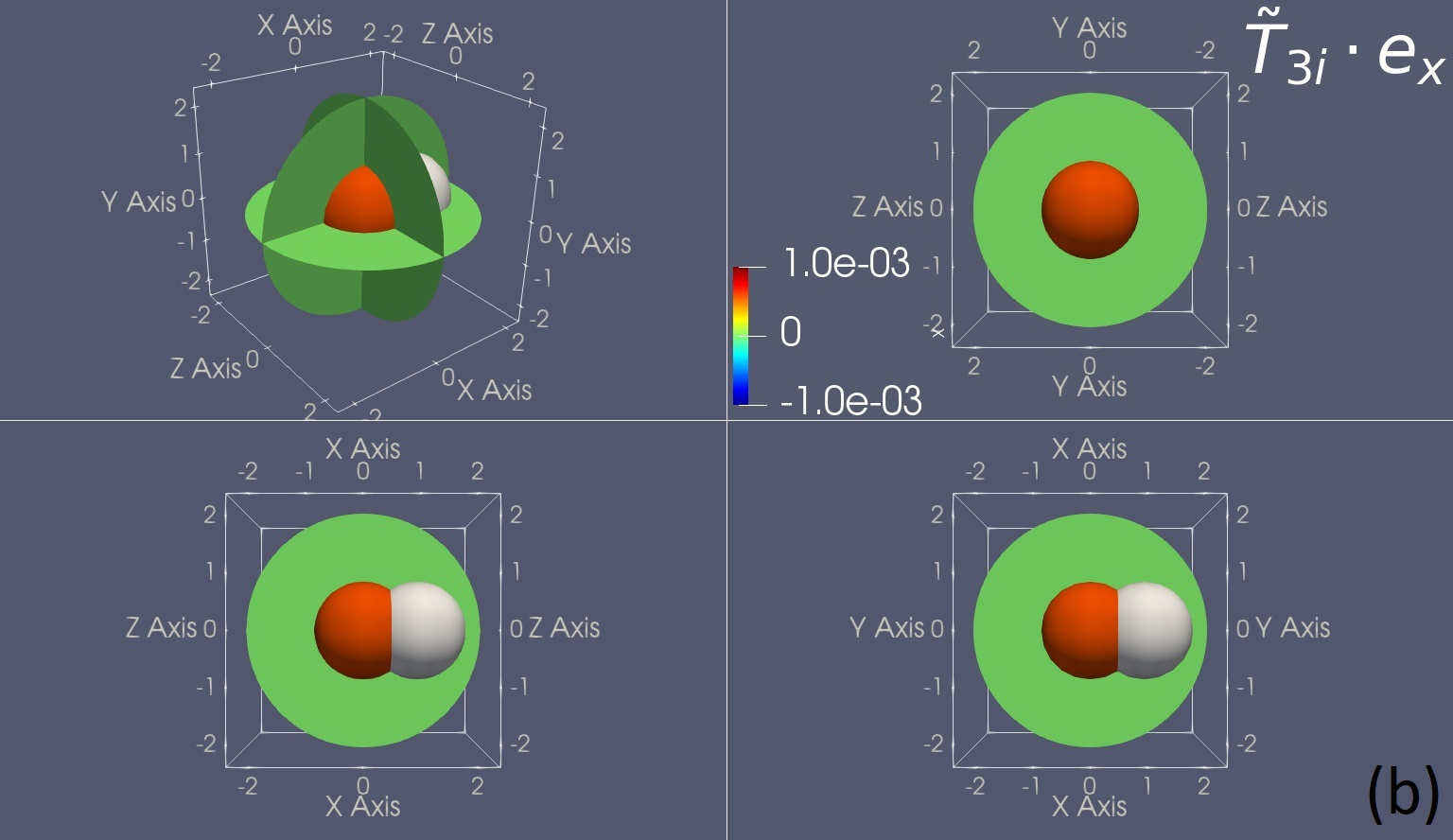}\vspace{0.5mm}
       \includegraphics[width=0.49\textwidth,keepaspectratio=true]{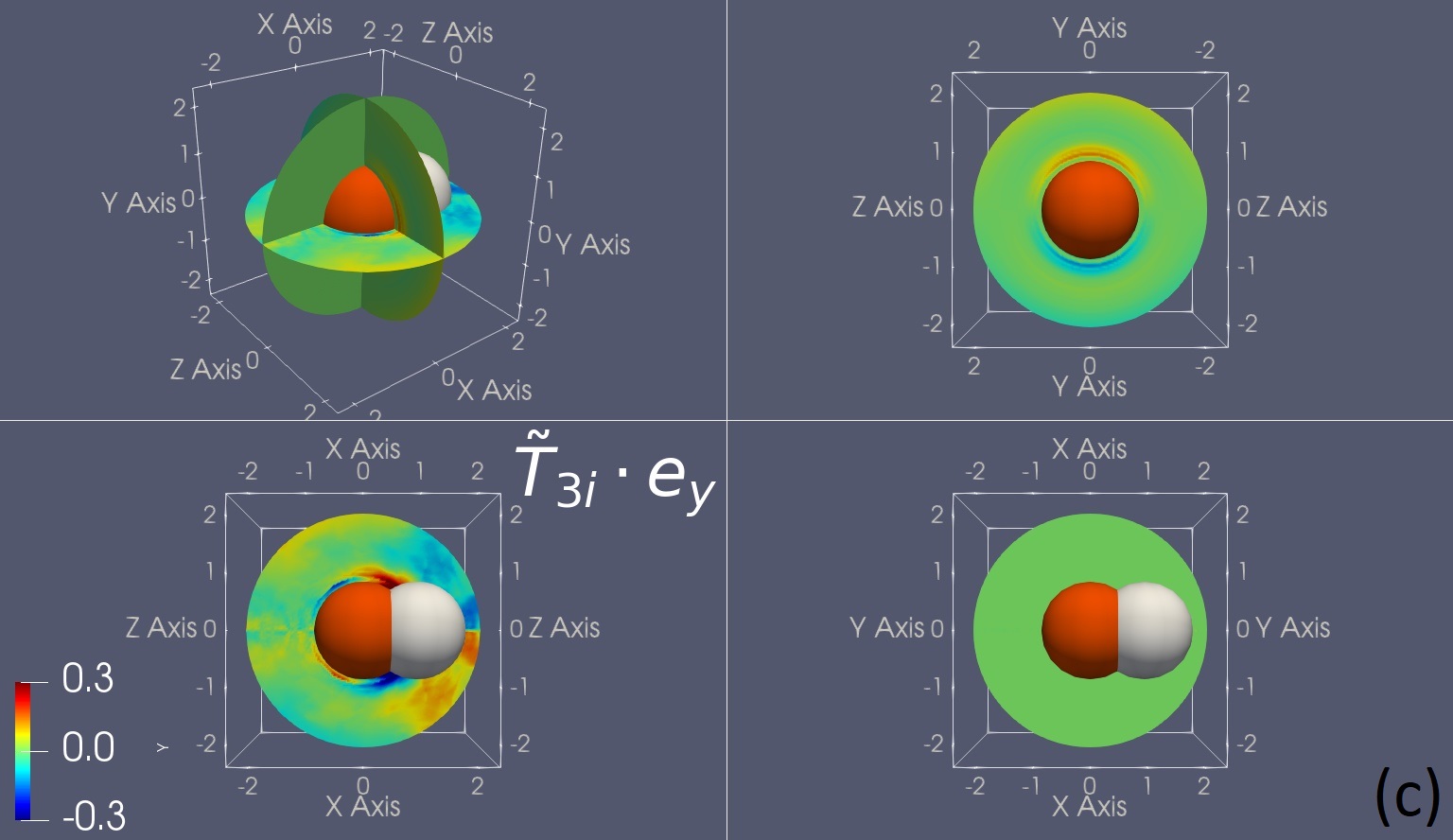}
       \includegraphics[width=0.49\textwidth,keepaspectratio=true]{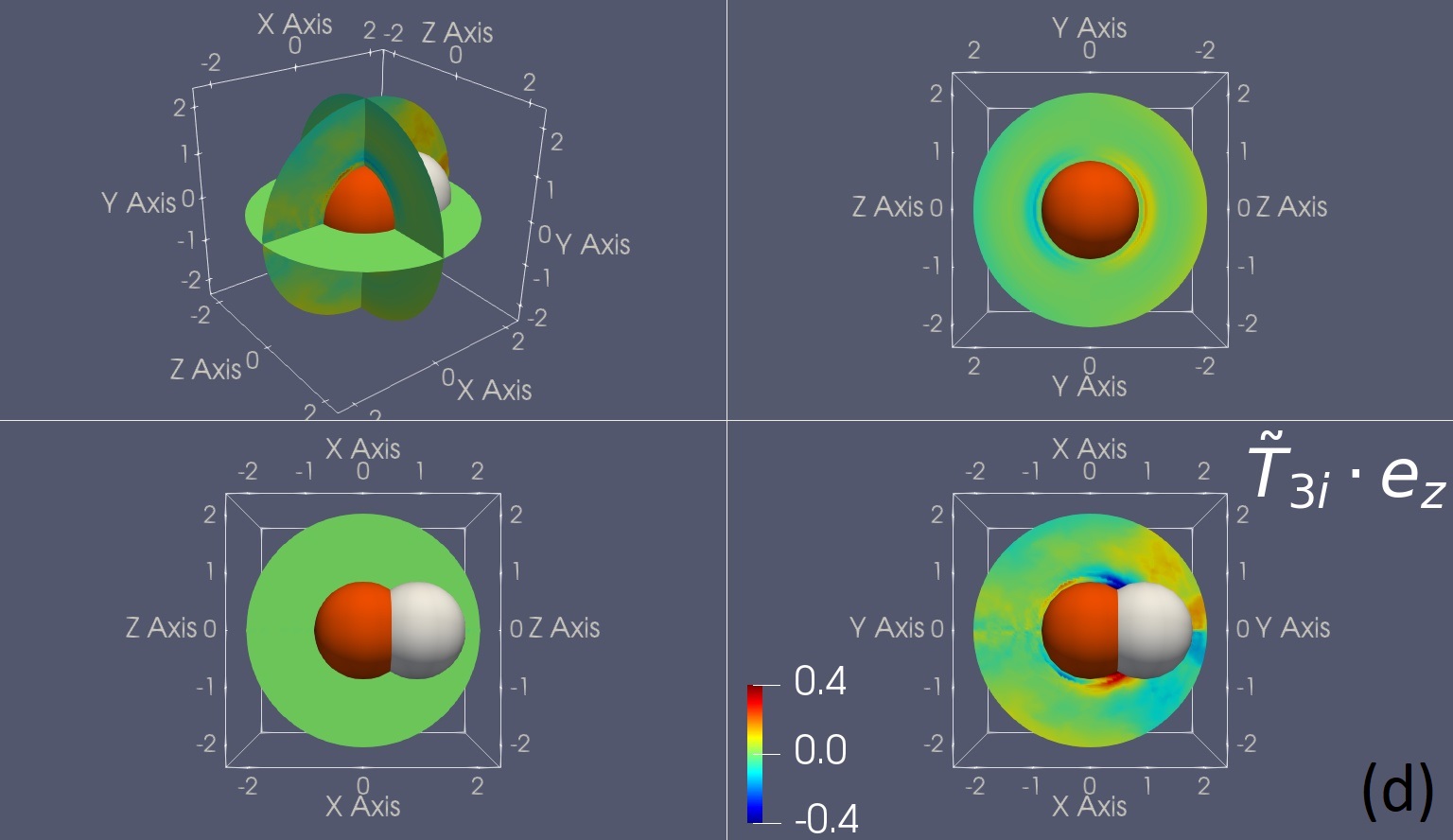}
    \end{subfigure}
    \caption{Trinary-interaction torque influence with downstream first neighbor (denoted with white) at (1.1, 0, 0) relative to reference particle (represented with red). (a) Magnitude of flucuations ($|\tilde{\bm{T}}_{3i}|$), (b) Torque perturbation along $x$ ($\tilde{\bm{T}}_{3i} \cdot \bm{e}_x$), (c) Torque perturbation along $y$ ($\tilde{\bm{T}}_{3i} \cdot \bm{e}_y$), and (d) Torque perturbation along $z$ ($\tilde{\bm{T}}_{3i} \cdot \bm{e}_z$).}
    \label{fig:torque_config_2}
\end{figure}

\begin{figure}
    \centering
    \begin{subfigure}[b]{\textwidth}
       \centering
       \includegraphics[width=0.49\textwidth,keepaspectratio=true]{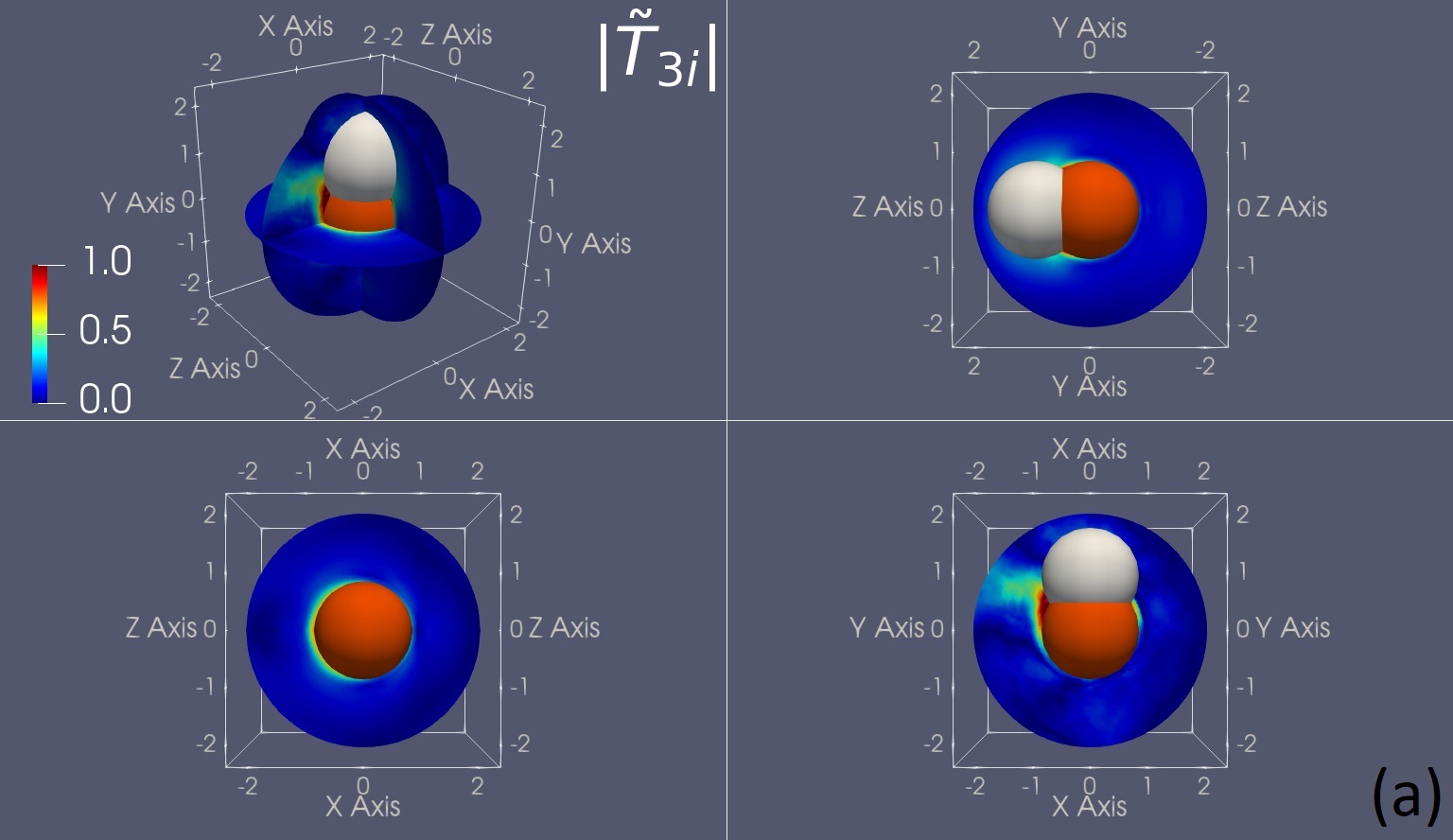}
       \includegraphics[width=0.49\textwidth,keepaspectratio=true]{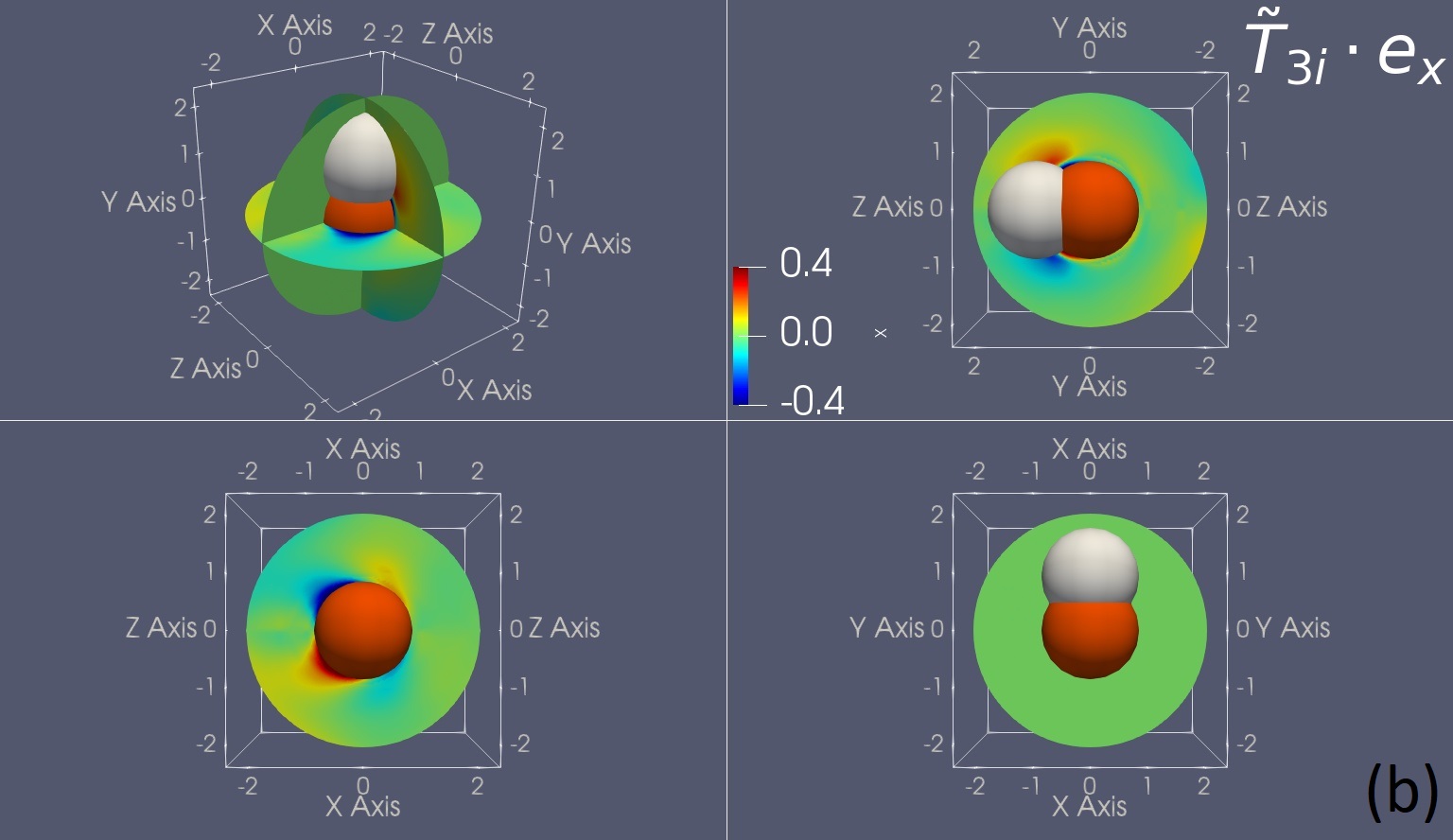}\vspace{0.5mm}
       \includegraphics[width=0.49\textwidth,keepaspectratio=true]{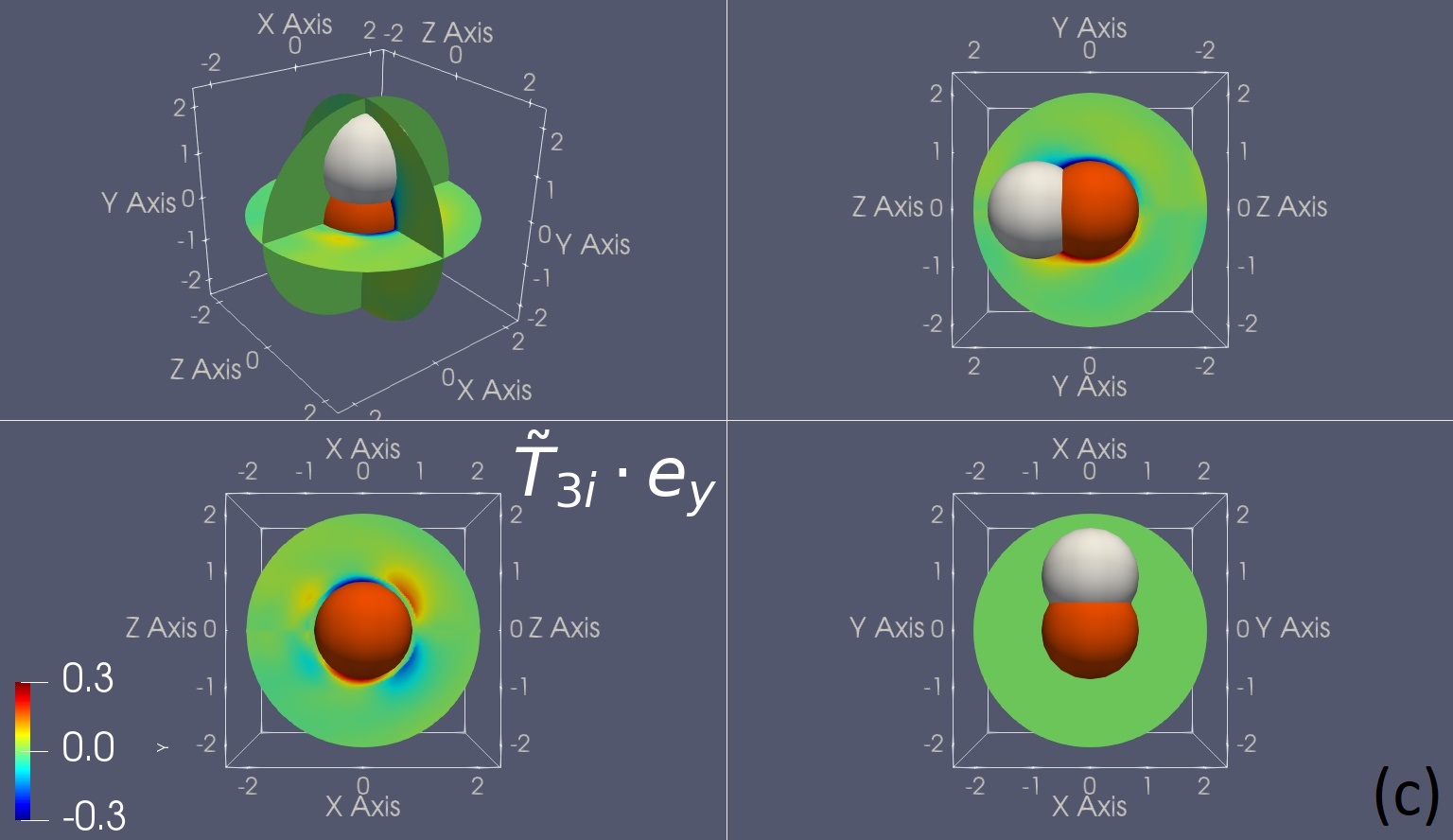}
       \includegraphics[width=0.49\textwidth,keepaspectratio=true]{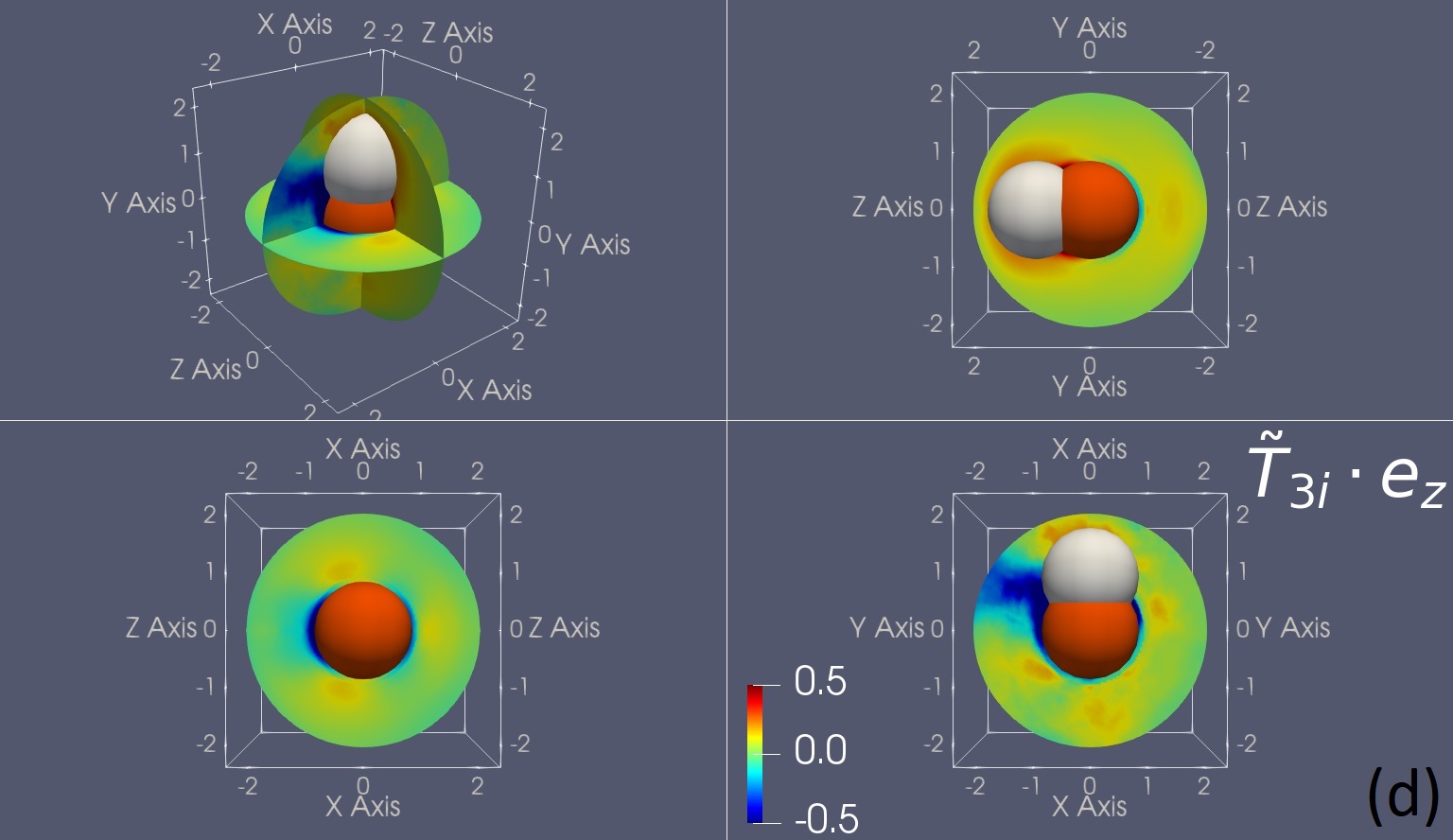}
    \end{subfigure}
    \caption{Trinary-interaction torque influence with transverse first neighbor (denoted with white) at (0, 1.1, 0) relative to reference particle (represented with red). (a) Magnitude of flucuations ($|\tilde{\bm{T}}_{3i}|$), (b) Torque perturbation along $x$ ($\tilde{\bm{T}}_{3i} \cdot \bm{e}_x$), (c) Torque perturbation along $y$ ($\tilde{\bm{T}}_{3i} \cdot \bm{e}_y$), and (d) Torque perturbation along $z$ ($\tilde{\bm{T}}_{3i} \cdot \bm{e}_z$).}
    \label{fig:torque_config_3}
\end{figure}

The torque fluctuation maps $\tilde{\bm{T}}_{3i}$ for $M_3=10$ are shown in Figures \ref{fig:torque_config_1}, \ref{fig:torque_config_2} \& \ref{fig:torque_config_3}. As the flow direction and the line of separation to the first neighbor are parallel in both the upstream (Figure \ref{fig:torque_config_1}) and downstream (Figure \ref{fig:torque_config_2}) configurations, the trinary model by construction should not produce any torque fluctuation along the streamwise direction for any positioning of the second neighbor, which can be observed to be satisfied in the torque maps. Unlike the trinary force maps, the trinary torque maps exhibit more coherent physics in regions that are in close proximity to both the reference particle and the first neighbor.

\section{Conclusions and Future Scope}
Euler-Lagrange (EL) methodology is an accurate and scalable approach that can be utilized to simulate particle-laden flow systems of many natural and industrial applications. The method achieves a desirable balance of accuracy and computational cost, primarily by considering the particles of the system as point-particles. The point-particle assumption comes with the requirement of closure models for particle forces and torques. The accuracy of point-particle closures determines the ability of an EL simulation to replicate actual flow dynamics. Standard EL approaches only consider mesoscale information of Reynolds number ($Re$) and particle volume fraction ($\phi$), and do not consider the deterministic influence induced by nearby particles (neighbors) in approximating force and torque of a particle of interest (reference particle). However, recent studies have elucidated the importance of neighbor-induced perturbations and several data-driven modeling efforts with aid of particle-resolved (PR) simulations have developed deterministic force/torque models that take into account the neighbor interactions. A consistent conclusion of these studies has been that to achieve robustness it is important to integrate underlying physics in the prediction process in order to overcome the paucity of PR simulation data.

Including the effect of neighbors in a deterministic way leads to a multi-body interaction problem. Past physics-based efforts have limited the interaction to only pairwise leading to a binary approximation. The current work presents a hierarchical approach that systematically considers terms of increasing complexity, starting from unary, to binary, to trinary interaction model. We also present neural network architectures that implement these hierarchical interactions to provide accurate predictions for a large range of mesoscale variables. While Galilean, translational and scale invariances are preserved by the theoretical formulation, the neural networks have been designed to preserve rotational and reflectional symmetries. These unique characteristics ensure that the systematic approach achieves robustness even with limited training data.

The current method demonstrates excellent universal predictions for $0.25 \le Re \le 250$ and $0 \le \phi \le 0.4$. The systematic approach of considering different levels of interactions separately has provided interpretability, and thereby yielding a thorough understanding of the models' applicability. Influence of each individual neighbor (binary-interaction), and the influence of a pair of neighbors (trinary-interaction) on a reference particle are visualized to understand the physics of neighbor perturbation and exemplify the interpretability of the hierarchical approach.   

The success of the current approach in achieving accurate predictions for particle distributions obtained from two different sources provides promise for the inclusion of other homogeneous particle distributions such as structured/ordered particle arrays in training and testing the current networks. The generalizable binary-interaction models achieved in this work might suggest that reasonably accurate binary-interaction models can be produced from $O(10^3)$ particle systems. Performing enhanced EL simulations using these rich models and evaluating their performance is an important future direction. Also, developing stochastic Euler-Euler (EE) models based on these improved EL simulations is also an exciting idea. 

\begin{acknowledgements}
This work was supported by the Office of Naval Research (ONR) as part of the Multidisciplinary University Research Initiatives (MURI) Program, under grant number N00014-16-1-2617. This work was also partly benefited from the U.S. Department of Energy, National Nuclear Security Administration, Advanced Simulation and Computing Program, as a Cooperative Agreement to the University of Florida under the Predictive Science Academic Alliance Program, under Contract No. DE-NA0002378. B.S. would like to thank Majid Allahyari for generously sharing particle-resolved simulation data generated using NekIBM. We also thank Arman Seyed-Ahmadi and Anthony Wachs for generously providing access to their data.
\end{acknowledgements}

\section*{Data availability}
The data and code that support the findings of this study will be made available upon publication of the manuscript at \href{https://github.com/siddanib/point-particle_closures_hierarchical_approach}{https://github.com/siddanib/point-particle\_closures\_hierarchical\_approach}. 

\appendix
\section{Neighbor-truncation error analysis}\label{app_A}
The influence of neighbors considered in binary and trinary interactions is expected to be different for each output quantity. Hence, it is important to quantify the improvement achieved by adding additional neighbors so that important deployment decisions can be made based on desired accuracy and computational cost associated with it. In this section we provide a neighbor-truncation analysis for binary and trinary force and torque models. The strategy of $(k=5)$-fold cross-validation is employed to obtain all results presented below.

\subsection{Binary models neighbor-truncation error}
As part of neighbor-truncation error analysis for the binary term we investigate the performance obtained using $M_2 = \{10,15,20,26\}$. The models are trained and tested with the same number of neighbors, i.e. $M_2$ is fixed. Training and testing performance of each individual dataset, in terms of $R^2$, achieved with different $M_2$ values for binary force model ($\tilde{\bm{F}}[2]$) are tabulated in Table \ref{tab:binary_force_neig_trunc}. It is clear that both training and testing performance increases with an increase in $M_2$ for every dataset. This suggests that the learnability and generalizability increase with increasing number of neighbors. The average test performance over all datasets for drag and lift components is mentioned below, and the values are presented for an ascending order of $M_2$. $ \Bar{R}^2_{\mathrm{Drag, test}} = \{ 0.601, 0.667, 0.701, 0.728\}$ and $\bar{R}^2_{\mathrm{Lift, test}} = \{ 0.623, 0.686, 0.705, 0.721 \}$. It can be observed that the net improvement achieved between consecutive values of $M_2$ is decreasing, and the final increment only being around 2-3\% for both components. This is an indication that adding more neighbors ($M_2 > 26$) will yield decreasing improvement for the binary force model.

It was earlier noticed in binary torque maps that the radius of influence is substantially smaller for all volume fractions when compared with corresponding binary force maps. Neighbor-truncation analysis of $\tilde{\bm{T}}[2]$ presented in Table \ref{tab:binary_torque_neig_trunc} confirms this observation. Test performance averaged over all datasets of transverse torque components yield $\Bar{R}^2_{\mathrm{Torque},\perp,\mathrm{test}} = \{0.672, 0.668,0.675,0.677\}$ for $M_2 = \{10,15,20,26\}$, indicating that improvement beyond $M_2=10$ is almost negligible. 
\begin{table}
    \caption{Neighbor-truncation error analysis of $\tilde{\bm{F}}[2]$ model presented for each dataset in terms of $R^2$.}
    \label{tab:binary_force_neig_trunc}
    \begin{center}
    \begin{ruledtabular}
    \begin{tabular}{cccccccccccccccccc}
    \rule{0pt}{4ex}$\langle Re \rangle$ & $\langle \phi \rangle$ & \multicolumn{4}{c}{$\tilde{\bm{F}}_i[M_2=10,2]$} & \multicolumn{4}{c}{$\tilde{\bm{F}}_i[M_2=15,2]$} & \multicolumn{4}{c}{$\tilde{\bm{F}}_i[M_2=20,2]$} &  \multicolumn{4}{c}{$\tilde{\bm{F}}_i[M_2=26,2]$} \\
     & & \multicolumn{2}{c}{Drag} & \multicolumn{2}{c}{Lift} & \multicolumn{2}{c}{Drag} & \multicolumn{2}{c}{Lift} & \multicolumn{2}{c}{Drag} & \multicolumn{2}{c}{Lift} & \multicolumn{2}{c}{Drag} & \multicolumn{2}{c}{Lift} \\
     & & Train & Test & Train & Test & Train & Test & Train & Test & Train & Test & Train & Test & Train & Test & Train & Test  \\[2ex] \hline
     \rule{0pt}{4ex}9.86 & 0.10 & 0.669 & 0.611 & 0.730 & 0.708 & 0.712 & 0.681 & 0.770 & 0.750 & 0.743 & 0.708 & 0.783 & 0.763 & 0.775 & 0.740 & 0.797 & 0.776  \\
     121.36 & 0.10 & 0.599 & 0.542 & 0.677 & 0.663 & 0.635 & 0.595 & 0.690 & 0.681 & 0.664 & 0.634 & 0.703 & 0.687 & 0.685 & 0.645 & 0.707 & 0.696 \\
     6.95 & 0.21 & 0.663 & 0.633 & 0.716 & 0.697 & 0.754 & 0.736 & 0.768 & 0.750 & 0.792 & 0.781 & 0.794 & 0.782 & 0.816 & 0.809 & 0.812 & 0.800 \\
     73.40 & 0.21 & 0.639 & 0.597 & 0.713 & 0.693 & 0.712 & 0.682 & 0.732 & 0.714 & 0.741 & 0.719 & 0.738 & 0.730 & 0.757 & 0.728 & 0.746 & 0.739 \\
     27.81 & 0.40 & 0.655 & 0.625 & 0.663 & 0.654 & 0.734 & 0.711 & 0.744 & 0.731 & 0.772 & 0.755 & 0.775 & 0.764 & 0.807 & 0.788 & 0.795 & 0.784 \\
     73.42 & 0.40 & 0.622 & 0.600 & 0.648 & 0.639 & 0.692 & 0.672 & 0.707 & 0.698 & 0.731 & 0.712 & 0.736 & 0.725 & 0.759 & 0.738 & 0.754 & 0.738 \\
     2.20 & 0.10 & 0.705 & 0.695 & 0.669 & 0.663 & 0.771 & 0.761 & 0.738 & 0.734 & 0.809 & 0.802 & 0.775 & 0.773 & 0.837 & 0.833 & 0.803 & 0.800 \\
     10.92 & 0.10 & 0.720 & 0.714 & 0.782 & 0.775 & 0.745 & 0.744 & 0.795 & 0.791 & 0.781 & 0.778 & 0.805 & 0.801 & 0.810 & 0.807 & 0.812 & 0.808 \\
     165.96 & 0.10 & 0.555 & 0.541 & 0.712 & 0.710 & 0.584 & 0.568 & 0.711 & 0.713 & 0.604 & 0.592 & 0.706 & 0.708 & 0.624 & 0.611 & 0.715 & 0.718 \\
     0.25 & 0.20 & 0.600 & 0.584 & 0.604 & 0.598 & 0.686 & 0.672 & 0.688 & 0.682 & 0.730 & 0.717 & 0.728 & 0.722 & 0.756 & 0.747 & 0.757 &  0.753 \\
     2.48 & 0.20 & 0.630 & 0.623 & 0.629 & 0.620 & 0.702 & 0.698 & 0.708 & 0.701 & 0.744 & 0.737 & 0.739 & 0.733 & 0.780 & 0.773 & 0.764 & 0.758 \\
     49.78 & 0.20 & 0.634 & 0.634 & 0.695 & 0.695 & 0.687 & 0.690 & 0.700 & 0.704 & 0.720 & 0.717 & 0.706 & 0.708 & 0.747 & 0.743 & 0.716 & 0.719  \\
     187.25 & 0.20 & 0.564 & 0.555 & 0.660 & 0.659 & 0.614 & 0.605 & 0.659 & 0.660 & 0.634 & 0.630 & 0.664 & 0.668 & 0.662 & 0.656 & 0.668 & 0.671 \\
     3.25 & 0.40 & 0.536 & 0.515 & 0.439 & 0.414 & 0.631 & 0.604 & 0.527 & 0.507 & 0.667 & 0.633 & 0.567 & 0.543 & 0.698 & 0.672 & 0.592 & 0.567  \\
     64.35 & 0.40 & 0.624 & 0.609 & 0.596 & 0.584 & 0.701 & 0.685 & 0.635 & 0.626 & 0.725 & 0.705 & 0.644 & 0.629 & 0.744 & 0.727 & 0.655 & 0.643  \\
     245.65 & 0.40 & 0.556 & 0.541 & 0.527 & 0.513 & 0.600 & 0.574 & 0.550 & 0.538 & 0.630 & 0.596 & 0.564 & 0.548 & 0.661 & 0.632 & 0.574 & 0.558  \\
    \end{tabular}
    \end{ruledtabular}
    \end{center}
\end{table}

\begin{table}
    \caption{Neighbor-truncation error analysis of $\tilde{\bm{T}}[2]$ model for each dataset in terms of $R^2$.}
    \label{tab:binary_torque_neig_trunc}
    \begin{center}
    \begin{ruledtabular}
    \begin{tabular}{cccccccccc}
     \rule{0pt}{4ex}$\langle Re \rangle$ & $\langle \phi \rangle$ & \multicolumn{2}{c}{$\tilde{\bm{T}}_i[M_2=10,2]$} &  \multicolumn{2}{c}{$\tilde{\bm{T}}_i[M_2=15,2]$} & \multicolumn{2}{c}{$\tilde{\bm{T}}_i[M_2=20,2]$} & \multicolumn{2}{c}{$\tilde{\bm{T}}_i[M_2=26,2]$}\\
     & & \multicolumn{2}{c}{Torque,$\perp$} & \multicolumn{2}{c}{Torque,$\perp$} & \multicolumn{2}{c}{Torque,$\perp$} & \multicolumn{2}{c}{Torque,$\perp$}\\
     & & Train & Test & Train & Test & Train & Test & Train & Test \\[2ex] \hline
     \rule{0pt}{4ex}9.86 & 0.10 & 0.745 & 0.747 & 0.745 & 0.748 & 0.751 & 0.748 & 0.752 & 0.751 \\
     121.36 & 0.10 & 0.306 & 0.281 & 0.298 & 0.275 & 0.307 & 0.285 & 0.308 & 0.285 \\
     6.95 & 0.21 & 0.825 & 0.821 & 0.824 & 0.824 & 0.827 & 0.826 & 0.827 & 0.825 \\
     73.40 & 0.21 & 0.494 & 0.484 & 0.493 & 0.481 & 0.496 & 0.484 & 0.496 & 0.486 \\
     27.81 & 0.40 & 0.657 & 0.642 & 0.652 & 0.641 & 0.654 & 0.643 & 0.657 & 0.647 \\
     73.42 & 0.40 & 0.431 & 0.431 & 0.428 & 0.429 & 0.434 & 0.432 & 0.435 & 0.434 \\
     2.20 & 0.10 & 0.893 & 0.893 & 0.896 & 0.896 & 0.898 & 0.898 & 0.899 & 0.898  \\
     10.92 & 0.10 & 0.791 & 0.786 & 0.792 & 0.789 & 0.796 & 0.793 & 0.798 & 0.794 \\
     165.96 & 0.10 & 0.527 & 0.532 & 0.510 & 0.518 & 0.532 & 0.542 & 0.533 & 0.541 \\
     0.25 & 0.20 & 0.892 & 0.891 & 0.892 & 0.892 & 0.896 & 0.895 & 0.896 & 0.895  \\
     2.48 & 0.20 & 0.894 & 0.892 & 0.894 & 0.892 & 0.897 & 0.895 & 0.898 & 0.896  \\
     49.78 & 0.20 & 0.711 & 0.715 & 0.713 & 0.717 & 0.709 & 0.711 & 0.713 & 0.717 \\
     187.25 & 0.20 & 0.586 & 0.584 & 0.571 & 0.571 & 0.593 & 0.593 & 0.595 & 0.595 \\
     3.25 & 0.40 & 0.817 & 0.813 & 0.819 & 0.812 & 0.816 & 0.809 & 0.820 & 0.813 \\
     64.35 & 0.40 & 0.690 & 0.688 & 0.680 & 0.674 & 0.694 & 0.689 & 0.694 & 0.685 \\
     245.65 & 0.40 & 0.555 & 0.554 & 0.528 & 0.524 & 0.561 & 0.561 & 0.565 & 0.564 \\
    \end{tabular}
    \end{ruledtabular}
    \end{center}
\end{table}

\subsection{Trinary models neighbor-truncation error}
The neighbor-truncation analysis of trinary models is carried out by fixing $M_2=26$. Only two different values of $M_3 = \{5,10\}$ are considered to reduce the computational cost associated with higher values of $M_3$. The performance of $\tilde{\bm{F}}[3]$ model is evaluated in Table \ref{tab:trinary_force_neig_trunc}. The dataset-averaged test performance for drag and lift are $\Bar{R}^2_{\mathrm{Drag,test}}=\{0.729, 0.740\}$ and $\Bar{R}^2_{\mathrm{Lift,test}}=\{0.739, 0.754\}$. Although the improvement appears to be only around 1-2\% when $M_3$ is increased from 5 to 10 it must be realized that this improvement is only under conditions of limited training data. The data paucity can be observed from the large difference between training and testing performance at both values of $M_3$ for all datasets.

As mentioned earlier trinary-interaction torque model is necessary for streamwise torque component prediction. Hence, it can be seen from Table \ref{tab:trinary_torque_neig_trunc} that $\tilde{\bm{T}}[3]$ has high sensitivity with respect to $M_3$. Dataset-averaged test performance for streamwise and transverse torque components of $\tilde{\bm{T}}[3]$ is $\Bar{R}^2_{\mathrm{Torque},\parallel,\mathrm{test}} = \{0.245, 0.408\}$ and $\Bar{R}^2_{\mathrm{Torque},\perp,\mathrm{test}} = \{0.798, 0.840\}$. 16\% improvement in prediction is noticed for streamwise component when $M_3$ is increased from 5 to 10. However, the substantial difference between training and testing performance indicate data paucity.

\begin{table}
    \caption{Neighbor-truncation error analysis for $\tilde{\bm{F}}[3]$ model in terms of $R^2$.}
    \label{tab:trinary_force_neig_trunc}
    \begin{center}
    \begin{ruledtabular}
    \begin{tabular}{cccccccccc}
    \rule{0pt}{4ex}$\langle Re \rangle$ & $\langle \phi \rangle$ &  \multicolumn{4}{c}{$\tilde{\bm{F}}_i[M_2=26,M_3=5,3]$} &  \multicolumn{4}{c}{$\tilde{\bm{F}}_i[M_2=26,M_3=10,3]$}\\
    & & \multicolumn{2}{c}{Drag} & \multicolumn{2}{c}{Lift} & \multicolumn{2}{c}{Drag} & \multicolumn{2}{c}{Lift} \\
     & & Train & Test & Train & Test & Train & Test & Train & Test \\[2ex] \hline
     \rule{0pt}{4ex}9.86 & 0.10 & 0.816 & 0.732 & 0.851 & 0.797& 0.842 & 0.762 & 0.862 & 0.803 \\
     121.36 & 0.10 & 0.766 & 0.641 & 0.789& 0.701& 0.787 & 0.650 & 0.805 & 0.701 \\
     6.95 & 0.21 & 0.832 & 0.808 & 0.841 & 0.813 & 0.853 & 0.817 & 0.862 & 0.827 \\
     73.40 & 0.21 & 0.797 & 0.729 & 0.809 & 0.764 & 0.818 & 0.745 & 0.833 &  0.772 \\
     27.81 & 0.40 & 0.818 & 0.794 & 0.816 & 0.799 & 0.837 & 0.809 & 0.841 & 0.823 \\
     73.42 & 0.40 & 0.787 & 0.749 & 0.779 & 0.751 & 0.807 & 0.769 & 0.817 & 0.786 \\
     2.20 & 0.10 & 0.846 & 0.834 & 0.824 & 0.811 & 0.854 & 0.838 & 0.842 & 0.826 \\
     10.92 & 0.10 & 0.821 & 0.810 & 0.849 & 0.838 & 0.826 & 0.815 & 0.861 & 0.851 \\
     165.96 & 0.10 & 0.660 & 0.605 & 0.757 & 0.732 & 0.679 & 0.615 & 0.759 & 0.732 \\
     0.25 & 0.20 & 0.760 & 0.749 & 0.762 & 0.754 & 0.769 & 0.758 & 0.773 & 0.764 \\
     2.48 & 0.20 & 0.786 & 0.777 & 0.773 & 0.765 & 0.791 & 0.782 & 0.791 & 0.784 \\
     49.78 & 0.20 & 0.757 & 0.747 & 0.771 & 0.768 & 0.772 & 0.754 & 0.795 & 0.787 \\
     187.25 & 0.20 & 0.681 & 0.660 & 0.715 & 0.701 & 0.698 & 0.661 & 0.731 & 0.711\\
     3.25 & 0.40 & 0.702 & 0.672 & 0.598 & 0.570 & 0.722 & 0.679 & 0.625 & 0.584 \\
     64.35 & 0.40 & 0.748 & 0.728 & 0.688 & 0.670 & 0.764 & 0.739 & 0.717 & 0.696 \\
     245.65 & 0.40 & 0.677 & 0.636 & 0.619 & 0.594 & 0.698 & 0.649 & 0.654 & 0.611 \\
    \end{tabular}
    \end{ruledtabular}
    \end{center}
\end{table}

\begin{table}
    \caption{Neighbor-truncation error analysis for $\tilde{\bm{T}}[3]$ model in terms of $R^2$.}
    \label{tab:trinary_torque_neig_trunc}
    \begin{center}
    \begin{ruledtabular}
    \begin{tabular}{cccccccccc}
    \rule{0pt}{4ex}
    $\langle Re \rangle$ & $\langle \phi \rangle$ &  \multicolumn{4}{c}{$\tilde{\bm{T}}_i[M_2=26,M_3=5,3]$} &  \multicolumn{4}{c}{$\tilde{\bm{T}}_i[M_2=26,M_3=10,3]$}\\
    & & \multicolumn{2}{c}{Torque,$\parallel$} & \multicolumn{2}{c}{Torque,$\perp$} & \multicolumn{2}{c}{Torque,$\parallel$} & \multicolumn{2}{c}{Torque,$\perp$} \\ 
     &  & Train & Test & Train & Test & Train & Test & Train & Test \\[2ex] \hline
     \rule{0pt}{4ex}
     9.86 & 0.10 & 0.475 & 0.326 & 0.906 & 0.862 & 0.602 & 0.469 & 0.933 & 0.899\\
     121.36 & 0.10 & 0.368 & -0.049 & 0.798 & 0.556 & 0.467 & 0.011 & 0.835 & 0.618\\
     6.95 & 0.21 & 0.523 & 0.371 & 0.923 & 0.893 & 0.699 & 0.610 & 0.951 & 0.929\\
     73.40 & 0.21 & 0.310 & 0.070 & 0.816 & 0.723 & 0.458 & 0.234 & 0.868 & 0.774\\
     27.81 & 0.40 & 0.409 & 0.244 & 0.870 & 0.812 & 0.658 & 0.544 & 0.937 & 0.897\\
     73.42 & 0.40 & 0.324 & 0.221 & 0.772 & 0.709 & 0.571 & 0.437 & 0.864 & 0.804\\
     2.20 & 0.10 & 0.500 & 0.449 & 0.953 & 0.943 & 0.678 & 0.630 & 0.969 & 0.962\\
     10.92 & 0.10 & 0.443 & 0.395 & 0.901 & 0.891 & 0.583 & 0.549 & 0.928 & 0.918\\
     165.96 & 0.10 & 0.104 & -0.013 & 0.757 & 0.702 & 0.181 & 0.047 & 0.793 & 0.741\\
     0.25 & 0.20 & 0.452 & 0.435 & 0.938 & 0.932 & 0.673 & 0.647 & 0.954 & 0.949\\
     2.48 & 0.20 & 0.455 & 0.437 & 0.937 & 0.933 & 0.665 & 0.653 & 0.958 & 0.954\\
     49.78 & 0.20 & 0.280 & 0.245 & 0.842 & 0.832 & 0.415 & 0.363 & 0.876 & 0.864\\
     187.25 & 0.20 & 0.136 & 0.080 & 0.754 & 0.729 & 0.192 & 0.120 & 0.793 & 0.758\\
     3.25 & 0.40 & 0.384 & 0.340 & 0.879 & 0.859 & 0.661 & 0.598 & 0.916 & 0.886\\
     64.35 & 0.40 & 0.263 & 0.242 & 0.758 & 0.739 & 0.451 & 0.408 & 0.806 & 0.777\\
     245.65 & 0.40 & 0.168 & 0.122 & 0.696 & 0.660 & 0.305 &  0.208 & 0.768 & 0.715\\
    \end{tabular}
    \end{ruledtabular}
    \end{center}
\end{table}

\clearpage
\bibliography{references}

\end{document}